\documentclass[a4paper,12pt]{article}
\usepackage[dvipdfm]{graphicx}
\usepackage[usenames,dvipsnames,svgnames,table]{xcolor}
\usepackage[margin=1in]{geometry}
\usepackage{amsmath,amssymb,xcolor,graphicx,adjustbox,natbib}
\usepackage{xr}

\usepackage{caption}
\usepackage{subcaption}
\usepackage{float}

\usepackage{pifont} 

\usepackage{{Bib/rasmus2}}
\usepackage[utf8]{inputenc}
\usepackage{bbm}
\usepackage[pagebackref=true,colorlinks,linkcolor=blue,citecolor=magenta]{hyperref} 
\def \RR{{\mathbb R}}

\def \ZZ{{\mathbb Z}}

\def \dd{\mathrm{d}}
\def \oo{\mathrm{o}}

\def \M{\mathcal{M}}

\def \x{\mathbf{x}}

\def \H{{\mathcal H}}

\def \b{\mathrm{b}}

\newcommand{\bbeta}{\boldsymbol{\beta}}
\newcommand{\btheta}{\boldsymbol{\theta}}
\newcommand{\bgamma}{\boldsymbol{\gamma}}
\newcommand{\bS}{\mathbf{S}}
\newcommand{\bV}{\mathbf{V}}
\newcommand{\bZ}{\mathbf{Z}}
\newcommand{\be}{\mathbf{e}}
\newcommand{\bc}{\mathbf{c}}
\newcommand{\bd}{\mathbf{d}}
\newcommand{\thirdparr}{\}}
\newcommand{\thirdparl}{\{}

\title{Composite likelihood inference for space-time point processes}

\author{Abdollah Jalilian, stat4aj@gmail.com\\ Razi University,
  Iran\\ Lancaster Ecology and Epidemiology Group, \\
  Lancaster University, United Kingdom \and Francisco Cuevas-Pacheco, francisco.cuevas@usm.cl\\
  Departamento de Matemática, \\ Universidad T{\'e}cnica Federico Santa Mar{\'i}a, Chile \and
	Ganggang Xu, gangxu@bus.miami.edu\\ University of Miami, United States of America \and
	Rasmus Waagepetersen, rw@math.aau.dk\\Skjernvej 4A, DK-9220 Aalborg\\Aalborg University\\ Denmark}

\begin{document}

\maketitle

\begin{abstract}
The dynamics of a  rain forest is extremely complex involving
births, deaths and growth of trees with complex interactions between trees, animals, climate, and environment. We consider
the patterns of recruits (new trees) and dead trees between rain forest
censuses. For a current census we specify regression models for the
conditional intensity of recruits and the conditional probabilities of death
given the current trees and spatial covariates. We estimate regression
parameters using conditional composite likelihood functions that only
involve the conditional first order properties of the
data. When constructing assumption lean estimators of covariance
matrices of parameter estimates we only need mild assumptions of decaying
conditional correlations in space while assumptions regarding correlations over time are avoided by exploiting conditional centering of
composite likelihood score functions.
Time series of point patterns from rain forest
censuses are quite short while each point pattern covers a fairly big
spatial region. To obtain asymptotic results we therefore use a central limit theorem
for the fixed timespan - increasing spatial domain asymptotic
setting. This also allows us to handle the challenge of using stochastic covariates constructed from past point patterns. Conveniently, it suffices to impose weak
dependence assumptions on the innovations of the space-time process.
We investigate the proposed methodology by simulation studies \rwrev{and an application} to rain forest data.\\[\bsl]
\end{abstract}


\section{Introduction}

This paper develops composite likelihood methodology for analysing a
discrete-time continuous-space time series of spatial point
patterns. Our primary motivation is the need to understand the complex spatio-temporal development of a rain forest ecosystem. Essentially, this process can be
characterized in terms of growth, recruitment and mortality
\citep{wolf2005fifty,wiegand2009recruitment,shen2013forest,kohyama2018definition}. Each of these processes depend on species specific factors (e.g.\ genetics, light
requirements, seed dispersal), inter- and intraspecies interactions
(e.g.\ competition), interactions with animals, as well as exogeneous factors such as climate or weather,
soil properties, and topography
\citep{ruger2009response,habel2019new,hiura2019long}.
We leave aside the aspect of tree
 growth and confine ourselves to considering recruitment and mortality.


Extensive
rain forest tree census data have been collected within a global
network of forest research sites including the Barro
Colorado Island plot with eight censuses collected with 5 year
intervals \citep{condit2019complete}. In the ecological literature
\cite[e.g.][]{hubbell2001local,ruger2009response,johnson2017abiotic,zhu2018density,zuleta2022individual}
there is much interest in the spatial patterns of trees that
were recruited or died between consecutive censuses (see Figure~\ref{fig:data}). In particular, biologists wants to assess how the intensity of recruits and how the death probabilities depend on various covariates including soil properties, topography, and covariates representing
the influence of trees from previous censuses. The latter type of
covariates can be regarded as stochastic or auto-regressive since they
depend on the process of recruitment and mortality up to the current
time interval. We model the intensity of recruits and the death probabilities using parametric log-linear and logistic regression models. This enables biologists to study the
impact of covariates through estimates of regression
parameters in these models.  

We estimate regression parameters using conditional composite
likelihood functions. This is computionally very efficient since the composite likelihoods can be maximized using
standard efficient logistic regression software.
We obtain asymptotic distributions of regression parameter
estimates using asymptotic results for sequences of conditionally centered random fields with
increasing spatial domain but fixed time horizon
\citep{jalilian:poinas:xu:waagepetersen:23}. The aforementioned
stochastic covariates do not violate conditional centering and are
easily accommodated by the asymptotic framework. This
enables us to construct variance estimators  and confidence intervals so that significance of covariates can be
assessed.

To enhance robustness
to model misspecification, we, inspired by approaches in econometrics
\citep{white:80,conley:10}, avoid 
parametric modeling of second-order properties or further
distributional characteristics of the spatial patterns of recruitment and mortality. Instead we  estimate  variance matrices of parameter estimates by exploiting conditional
centering of the conditional composite likelihood score functions and model free
estimators \citep{conley:10,coeurjolly:guan:14} of conditional
variance matrices for the score functions. For the space-time
correlation structure we only need mild assumptions of spatial
decay of correlations for recruits and death events in each census
interval conditional on the previous state of the forest. The
estimation of variance matrices is computationally efficient. This is
because it only involves pairs of recruits or of dead trees that are
close in space within each of the time intervals rather than all pairs of points across all spatial distances and all time intervals.


Logistic regression models for mortality are used
extensively in the ecological literature
\citep{comita2009local,johnson2017abiotic,zhu2018density,zuleta2022individual}
with various approaches to handling spatial correlation including block
bootstrap or random effects associated with grid cells
partitioning the study region. \cite{rathbun1994space} and
\cite{hubbell2001local} instead model spatial correlation between
deaths using auto-logistic models and implement approximate maximum
likelihood estimation using computationally heavy Markov chain Monte
Carlo (MCMC) methods. 

Regarding recruits, \cite{rathbun1994space} consider a single generation
 and explicitly
model dependence between recruits using a spatial Cox point process and implement parameter estimation using MCMC. \cite{wiegand2009recruitment}
and \cite{getzin2014stochastically}  use spatial point process summary
statistics to investigate associations between recruits and adults and
test independence by randomization of the recruits. Regression
modeling is not used in these papers. \cite{ruger2009response} use a negative binomial regression to model
effects of light availability on grid cell
counts of recruits implementing inference using
MCMC. \cite{brix2001space} consider a
multi-type spatio-temporal log Gaussian Cox process for modeling
weed recruits on a barley field and use minimum contrast parameter
estimation. However, their approach does not accommodate covariates. Within rain forest ecology, \cite{may:huth:wiegand:15} use approximate Bayesian computation for a discrete time - continuous space model based on ecological neutral theory \citep{hubbell2001unified}. 

In contrast to previous approaches to analyzing recruitment and
mortality we avoid potentially restrictive assumptions regarding the
space-time correlation structure, we avoid dependence on choice of grids
for random effects or counts, and we avoid computationally intensive bootstrap,
approximate Bayesian, or MCMC methods. We demonstrate the usefulness and validity of our approach by application to simulated and rain forest data.

\section{Space-time model for rain forest census data}

We consider marked spatial point pattern datasets originating from
censuses that record location $u=(u_1,u_2)\in\RR^2$, species
$s\in\{1,\ldots,p\}$ and possibly further marks $m\in\mathcal{M}$ for
all trees in a research plot.  The mark $m$ for a tree could \rwrev{represent size} in terms of diameter at breast height and we assume for specificity that $\mathcal{M}=\R_+$. Such
data can be viewed as a time series of  multivariate marked space-time
point processes, $X=\{X_t\}_{t \in T}$, where $X_t=(X_t^{(1)},\ldots,X_t^{(p)})$  and $X^{(s)}_t$ is
the marked point process consisting of marked points $x=(u,m)$ at time
$t \in T$ for species $s$ \citep{diggle2013statistical,gonzalez2016spatio}.
The distribution of the point process $X$ may depend on a space-time
process $Z=\{Z_t\}_{t \in \RR}$, where $Z_t=\{\bZ_t(u)\}_{u \in \RR^2}$
and $\bZ_t(u) = (Z_{t}^{(1)}(u),\ldots, Z_{t}^{(q)}(u))^\T$ is a vector of
$q \ge 1$ environmental covariates at location $u\in\RR^2$ at time
$t$.

We assume that the time index set $T$ consists of equidistant time
points $t_k=\Delta k$, $k=0,1,\ldots,$ for some $\Delta >0$, where we
henceforth take $\Delta=1$. For any $k \ge 0$, we let the `observation history'
$\H_{k}$ denote the information given by $X_0,\ldots,X_{k}$ and
$Z_0,\ldots,Z_{k}$. 
In practice we only observe $X$ and $Z$ within $W = \widetilde W \times \M$ for a bounded $\widetilde W\subset\RR^2$ (typically a
rectangle) and for a finite number $K+1$ of observation times
$0,1,\ldots,K$. More precisely, for an observation time $0\le k \le K$, we
observe those marked points $x=(u,m)$ in $X_{k}^{(s)}$ and covariate
vectors $\bZ_{k}(v)$ where $u,v \in \widetilde W$. 

We focus on statistical modeling of recruits and deaths of trees for a single species but our models
for recruit intensities and death probabilities may in general depend
on existing trees of other species. Without loss of generality we consider the first species $s=1$ and for two consecutive observation times $k-1$ and $k$ we let
$B_{k}=X_{k}^{(1)}\backslash X_{k-1}^{(1)}$ and $D_{k}=X_{k-1}^{(1)}\backslash
X_{k}^{(1)}$  denote the recruitment and mortality
processes of species $s=1$ over the interval $]k-1,k]$.

The spatial pattern of recruits $B_{k}$  often exhibit  clustering around parent plants (plants that reached reproductive size at the previous census)  due to seed dispersal and favorable soil conditions  \citep{wiegand2009recruitment}. Moreover, when regeneration occurs in canopy gaps, recruits tend to aggregate and be positively correlated with dead trees  \citep{wolf2005fifty}.
The spatial pattern of deaths $D_{k}$
is influenced by biotic and abiotic factors such as intra- and
inter-specific interactions as well as environmental factors
\citep{shen2013forest}. The negative density-dependent
mortality hypothesis for example implies  clustering of dead individuals and
repulsion between surviving and dead individuals
\citep[][p.~210]{wiegand2013handbook}. 

In the next section we specify models for the intensity functions of
recruits $B_{k}$ and probabilities of death for trees $x\in
X_{k-1}^{(1)}$ over intervals
$]k-1,k]$, \rwrev{conditional on  $\H_{k-1}$}. We do not
impose assumptions regarding the dependence structure of the recruit
and mortality processes until Sections~\ref{sec:estimationcovmatrix}
and \ref{sec:asymptotics} where spatially decaying
correlations and spatial mixing are
needed for the conditional distributions of recruits and deaths at
time $k$ conditional on $\H_{k-1}$, $k=1,\ldots,K$.

\subsection{Models for recruit intensity and death probabilities}\label{sec:recruitmodel}

Given $\H_{k-1}$, we model the recruits $B_{k}$ as
an inhomogeneous point process on $\R^2 \times \M$ with  log linear intensity function
of the form
\begin{equation}\label{eq:recruits_model}
  \zeta_k(x|\H_{k-1}) = f(m)\exp\left[\beta_{\b 0,k}  +\bZ_{k-1}^\T(u)\bbeta_{\mathrm{b}}
      +   \bc_{k-1}(x)^\T \bgamma_\b \right ], \quad x=(u,m) \in \R^2 \times \M,
\end{equation}
where $f$ is a probability density for the marks,  $\beta_{\b 0,k}$ is a time-dependent intercept, $\bbeta_{\mathrm{b}}\in\RR^{q}$ and $\bgamma_{\b} \in \RR^{p}$ are
vectors of regression parameters, and
$\bc_{k-1}(x)=(c^l_{k-1}(x))_{l=1}^p$ where $c^{l}_{k-1}(\cdot)$ represents the `influence' of
trees of species $l$ on the intensity of recruits. In general, the
intensity $\zeta_k(\cdot|\H_{k-1})$ may not capture all sources of
variation of the recruits. It is therefore \rwrev{important} to recognize the possibility of stochastic dependence between recruits conditional on $\H_{k-1}$. 
In \rwrev{forest ecology, the marks of recruits}
are often quite similar and in the following we \rwrev{focus} on
estimation of the log linear part treating $f(m)$ as a known
density. 

For $x =(u,m) \in X_{k-1}^{(1)}$ we define $I_{k}(x)=1[x \in D_k]$ to be an indicator of death. The $I_{k}(x)$, $x \in X_{k-1}^{(1)}$,
are Bernoulli random
variables with death probabilities
$p_k(x|\H_{k-1})=P(I_{k}(x)=1|\H_{k-1})$. We do not assume
that the death indicators are independent.
We model the death probabilities by logistic regressions
$p_k(x|\H_{k-1}) = \exp\left [ \eta_k(x|\H_{k-1}) \right ] / \thirdparl
1
    + \exp\left [ \eta(x|\H_{k-1}) \right ] \thirdparr$ with
\begin{equation}\label{eq:deathseta}
  \eta_k(x|\H_{k-1}) =  \beta_{\dd 0,k}+\alpha m+\bZ_{k-1}^\T(u) \bbeta_{\mathrm{d}} +  \bd_{k-1}(x)^\T \bgamma_{\dd},
\end{equation}
where  $\beta_{\dd 0,k}$ is a time-dependent intercept, $\alpha$ is a regression parameter for the mark $m$, $\bbeta_{\mathrm{d}}\in\RR^{q}$ and $\bgamma_{\dd} \in \rwrev{\RR^{p}}$ are
vectors of regression parameters, and
$\bd_{k-1}(x)=(d^{(l)}_{k-1}(x))_{l=1}^p$ where $d^{(l)}_{k-1}(x)$ represents the `influence' of
trees of species $l$ on death of a tree $x$.

	Unlike the spatial covariates $\bZ_{k-1}(u)$ which are commonly assumed to be deterministic, we permit the influence covariates $\bc_{k-1}$ and $\bd_{k-1}$ to be stochastic being constructed (next subsection) from past marked point patterns $X_{k-1}$.

\subsection{Stochastic covariates for influence of existing trees}\label{sec:specificmodels}
\rwrev{Recruits may be positively dependent} on previous conspecific trees since
recruits arise from parent tree seed dispersal. Conversely, the impact of the remaining species could be negative due to
competition for light and other resources. There is a rich ecological
literature on models for seed dispersal kernels
\citep{nathan:etal:12,bullock:etal:19,PROENCAFERREIRA2023} and
competition indices \cite[e.g.][]{burkhart2012,britton:richards:hovenden:23}.

\rwrev{We assume that the influence of an existing conspecific forest stand
on a recruit  $x=(u,m)$ is a function of the distance from $x$ to the
nearest neighbour in the forest stand. That is,}
\begin{equation}\label{eq:seedkernel}   c_{k-1}^1(x)=  \exp \thirdparl
	- [ d(x,X_{k-1}^{(1)}) ]/\psi_1 )^{2} \thirdparr, 
\end{equation}
with $d(x,X^{(1)}_{k-1}) = \min_{(u',m') \in X^{(1)}_{k-1}} \|u - u'\|/m'$
being a mark-weighted spatial distance between the marked point  $x=(u,m)$ and the point
pattern $X^{(1)}_{k-1}$, while $\psi_1>0$ controls the range of
effect of the existing conspecific trees.  Thus, the influence of the existing trees is determined by a
Gaussian dispersal kernel \citep{PROENCAFERREIRA2023} placed at the
location of the nearest existing tree.

\rwrev{For the impact of competition from existing trees we follow \cite{burkhart2012} and use for
$c_{k-1}^{l}(x)$, $l\neq 1$, and $d_{k-1}^{l}(x)$, $l=1,\ldots,p$,
$x=(u,m)$, indices of the form}
\begin{equation}\label{eq:ci} \sum_{(u',m') \in X_{k-1}^{(l)}\setminus \{(u,m)\}}
	\frac{m'}{m} \exp[-(\|u-u'\|/\kappa_l)^2] , \quad \kappa_l >0.
\end{equation}
For the data example in Section~\ref{sec:bciapplication}, $m$ represents diameter at breast height which does not vary much between recruits. Further, diameter at breast height is missing for a large proportion of the dead trees. We therefore in Section~\ref{sec:bciapplication} modified \eqref{eq:ci}
by omitting division by $m$.

\rwrev{A huge} variety of plausible models for influence of conspecific trees and competition could be proposed and
compared e.g.\ in terms of the resulting maximized composite
likelihoods. However, the exact choice of model for the influence of
existing trees is not our primary focus and we leave it to
future users to investigate \rwrev{further} models. 

\rwrev{Using stochastic covariates such as $\bc_k(\cdot)$ and
  $\bd_k(\cdot)$ in the models~\eqref{eq:recruits_model}
  and~\eqref{eq:deathseta} is conceptually straightforward but
  challenging from  a theoretical point of view}. For example, the
resulting intensity in~\eqref{eq:recruits_model} becomes stochastic,
bearing a resemblance to the conditional intensity of the Hawkes
process. However, the conditional
intensity~\eqref{eq:recruits_model} allows for a more flexible
dependence structure on the past $\H_{k-1}$. To address the
theoretical challenges, we propose in
Sections~\ref{sec:estimationcovmatrix} and \ref{sec:asymptotics} a
framework based on  conditional centering for estimating functions. 

\rwrev{
\subsection{Campbell formulas and pair correlation function}

In Sections~\ref{sec:cle} and \ref{sec:estimationcovmatrix}, for recruits $B_k$ conditional on $\H_{k-1}$, we use  the so-called first and
second order Campbell
formulas
\begin{align*}  & \EE [ \sum_{x \in B_k} h_1(x) |\H_{k-1}]= \int h_1(x)
   \zeta_k(x|\H_{k-1}) \dd x \\ &   \EE[
  \sum_{\substack{x,x' \in B_k:\\x \neq x'}} h_2(x,x') |\H_{k-1}] = \int\int  h_2(x,x')
  \zeta_k(x|\H_{k-1}) \zeta_k(x'|\H_{k-1}) g_{B_k}(x,x') \dd x \dd x' \end{align*}
for any non-negative functions $h_1$ and $h_2$ and where
$g_{B_k}(\cdot,\cdot)$ is the so-called pair correlation function of
$B_k$ conditional on $\H_{k-1}$ (the second equation is actually the
defining equation for the pair correlation function). We also use the
Campbell formulas for a Poisson process $Y_k$ of intensity
$\rho_0(\cdot)$ and independent of $\H_{k-1}$. The Campbell formulas
then become
\[ \EE  \sum_{x \in Y_k} h_1(x)= \int h_1(x) \rho_0(x) \dd x \text{ and } \EE
  \sum_{\substack{x,x' \in Y_k:\\x \neq x'}} h_2(x,x') = \int\int  h_2(x,x')
  \rho_0(x)\rho_0(x') \dd x \dd x'\]
since the pair correlation function of a Poisson process is one. \cite{moller2003statistical} provide more
details regarding Campbell formulas and pair correlation functions.}

\section{Composite likelihood estimation}\label{sec:cle}
Given observations $X_{k}$ and $Z_{k}$, $k=0,\ldots,K$, we infer
regression parameters using estimating functions derived from
composite likelihoods for the recruit  and death patterns $B_{k}$ and
$D_{k}$, $k=1,\ldots,K$. Let $\btheta_{\mathrm{b}}=(\beta_{\b
  0,1},\ldots,\beta_{\b 0,K},\bbeta_\b^\T,\bgamma_{\b}^\T)^\T$ and
$\btheta_{\mathrm{d}}=(\beta_{\dd 0,1},\ldots,\beta_{\dd
  0,K},\alpha,\bbeta_\dd^\T,\bgamma_{\dd}^\T)^\T$ denote the parameter
vectors for the recruit and mortality models. Since the models for recruits
and deaths do not share parameters, we construct separate estimating
functions for $\btheta_\b$ and $\btheta_\dd$. \rwrev{
The proposed estimating functions are unbiased,
leading to  consistent estimators of $\btheta_\b$ and
$\btheta_\dd$.} Background on composite likelihood for intensity
function estimation can be found in \cite{moller2017some}.
\subsection{Composite likelihoods for recruits at time $k$}
For the  recruits $B_k$, we  consider the following
conditional composite log likelihood
\begin{align*}
    \sum_{x \in B_{k} \cap W} \log \zeta_k(x|\H_{k-1}) -
  \int_{W} \zeta_k(x|\H_{k-1}) \dd
  x \label{eq:poisscomploglik} 
\end{align*}
 which \rwrev{would be the log likelihood if $B_k$ was a Poisson
process given $\H_{k-1}$. For a Poisson process, points
occur independently of each other.} Hence for the estimation of
$\btheta_{\mathrm{b}}$ we ignore possible dependencies between
recruits that are not explained by $\H_{k-1}$. Since
$\int_{\mathcal{M}}f(m) \dd m=1$,  the integral over $W$ reduces to an
integral over $\widetilde W$ involving just the log-linear part of
$\zeta_k(\cdot|\H_{k-1})$. The score function (gradient) is
\begin{equation}\label{eq:poisscomploglik} \sum_{x \in B_{k} \cap W}  \frac{\nabla \zeta_k(x|\H_{k-1})}{\zeta_k(x|\H_{k-1}) }  - \int_{W}  \nabla\zeta_k(x|\H_{k-1}) \dd x , \end{equation}
where $\nabla\zeta_k(x|\H_{k-1})$ denotes the gradient of
$\zeta_k(x|\H_{k-1})$ with respect to $\btheta_{\mathrm{b}}$. By the
\rwrev{first order}
Campbell formula, the score function is conditionally centered meaning
that it has expectation zero given
$\H_{k-1}$.

  In practice we need to estimate the integral in the score function~\eqref{eq:poisscomploglik}.
Following \cite{waagepetersen2008estimating} and \cite{baddeley2014logistic},
consider for each $k=1,\ldots,K$ a dummy Poisson point process $Y_{k}$ on
$W$ independent of $B_k$ and $\H_{k-1}$ and with known intensity function
$\rho_0(x)=f(m)\rho(u)$ for $x=(u,m)$.
\rwrev{By the first order Campbell formula for the union $B_k \cup Y_k$ with intensity $\zeta_k(\cdot|\H_{k-1}) + \rho_0(\cdot)$, the integral $\int_{W}
\nabla\zeta_k(x|\H_{k-1}) \dd x$ can be estimated unbiasedly by
$  \sum_{x \in (B_{k} \cup Y_{k}) \cap W } \nabla \zeta_k(x|\H_{k-1})/
\left [\zeta_k(x|\H_{k-1}) + \rho_0(x) \right ]$}. Crucially, \rwrev{after replacing the
integral with the estimate,} the resulting approximate score function
\begin{equation}\label{eq:estmeq}
  \be_{\b,k}(\btheta_{\mathrm{b}}) =  \sum_{x \in (B_{k} \cup Y_{k})
    \cap W} \left[ \frac{\nabla \zeta_k(x|\H_{k-1})}{\zeta_k(x|\H_{k-1}) } 1[x \in B_k] - \frac{ \nabla\zeta_k(x|\H_{k-1})}{\zeta_k(x|\H_{k-1}) + \rho_0(x)} \right ]
\end{equation}
is still conditionally centered. As explained in \cite{waagepetersen2008estimating} and
\cite{baddeley2014logistic}, \eqref{eq:estmeq} is formally equivalent to a logistic
regression score function and parameter estimates can be obtained
using \rwrev{standard \texttt{glm} software}. 

\subsection{Composite likelihood for deaths at time $k$}\label{sec:scoredeath}

For the mortality process $D_k$ we ignore possible dependencies between deaths and use the Bernoulli composite log likelihood function
\begin{align*}
  &  \sum_{x \in X_{k-1}^{(1)} \cap W} \left\thirdparl I_{k}(x) \log
    p_k(x|\H_{k-1}) + [1-I_k(x)] \log [1 -p_k(x|\H_{k-1} ] \right \thirdparr \\
     = &\sum_{x \in X_{k-1}^{(1)} \cap W } \left [ I_{k}(x) \eta_k(x|\H_{k-1}) - \log\left\{1 +
   \exp \left[\eta_k(x|\H_{k-1}) \right] \right\} \right]
\end{align*}
with conditionally centered composite score function
\begin{equation}\label{eq:scoredeath}
 \be_{\mathrm{d},k}(\btheta_{\mathrm{d}}) = \sum_{x \in X_{k-1}^{(1)} \cap W } \nabla \eta_k(x|\H_{k-1}) \left[ I_{k}(x) -  p_k(x|\H_{k-1})   \right],
\end{equation}
where $\nabla \eta_{\rwrev{k}}(x|\H_{k-1})$ denotes the gradient of $\eta_k(x|\H_{k-1})$ with respect to $\btheta_{\mathrm{d}}$.

\subsection{\rwrev{Conditional likelihoods and estimating functions
    based on all data}}

\rwrev{Log composite likelihoods based on all generations of recruits and
deaths are obtained by adding the one generation log composite
likelihoods derived in the previous paragraphs. This results in
estimating functions $\be_{\mathrm{o}}(\btheta_{\mathrm{o}}) =  \sum_{k=1}^{K}
  \be_{\mathrm{o},k}(\btheta_{\mathrm{o}})$, $\mathrm{o}=\mathrm{b},\mathrm{d}$. An estimate
$\widehat\btheta_{\mathrm{o}}$ of $\btheta_{\mathrm{o}}$ is obtained by
  solving $\be_{\mathrm{o}}(\btheta_{\mathrm{o}}) = 0$.
Conditional centering of \eqref{eq:estmeq} and \eqref{eq:scoredeath},
 $\EE [\be_{\b,k}(\btheta_{\mathrm{b}})|\H_{k-1}]=0$ and $\EE [\be_{\dd,k}(\btheta_{\dd})|\H_{k-1}]=0$,
implies that $\be_{\b}(\cdot)$ and $\be_{\dd}(\cdot)$ are unbiased
, $\EE \be_{\mathrm{o}}(\btheta_{\mathrm{o}}) =0$, $\oo=\b,\dd$.
}

\section{Approximate covariance \rwrev{matrices} of parameter
  estimates}\label{sec:estimationcovmatrix}
\rwrev{
According to standard estimating function theory \cite[e.g.][and Section~\ref{sec:asymptotics}]{sorensen:99}, the approximate covariance matrix of $\widehat \btheta_{\mathrm{o}}$ is given by the inverse of the Godambe matrix $\bS_{\oo}(\btheta_{\oo}) \bV_{\oo}(\btheta_{\oo})^{-1} \bS_{\oo}(\btheta_{\oo})^\T,$
where $ \bV_{\oo}(\btheta_{\oo})  = \Var \be_{\oo}(\btheta_{\oo})$ and $\bS_{\oo}(\btheta_{\oo}) = -\EE \frac{\dd}{\dd \btheta_{\oo}^\T} \be(\btheta_{\oo})$
are the variability and sensitivity matrices. 

By iterated conditioning, $ \bS_{\oo}(\btheta_{\oo}) = - \sum_{k=1}^K \EE \EE \big[ \frac{\dd}{\dd \btheta_{\mathrm{o}}^\T}
\be_{\mathrm{o},k}(\btheta_{\mathrm{o}}) | \H_{k-1} \big]$.
Moreover, conditional \rwrev{centering of $\be_{\mathrm{o},k}(\btheta_{\mathrm{o}})$ and $\be_{\mathrm{o},k'}(\btheta_{\mathrm{o'}})$} 
 implies that $\Cov\big[ \be_{\mathrm{o},k}(\btheta_{\mathrm{o}}), \be_{\mathrm{o},k'}(\btheta_{\mathrm{o'}})
\big] = 0$, $\mathrm{o}=\b,\dd$, whenever $k \neq k'$. 
This is very appealing since we avoid assumptions
regarding the correlation structure across time for the space-time
point process. It follows that $ \Var\,
\be_{\mathrm{o}}(\btheta_{\mathrm{o}}) = \sum_{k=1}^K \Var\,
\be_{\mathrm{o},k}(\btheta_{\mathrm{o}})$, $\mathrm{o}=\b,\dd$,
and again by conditional centering, $\Var\, \be_{\mathrm{o},k}(\btheta_{\mathrm{o}}) = \EE \Var \big[
  \be_{\mathrm{o},k}(\btheta_{\mathrm{o}}) | \H_{k-1} \big]$.

To estimate the approximate covariance matrix of $\widehat \btheta_{\oo}$ we thus
need estimates of the conditional expectations $-\EE \big[
\frac{\dd}{\dd \btheta_{\mathrm{o}}^\T}
\be_{\mathrm{o},k}(\btheta_{\mathrm{o}}) | \H_{k-1} \big]$ and the
conditional variances $\Var \big[
\be_{\mathrm{o},k}(\btheta_{\mathrm{o}}) | \H_{k-1} \big]$. 
For the estimation of the conditional
variances we assume that spatial correlation decays as a function of distance for recruits and
deaths at each time $k$ given the past $\H_{k-1}$. }

\subsection{\rwrev{Conditional expectation} and variance for recruits}

\rwrev{By the first order Campbell formula used for $B_k$,}
\[ - \EE \big[ \frac{\dd}{\dd \btheta_{\mathrm{b}}^\T} \rwrev{\be_{\b,k}}(\btheta_{\mathrm{b}}) | \H_{k-1}
  \big] =  \int_{W } \frac{\nabla\zeta (x|\H_{k-1})
    \big[\nabla\zeta (x|\H_{k-1}) \big]^\T \rho_0(x)}{\zeta
    (x|\H_{k-1}) \big[\zeta (x|\H_{k-1}) + \rho_0(x) \big]} \dd x ,\]
\rwrev{which by the first order Campbell formula for  $B_k \cup Y_k$  can be estimated unbiasedly by}
\[  \sum_{x \in B_k \cup Y_k} \frac{\nabla\zeta (x|\H_{k-1})
    \big[\nabla\zeta (x|\H_{k-1}) \big]^\T \rho_0(x)}{\zeta
    (x|\H_{k-1}) \big[\zeta (x|\H_{k-1}) + \rho_0(x) \big]^2} =
  \sum_{x \in B_k \cup Y_k}
  \frac{h_k(x)h_k(x)^\T}{\zeta_k(x|\H_{k-1})\rho_0(x)}  \]
where $ h_{k}(x) = \nabla \zeta_k(x|\H_{k-1})
\rho_0(x)/[\zeta_k(x|\H_{k-1}) + \rho_0(x)]$. 

\rwrev{By the first and second order Campbell formulas, the variance}
of  \eqref{eq:estmeq} is
\begin{align*}
\Var \big[ \be_{\rwrev{\b,k}} (\btheta_{\mathrm{b}}) | \H_{k-1} \big] 
= \bS_{\mathrm{b}}(\btheta_{\mathrm{b}}) + \int_{W^2} 
 h_k(x)h_k(x')^\T\left[ g_{B_{k}}(x,x') -1 \right] \dd x \dd x',
\end{align*}
where $g_{B_{k}}$ 
is the pair correlation function of
$B_k$ given $\H_{k-1}$. Following 
\cite{coeurjolly:guan:14} we estimate the last term in the variance by
\begin{align*}
\sum_{\substack{x,x'  \in  B_{k} \cup Y_k:\\\rwrev{x \neq x'}}} k(x,x')\frac{h_{k}(x) h_{k}(x')^{\T}}{\zeta_k(x|\H_{k-1}) \zeta_k(x'|\H_{k-1})}\phi(x)\phi(x'),
\end{align*}
where $k((u,m),(u',m'))=1[\|u-u'\| \le \omega]$ is a uniform kernel function
depending on a truncation distance $\omega$ to be chosen by the user, and
$\phi(x)=1$ if $x \in B_k$ and
$\phi(x)=-\rho_0(x)/\zeta_k(x|\H_{k-1})$ if $x \in Y_k$. 

The
underlying assumption of the estimator is that correlation vanishes
for large spatial lags in the sense that $g_{B_k}(x,x') \approx 1$
when \rwrev{the spatial distance between $x$ and $x'$ is large. Hence, the kernel
function eliminates} pairs of distant points which are
uncorrelated (meaning $g_{B_K}$ close to one) and only add noise to the estimate. The
unknown regression parameters appearing in $\zeta_k(\cdot|\H_{k-1})$ and
$h_k(\cdot)$ are replaced by their composite likelihood estimates. 

Crucially,
we avoid specifying a model for the pair correlation function $g_{B_k}$
which makes our variance estimate less prone to model
misspecification. In contrast to non-parametric kernel
  estimates of $g_{B_k}$ we also avoid assuming isotropy. If isotropy for $g_{B_k}$ is preferred, shape-constrained non-parametric estimators \citep{hessellund2022semiparametric,xu2023semiparametric} may be plugged in for $g_{B_k}$ in $\Var \big[ \be_{\rwrev{\b,k}} (\btheta_{\mathrm{b}}) | \H_{k-1} \big]$.
\subsection{\rwrev{Conditional expectation} and variance for death score}

\rwrev{When conditioning on $\H_{k-1}$, the index set $X_{k-1}^{(1)}$ for the
sum in \eqref{eq:scoredeath} becomes
non-random. Therefore, by standard computations for
expectations and variances of sums,}
\begin{align*}  
  -  \EE \big[ \frac{\dd}{\dd \btheta_\dd^\T} \be_{\rwrev{\dd,k}}(\btheta_\dd) \big|
  \H_{k-1} \big] &=  \sum_{x \in X_{k-1}^{(1)}\cap W} \nabla \eta_k(x|\H_{k-1})\nabla \eta_k(x|\H_{k-1})^\T  \frac{ p_k(x|\H_{k-1})^2 }{\exp_k(  \eta_k(x|\H_{k-1}))} \:\:\: \text{ and }\\
 \Var(\be_{\dd,k}(\btheta_{\mathrm{d}}) | \H_{k-1}) &= \!\! \sum_{x,x^{\prime}
  \in X^{(1)}_{k-1} \cap W} 
  \nabla \eta_k(x|\H_{k-1}) \left( \nabla \eta_k(x^{\prime}|\H_{k-1}) \right)^{\T} \Cov[I_{k}(x), I_{k}(x^{\prime})|\H_{k-1}].
\end{align*}
In the spirit of \cite{conley:10}, we \rwrev{ estimate the conditional variance by
  \[ \sum_{x,x^{\prime} \in X^{(k)}_{k-1}} 
  \nabla \eta_k(x|\H_{k-1})  \nabla \eta_k(x^{\prime}|\H_{k-1})^{\T} k(x,x^{\prime}) \left[ I_{k}(x) - p_k(x|\H_{k-1}) \right] \left[ I_{k}(x^{\prime}) - p_k(x^{\prime}|\H_{k-1}) \right], \]}
where as in the previous section, $k(\cdot,\cdot)$ is a uniform kernel used
to avoid contributions from pairs of distant points in $X_{k-1}$, and where the
unknown parameters in $p_k(\cdot|\H_{k-1})$ and $\nabla
\eta_k(\cdot|\H_{k-1})$ are replaced by their composite likelihood
estimates. \rwrev{Similar to the previous section we avoid modeling of
  the conditional covariance $\Cov[I_{k}(x),
  I_{k}(x^{\prime})|\H_{k-1}]$ and just need that the covariance
  vanishes when the spatial distance between $x$ and $x'$ increases.}
\section{Asymptotic distribution of parameter estimates}\label{sec:asymptotics}
The key elements in establishing asymptotic normality are a first order Taylor expansion of the estimating
function and asymptotic normality of the estimating
function. In addition various regularity conditions are needed \cite[e.g.][]{sorensen:99}. Given asymptotic normality of the estimating
function, the further conditions and derivations needed are quite
standard \cite[see for example][for a case with all details
provided]{waagepetersen:guan:09}. 
\rwrev{In the following we give a sketch of the
  asymptotic approach. We do not present the standard technical
  assumptions and just focus on the essential peculiarities for our setting.}

In the space-time context, several asymptotic regimes are possible. One option is
increasing $K$, i.e.\ accumulating information over time. Another is
increasing spatial domain $W$. A combination of these is also possible \citep{jalilian:poinas:xu:waagepetersen:23}. In our setting, $K$ is moderate (seven for the specific data example) and
increasing domain asymptotics therefore seems more relevant than increasing $K$. 
 Consider a sequence of observation windows $W_n$, $n=1,2,\ldots$
and add the subindex $n$ to \rwrev{$\be_{\b}(\btheta_{\b})$ and $\be_{\dd}(\btheta_{\dd})$} when the observation
window $W_n$ is considered. We further divide $\R^2$ into unit squares $C(z)=[z_1-1/2,z_1+1/2[ \times [z_2-1/2,z_2+1/2[$ for $z=(z_1,z_2) \in \Z^2$. \rwrev{Then
 $ \be_{\oo,n}(\btheta)= \sum_{z \in \Z^2}
  \sum_{k=1}^K E_{\oo,k,n}(z)$, $\oo=\b,\dd$, 
where
\[ E_{\b,k,n}(z)=\sum_{x \in (B_{k} \cup Y_{k}) \cap [C(z) \times \M] \cap W_n} \left[ \frac{\nabla \zeta_k(x|\H_{k-1})}{\zeta_k(x|\H_{k-1}) } 1[x \in B_k] - \frac{ \nabla\zeta_k(x|\H_{k-1})}{\zeta_k(x|\H_{k-1}) + \rho_0(x)} \right ] \]
and
\[E_{\dd,k,n}(z)=\sum_{x \in X_{k-1}^{(1)} \cap [C(z)\times \M] \cap W_n } \nabla \eta_k(x|\H_{k-1}) \left( I_{k}(x) -  p_k(x|\H_{k-1})   \right )  \]
are the contributions to $\be_{\b,n}(\btheta_{\b})$ and $\be_{\dd,n}(\btheta_{\dd})$ arising from the intersections of time $k$ recruits and deaths
with $[C(z) \times \M ]\cap W_n$.}

For fixed $K$, \rwrev{$\be_{\oo,n}(\btheta_{\oo})$}  can be viewed as a sum of purely spatially indexed variables \rwrev{$E_{\oo,n}(z)=\sum_{k=1}^K E_{\oo,k,n}(z)$.} It is, however, difficult to control the spatial dependence structure of these variables since spatial dependence may propagate over time. Instead, for $n \rightarrow \infty$, we invoke case (i) of the central limit theorem established in \cite{jalilian:poinas:xu:waagepetersen:23} for the sequence of
conditionally centered random fields \rwrev{$E_{\oo,k,n}=\{E_{\oo,k,n}(z)\}_{z \in
  \Z^2}$, $k=1,\ldots,K$.}

Regarding spatial dependence it then suffices to assume for
each $k=1,\ldots,K$, weak spatial dependence ($\alpha$-mixing) of $B_k$ and $D_k$ conditional on $\H_{k-1}$. For
instance, such conditional weak dependence is trivially
satisfied for $B_k$ if $B_k$ is a Poisson process conditional on $\H_{k-1}$
and could also be established if for example $B_k$ is a
Poisson-cluster process conditional on $\H_{k-1}$. Note that even in
the simple case of $B_k$ being conditionally a Poisson process and
deaths being conditionally independent, the
aggregated processes $E_{\oo,n}=\{E_{\oo,n}(z)\}_{z \in \Z^2}$,
$\oo=\b,\dd$ (with $E_{\oo,n}(z)$ defined above), have non-trivial
spatial dependence structures. We refer to  \cite{jalilian:poinas:xu:waagepetersen:23} for further technical details and discussion of assumptions.

According to the central limit theorem, 
$\bV_{\oo,n}^{-1/2}(\btheta_{\oo}^*) \be_n(\btheta_{\oo}^*)$ converges to a standard Gaussian
vector where $\btheta_{\oo}^*$ denotes the true value of $\btheta_{\oo}$, $\oo=\b,\dd$. Using standard arguments, the convergence in
distribution of $\bV_{\oo,n}(\btheta_{\oo}^*)^{-1/2} \bS_{\oo,n}(\btheta_{\oo}^*) (\widehat \btheta_{\oo,n} - \btheta_{\oo}^*)$ to a standard Gaussian vector is obtained. Hence, $(\widehat \btheta_{\oo,n} - \btheta_{\oo}^*)$ is approximately
Gaussian distributed with covariance matrix $\bS_{\oo,n}(\btheta_{\oo}^*)^{-1}
\bV_{\oo,n}(\btheta_{\oo}^*) \bS_{\oo,n}(\btheta_{\oo}^*)^{-1}$. Plugging in our estimates for
$\bS_{\oo,n}(\btheta_{\oo}^*)$ and $\bV_{\oo,n}(\btheta_{\oo}^*)$, we
obtain standard
errors and confidence intervals for the various parameters. The
sensitivity $\bS_{\oo,n}(\btheta_{\oo}^*)$ and the variance
$\bV_{\oo,n}(\btheta_{\oo}^*)$ are roughly proportional to the \rwrev{spatial
window
size $|\widetilde W_n|$ so the parameter estimation variance is
asymptotically inversely proportional to $|\widetilde W_n|$.}

\section{Simulation study}
\rwrev{
We conduct a simulation study to assess the performance of our
methodology \rwrev{and to benchmark it (Section~\ref{sec:inla})
  against existing methodology for space-time point and binary
  processes.} To emulate the expanding window asymptotics we consider
two observation windows $\widetilde W_{1} = [0, 500]\times[0, 250]$
and $\widetilde W_{2} =
[0, 1000]\times[0, 500]$. We simulate two types
of tree species, $p=2$, and estimate the parameters for the first
species. For sake of simplicity we disregard effects of marks which
are just fixed at an arbitrary value $1$. The covariate vector for recruit
intensities and death probabilities (Section~\ref{sec:recruitmodel})
is specified as $\bZ_{k-1}(u) = (Z^{(1)}(u), Z^{(2)}(u))^\T$, $u \in
\widetilde W_2$, where $Z^{(1)}(u)$ and $Z^{(2)}(u)$ are zero mean Gaussian
random fields 
(Figure~\ref{fig:covariates} in the supplementary
material). 

For the influence of the existing trees on recruits, we deviate a bit
from the description in Section~\ref{sec:specificmodels} and for both
species let $\bc_{k-1}(x)$ have components $c^l_{k-1}(x) = \exp \thirdparl
	- [ d(x,X_{k-1}^{(l)}) ]/\psi )^{2} \thirdparr$, $l=1,2$,  
as in \eqref{eq:seedkernel} with $\psi=6$. For the death probabilities we for both species  let
$\bd_{k-1}(x)$ have components $d^l_{k-1}(x) = \sum_{(u',m') \in
  X_{k-1}^{(l)}\setminus \{(u,m)\}} \frac{m'}{m}
\exp[-(\|u-u'\|/\kappa)^2]$, $l=1,2$, as in \eqref{eq:ci} 
with $m=m^{\prime} = 1$ and $\kappa = 10$ so that the practical
range of influence of an existing tree is less than $20$m. 

We simulate $K=10$ generations of recruits and deaths. For each time
step $1 \leq k \leq K$, the recruits for both species are simulated from a log
Gaussian Cox process \citep{moller1998log} with intensity function given by
\eqref{eq:recruits_model} with intercept
 $\beta_{\b0,k}=\beta_{\b0}=-6.32$, $(\beta_{\b1},\beta_{\b2})=(0,0.1)$, 
 and log-pairwise interaction function \rwrev{$\log g_{B_k}$} given by the Mat{\'e}rn
 covariance function \rwrev{(supplementary Section 1)} with variance, smoothness and correlation scale
 parameters $\sigma^2=1,\nu=1.75$ and $\xi=4$. For the first
 species $(\gm_{\b1}, \gm_{\b2})=(0.1,
 -2)$, while for the second species $(\gm_{\b1}, \gm_{\b2})=(-2,
 0.1)$. The initial point patterns $X^{(1)}_0$
 and $X^{(2)}_0$ are generated from the same log Gaussian Cox
 process but with $\gm_{\b1}=\gm_{\b2}=0$.
 For the death indicators 
 we for time $1\le k \le K$ use correlated logistic models. For both species we let
 $\beta_{\dd0,k}=\beta_{\dd0}=-0.25$,
 $(\beta_{\dd1},\beta_{\dd 2})=(0.25,0)$,
 $(\gm_{\dd1}, \gm_{\dd2})=(-0.25, 0.25)$, specify
 $\eta_k(\cdot|\H_{k-1})$ by \eqref{eq:deathseta}, and proceed as
 follows for $l=1,2$.
 \begin{enumerate}
  \item First, we simulate a zero mean Gaussian random field $U_k$, with a Mat{\'e}rn covariance function with parameters $\sigma^2=1,\nu=0.5$ and $\xi=7$.
  \item Then, for $x=(u,m) \in X^{(l)}_{k-1}$,  we compute $p_k(u) = \Phi^{-1}[U_k(u)]$ with $\Phi$ the standard normal cumulative distribution function and the logistic variable $\tau(u) = \log\left[ \frac{p_k(u)}{1 - p_k(u)} \right]$.
 \item Finally, $I_k(x)= 1[\tau(u) \leq \eta_k(u|\mathcal{H}_{k-1})]$, $x=(u,m) \in X^{(l)}_{k-1}$.
\end{enumerate} 
}

We generate $1000$ \rwrev{bivariate} space-time point patterns and compute parameter
estimates and estimates of parameter estimate covariance matrices for
each simulated space-time pattern. Figure~\ref{fig:counting_values} in the supplementary material shows the evolution of the numbers of recruits and deaths for the
windows $\widetilde W_{1}$ and $\widetilde W_{2}$.

Figures~\ref{fig:recruitsestimates} and
\ref{fig:deathestimates} in the supplementary material show
kernel density estimates of the $1000$ simulated regression parameter estimates. For
both window sizes the parameter distributions are close to normal and
with bias close to zero.
Moreover, the kernel density estimates and the numbers in
Table~\ref{tab:variance_table} show that according to asymptotic
theory, parameter estimation variance decreases at a rate inversely
proportional to window size (variances four times larger for
$\widetilde W_1$
than for $\widetilde W_2$). 
\begin{table}[h]
  \caption{Second and fifth rows: variances of the parameter estimates for the recruits and deaths for $\widetilde W_2$. Third and sixth rows: ratios of variances for $\widetilde W_1$ and $\widetilde W_2$.} \label{tab:variance_table}
  \centering
  \begin{tabular}{c|ccccc}
    & $\beta_{0{\rm{b}}}$ & $\beta_{1{\rm{b}}}$ & $\beta_{2{\rm{b}}}$ & $\gamma_{1{\rm{b}}}$ & $\gamma_{2{\rm{b}}}$ \\
 Var.\ $\widetilde W_{2}$ & 1.3e-03 & 3.2e-03 & 3.8e-03 & 9.9e-03 & 1.7e-02 \\
    Var.\ $\widetilde W_1$/ Var.\ $\widetilde W_{2}$ & 3.89 & 4.37 & 3.95 & 3.81 & 3.79 \\ \hline 
    & $\beta_{0{\rm{d}}}$ & $\beta_{1{\rm{d}}}$ & $\beta_{2{\rm{d}}}$ & $\gamma_{1{\rm{d}}}$& $\gamma_{2{\rm{d}}}$\\
 Var.\ $\widetilde W_{2}$ & 7.6e-02 & 7.6e-02 & 8.4e-02 & 2.6e-02 & 2.6e-02 \\
 Var.\ $\widetilde W_1$/ Var.\ $\widetilde W_{2}$ & 3.95 & 3.85 & 3.55 & 4.05 & 4.25
  \end{tabular}
\end{table}

\begin{figure}[!htb]
\centering
\begin{tabular}{ccccc}
   $\beta_{0\b}$ & $\beta_{1\b}$ & $\beta_{2\b}$ & $\gm_{1\b}$ & $\gm_{2\b}$ \\
        \adjustbox{valign=m,vspace=1pt}{\includegraphics[width=0.2\linewidth]{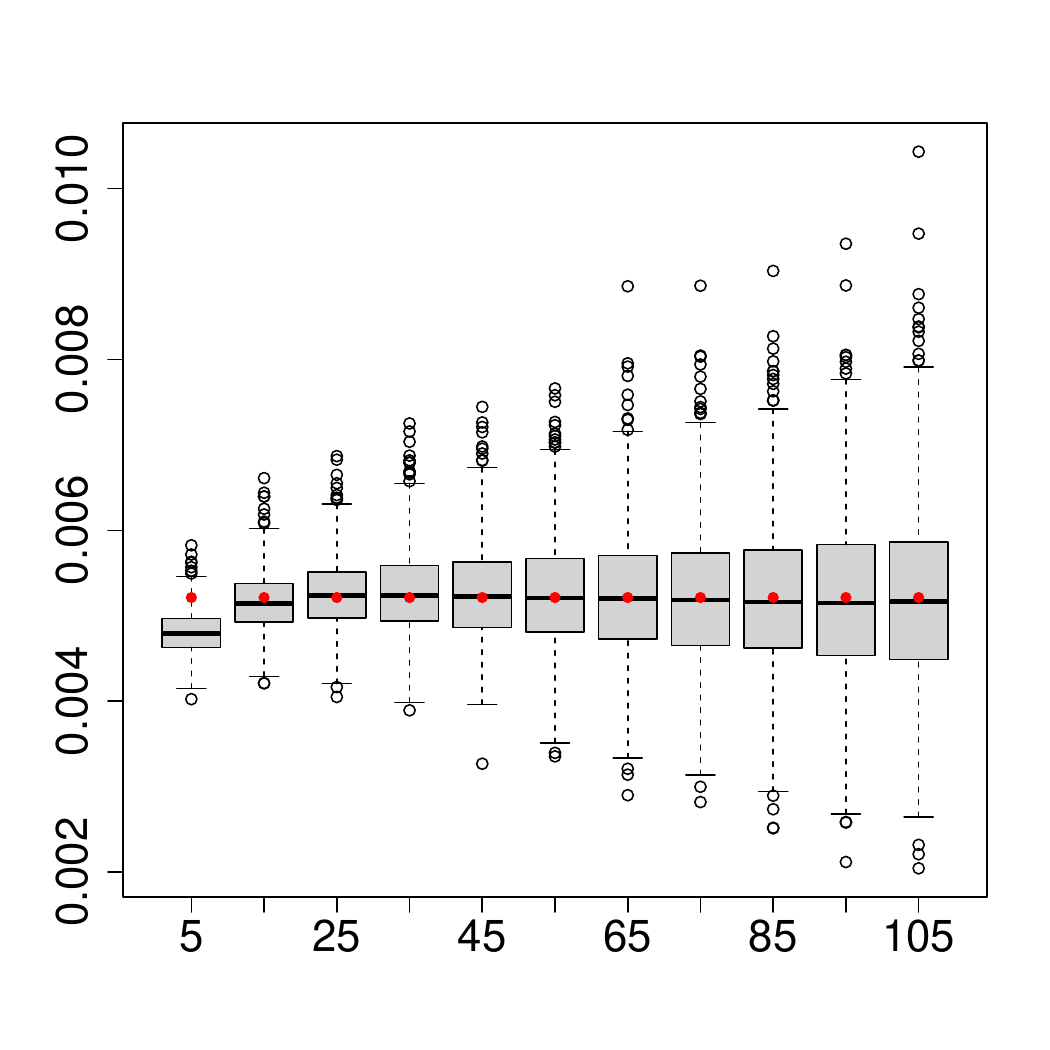}} & \adjustbox{valign=m,vspace=1pt}{\includegraphics[width=0.2\linewidth]{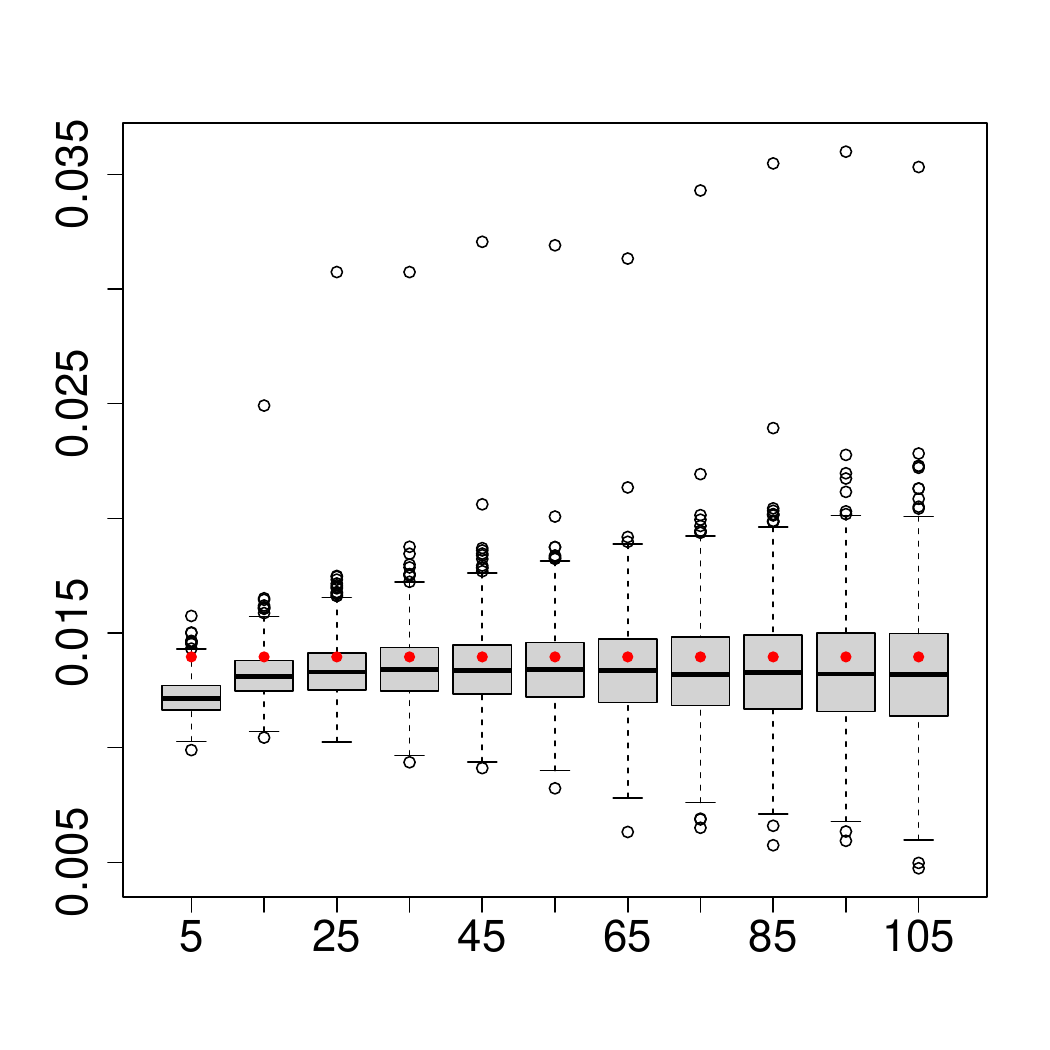}} &
        \adjustbox{valign=m,vspace=1pt}{\includegraphics[width=0.2\linewidth]{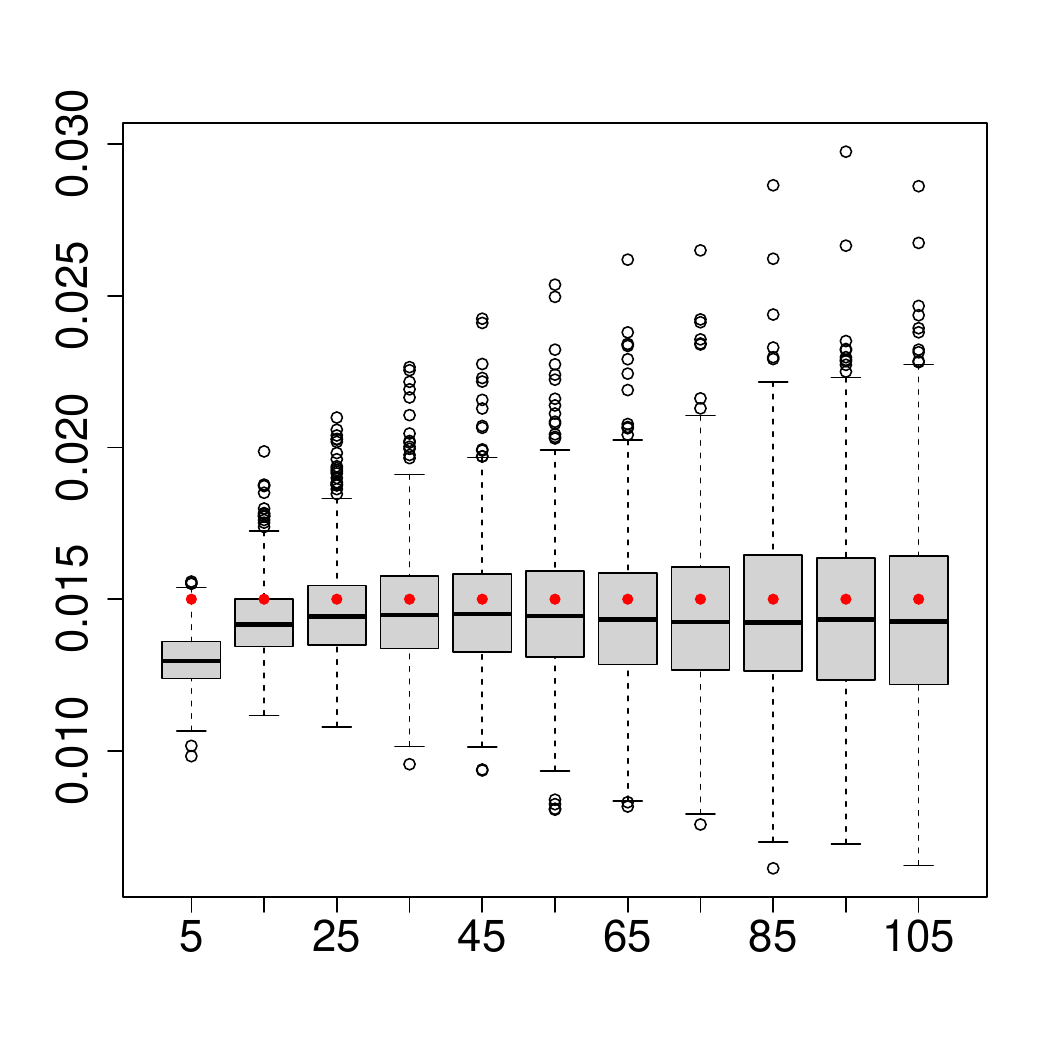}} &
        \adjustbox{valign=m,vspace=1pt}{\includegraphics[width=0.2\linewidth]{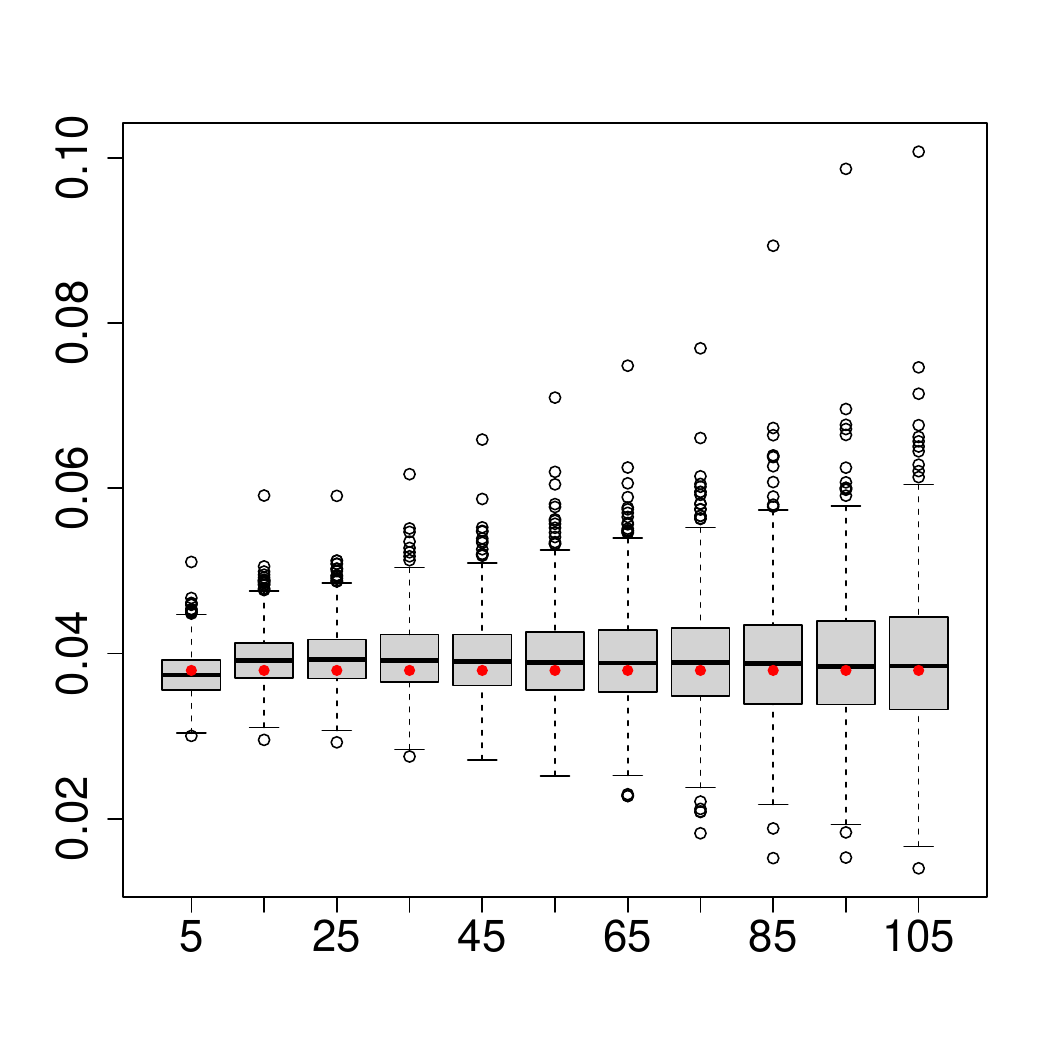}} &
        \adjustbox{valign=m,vspace=1pt}{\includegraphics[width=0.2\linewidth]{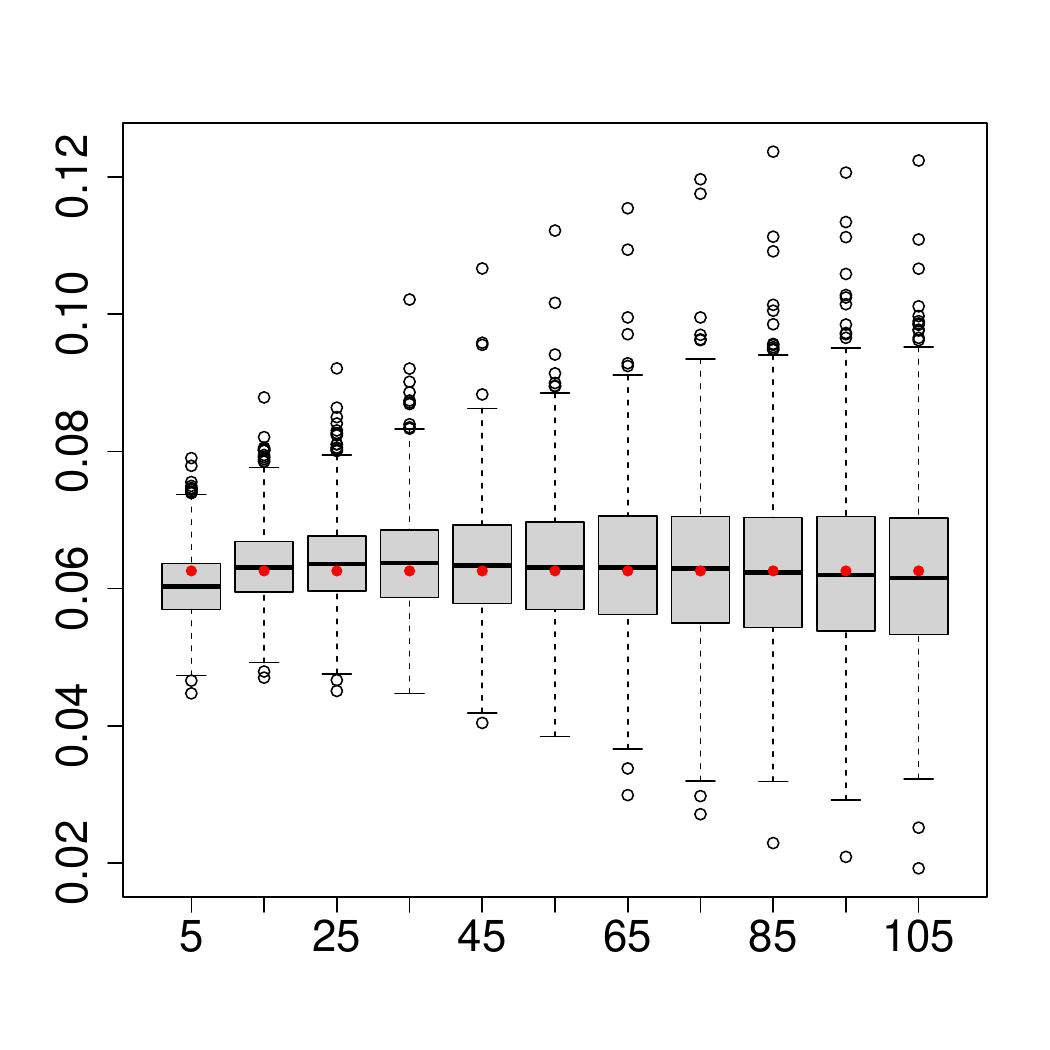}} \\
        \adjustbox{valign=m,vspace=1pt}{\includegraphics[width=0.2\linewidth]{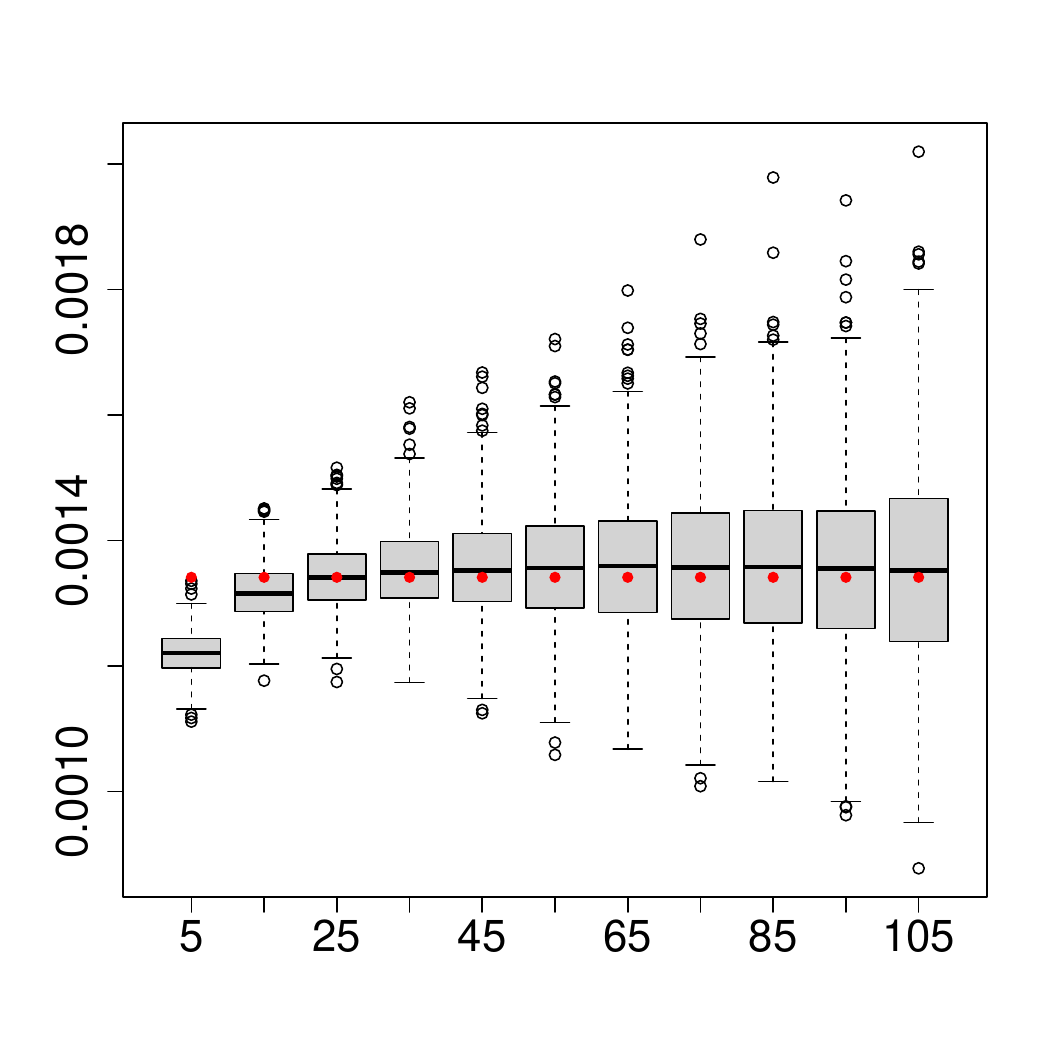}} &    \adjustbox{valign=m,vspace=1pt}{\includegraphics[width=0.2\linewidth]{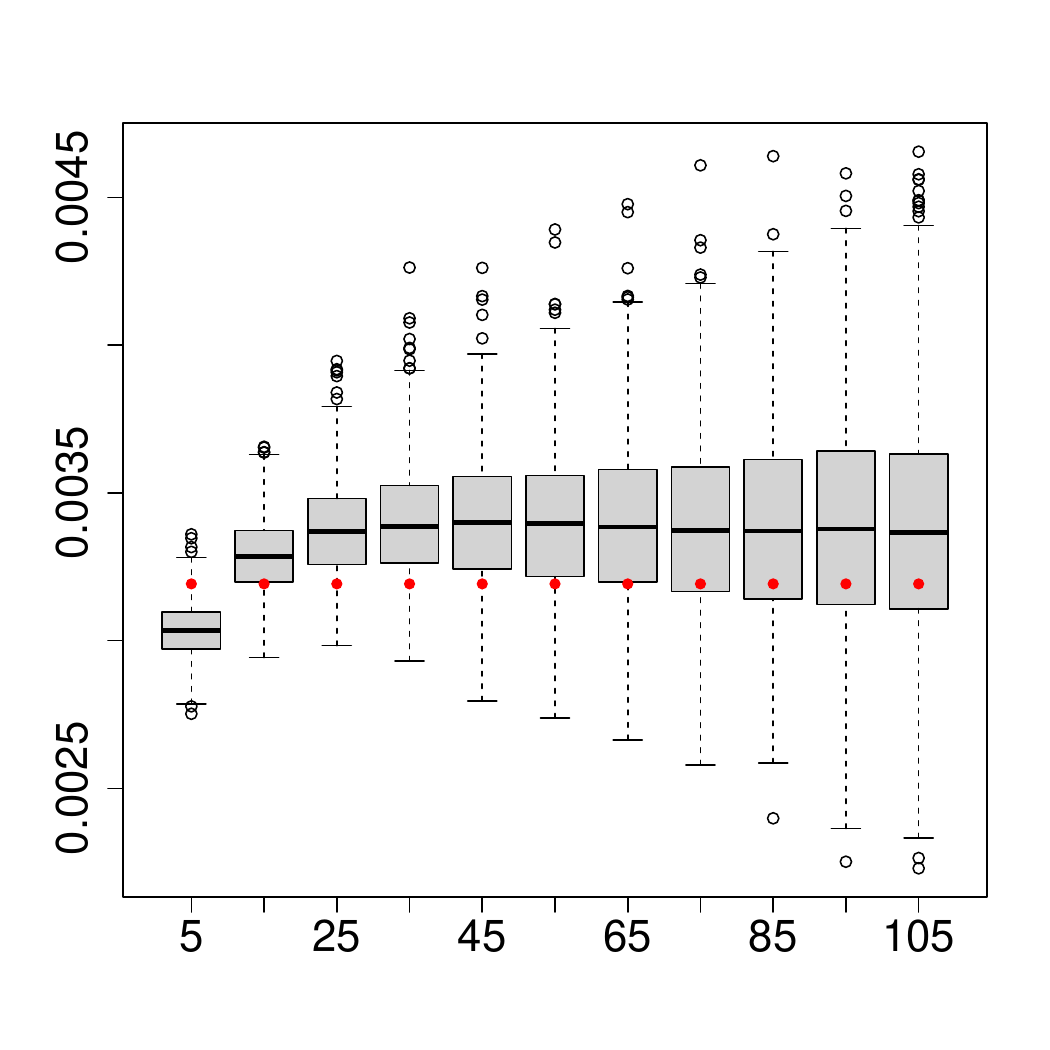}} &    \adjustbox{valign=m,vspace=1pt}{\includegraphics[width=0.2\linewidth]{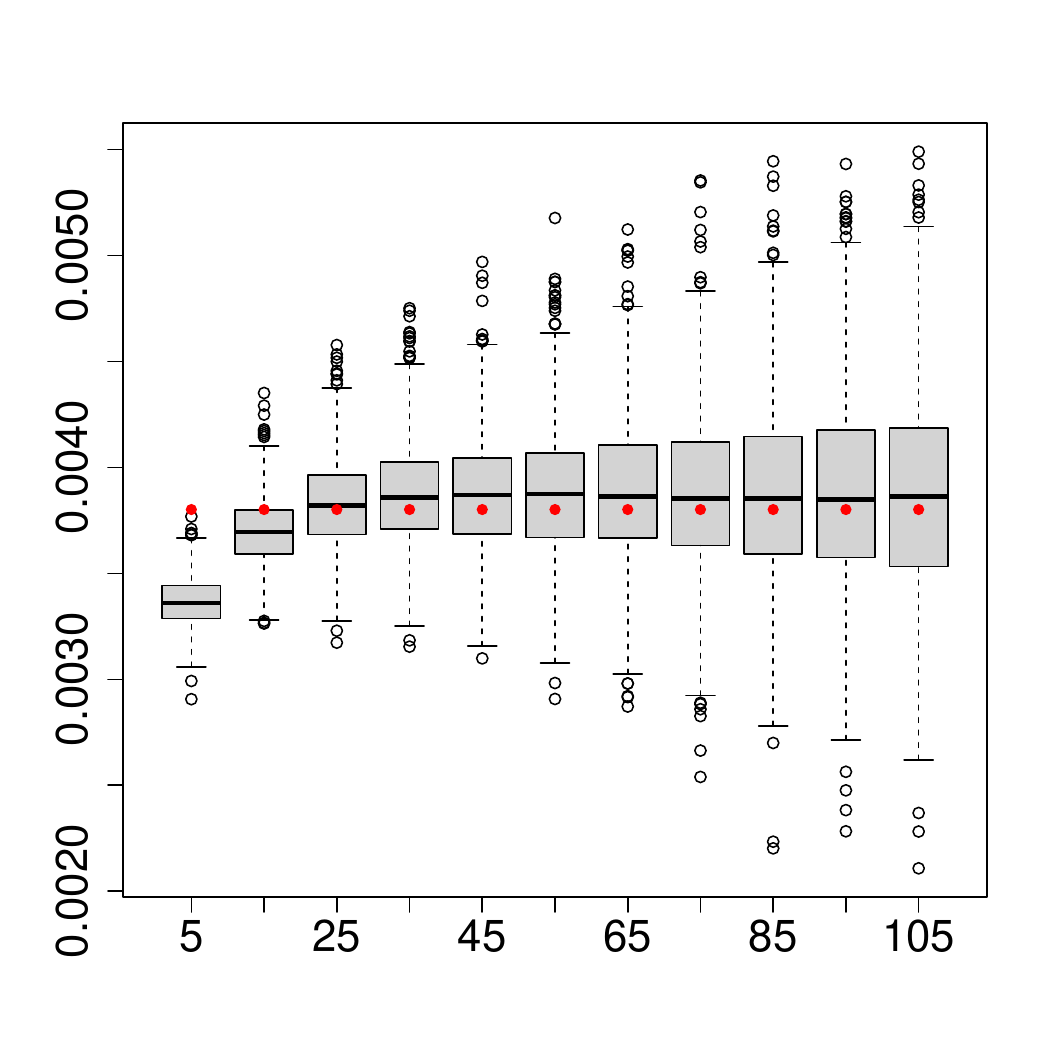}} & \adjustbox{valign=m,vspace=1pt}{\includegraphics[width=0.2\linewidth]{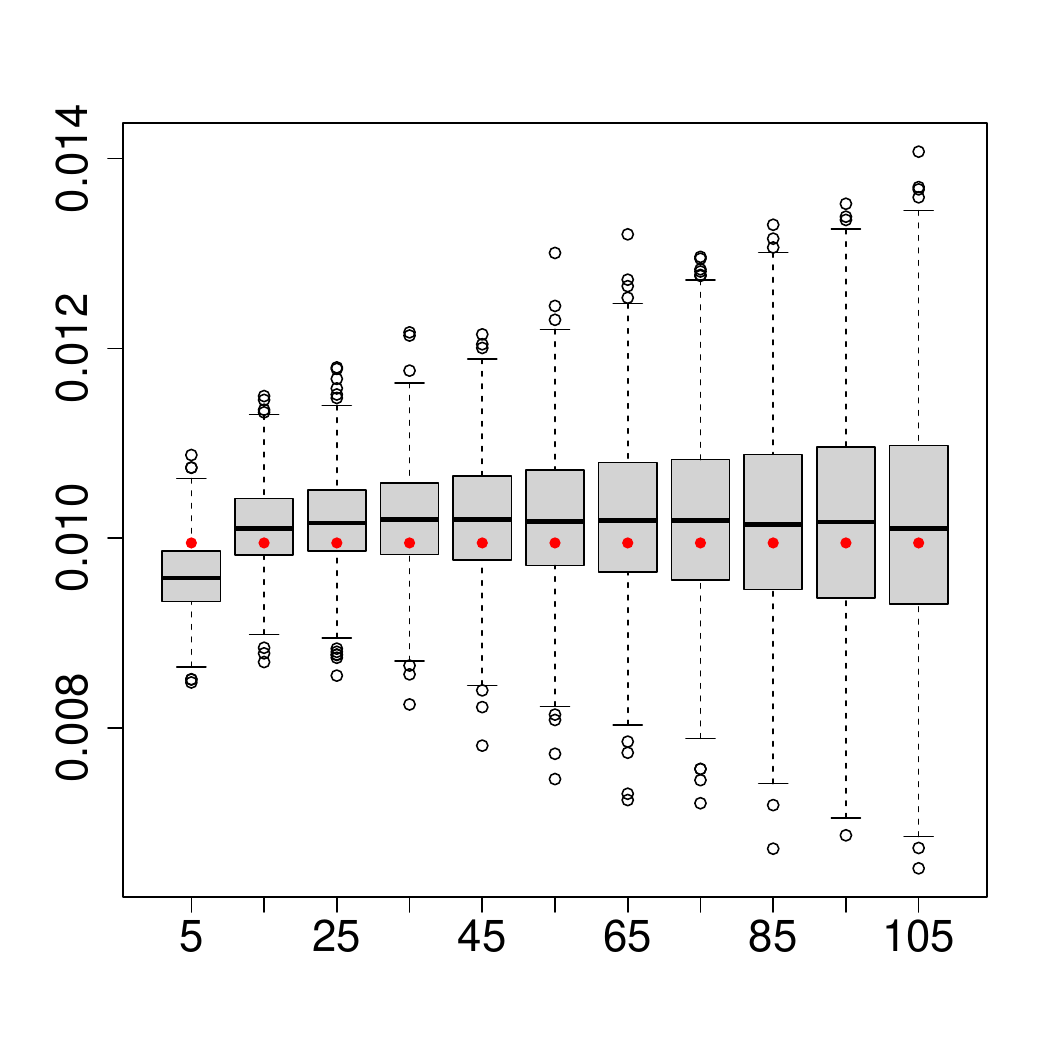}} &
        \adjustbox{valign=m,vspace=1pt}{\includegraphics[width=0.2\linewidth]{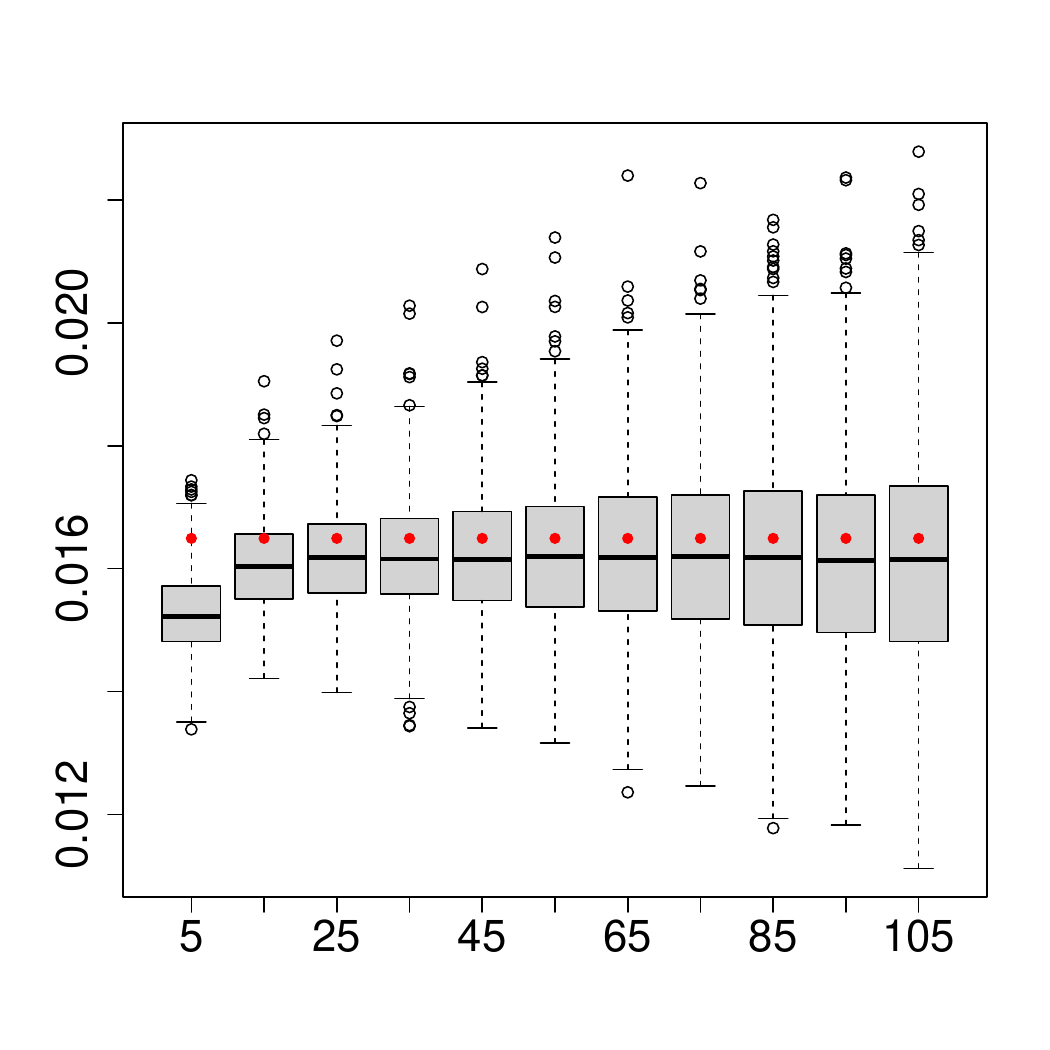}}
  \end{tabular}\vspace*{\bsl}
\caption{Boxplots of the estimated variances for the estimated recruit
  parameters for $\widetilde W_{1}$ (upper row) and $\widetilde W_{2}$ (lower row) for different
  truncation distances. The red
  dots show the empirical variance of the simulated parameter
  estimates.}\label{fig:estvarrecruits}
\end{figure}

Figure~\ref{fig:estvarrecruits} shows boxplots of
estimates (Section~\ref{sec:estimationcovmatrix}) of the variances of
recruit parameter estimates
for different choices of truncation distances equally spaced between 5 and
155m. For small truncation distances, the estimates are strongly biased downwards and the bias decreases for larger truncation distances. The medians
of the variance estimates are stable for truncation distances greater
than $30$ and the variance increases as the truncation distance
increases. The variances of the variance estimates are much reduced
when increasing the window from $\widetilde W_1$ to $\widetilde W_2$ (note the different
limits on the $y$-axes) while the bias relative to the variance seems
a bit larger for $\widetilde W_2$. 
The plots for the death parameters in Figure~\ref{fig:estvardeaths} in the supplementary material are similar to
Figure~\ref{fig:estvarrecruits} and with similar comments.

\begin{figure}[!htb]
\centering
\begin{tabular}{cc}
        \adjustbox{valign=m,vspace=1pt}{\includegraphics[width=0.35\linewidth]{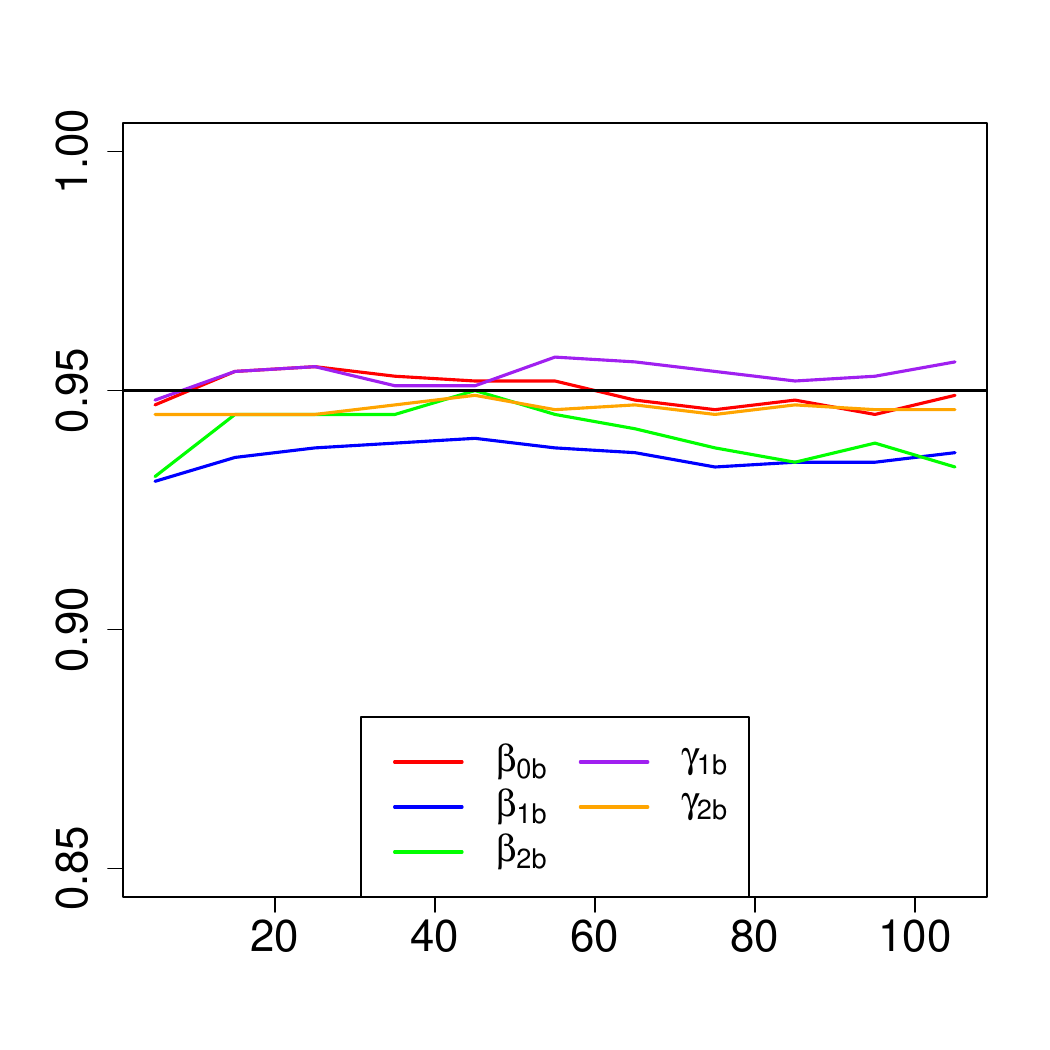}}
    &
        \adjustbox{valign=m,vspace=1pt}{\includegraphics[width=0.35\linewidth]{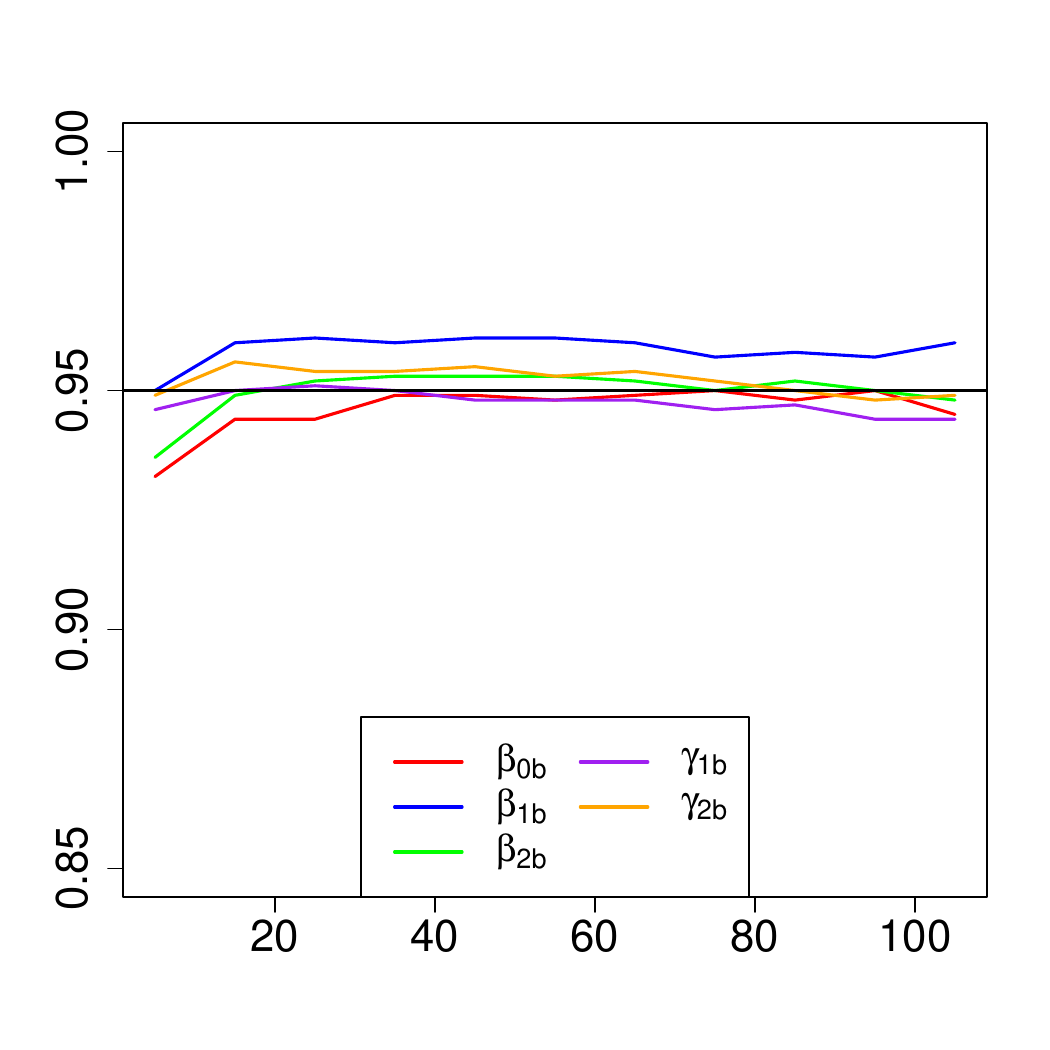}}
    \\
        \adjustbox{valign=m,vspace=1pt}{\includegraphics[width=0.35\linewidth]{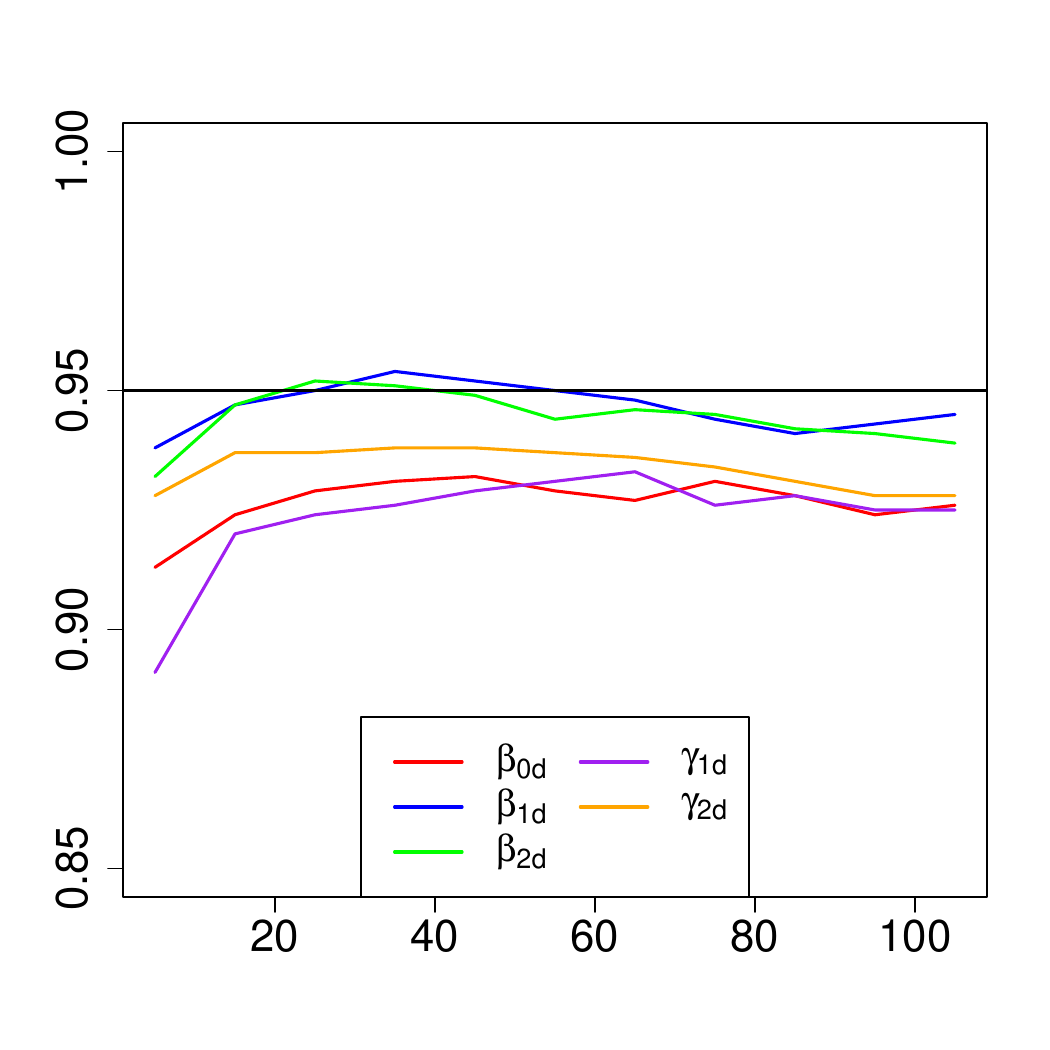}}
    &
        \adjustbox{valign=m,vspace=1pt}{\includegraphics[width=0.35\linewidth]{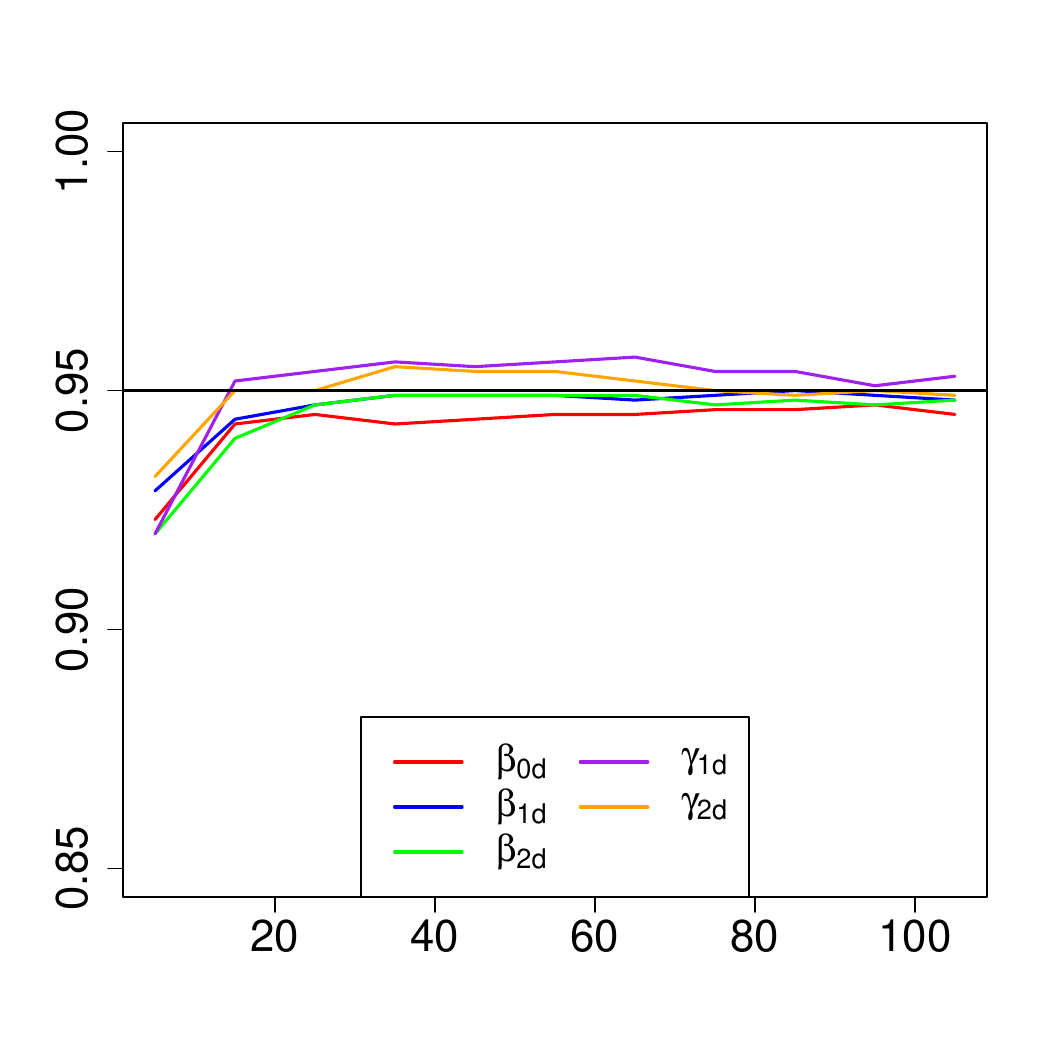}}
    \\
  \end{tabular}\vspace*{\bsl}
  \caption{Coverage probabilities for the recruits (top) and the deaths (bottom) regression parameter confidence intervals obtained with varying truncation distances and windows $\widetilde W_1$ (left) and $\widetilde W_2$ (right).}\label{fig:coverage}
\end{figure}

The plots in Figure~\ref{fig:coverage} show coverage probabilities
over the 1000 simulations of 95\% confidence intervals based on the
asymptotic normal distribution of the parameter estimates with
variances estimated following Section~\ref{sec:estimationcovmatrix}. For the recruits parameters, the coverage probabilities are quite close to the nominal 95\% over a wide range of truncation distances. For the death parameters, the coverage probabilities are a bit less satisfactory in case of $\widetilde W_1$ but quite close to the nominal 95\% in case of the bigger window $\widetilde W_2$.
Overall, inference based on the asymptotic normal distribution of
parameter estimates and the proposed estimates of parameter estimate
variances seems reliable at least when the observation window is
sufficiently large. Coverage probabilities of confidence
intervals are fairly stable across a wide range of truncation
distances indicating an appealing robustness to the choice of
truncation distance for variance estimation.

\subsection{\rwrev{Comparison with INLA}}\label{sec:inla}

\rwrev{The INLA (integrated nested Laplace approximation) package
  \citep{rue2009approximate,lindgren2011explicit} has gained huge
  popularity as a versatile and computationally efficient tool for
  analysing space-time data. INLA implements Bayesian inference for
  latent Gaussian random field models and thus requires a full specification of the data generating
  mechanism. For the recruits we use INLA to fit a log Gaussian Cox
  process with the same space-time correlation structure and
  covariates that were used to generate the simulated data sets. More
  precisely, INLA fits a Poisson log normal model to counts of points
  within cells of a partition of the observation window.
  Specifically, we consider counts within disjoint $10\times 10$m
  squares corresponding to a $50 \times 100$ grid on the large window
  $\widetilde W_2$. For the
  deaths we use INLA to implement a logistic regression with
  space-time correlated random effects. For the prior distributions we use the default non-informative priors specified by INLA. More details regarding the INLA method are given in supplementary Section~\ref{sec:inla}.

For the considered space-time setting, INLA is quite time consuming
and we therefore only consider the window $\widetilde W_2$ and only
100 simulated data sets. The mean computing time (Intel E5-2680 v4,
152GB RAM) for a simulated data set is
21.6 minutes (standard deviation 5.38) for recruits and 19.7 minutes
(standard deviation 4.3) for deaths. Our method is much faster with
mean computing time 13.7 seconds (standard deviation 3.8) for recruits
and 4.63 seconds (standard deviation 2.28) for deaths (including all 11 considered truncation distances). Supplementary Figure~\ref{fig:INLA} compares the composite likelihood estimates with the INLA posterior mean estimates. For 6 out of 10 parameters there is close agreement between the two types of estimates. However, the INLA posterior means for the remaining 4 parameters show considerable bias. As detailed in the supplementary material, this seems to be due to discretization error when a covariate is coarsened to the $50\times 100$ grid.}
\section{The Barro Colorado Island data}\label{sec:bciapplication}

\begin{figure}[!htb]
\centering
\begin{tabular}{cc}
1983 & 1985\\
\includegraphics[width=0.45\textwidth]{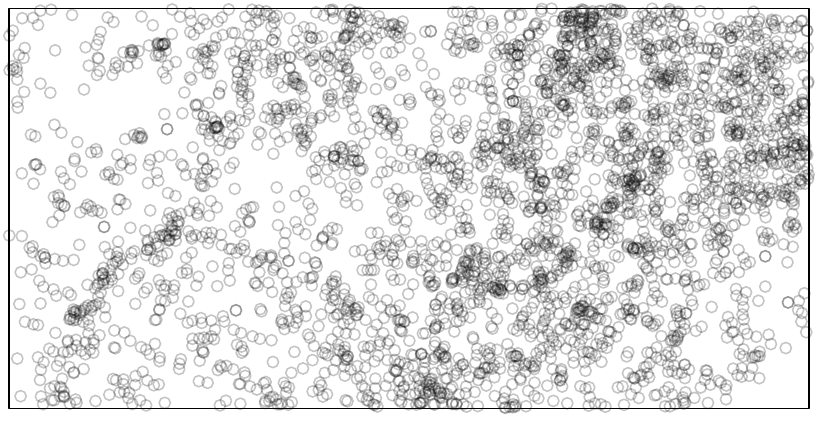}
  &\includegraphics[width=0.45\textwidth]{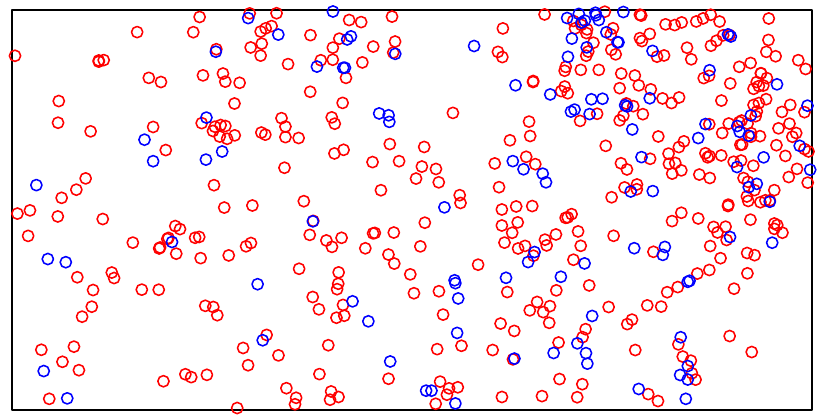}\\
1990 & 1995 \\
\includegraphics[width=0.45\textwidth]{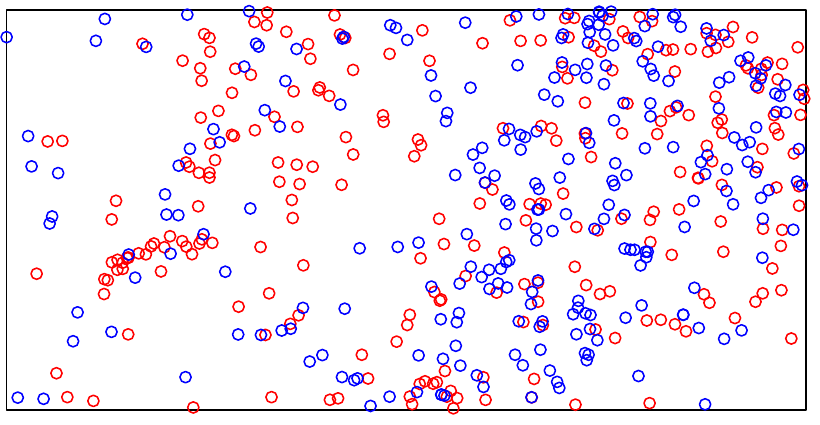}
  &\includegraphics[width=0.45\textwidth]{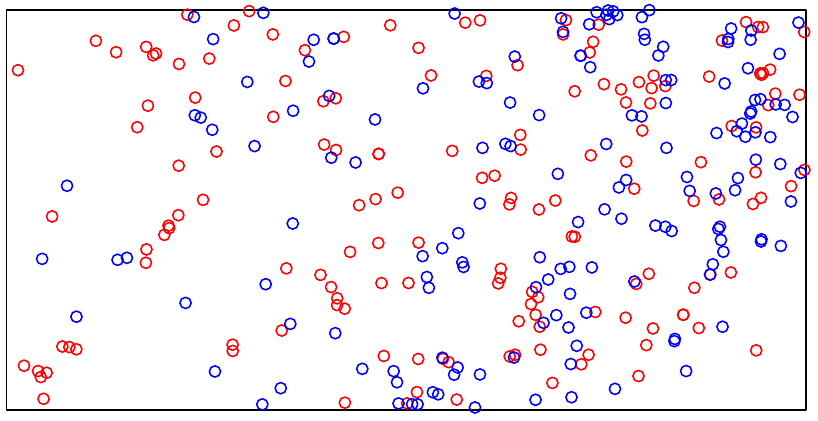}\\
2000 & 2005\\
\includegraphics[width=0.45\textwidth]{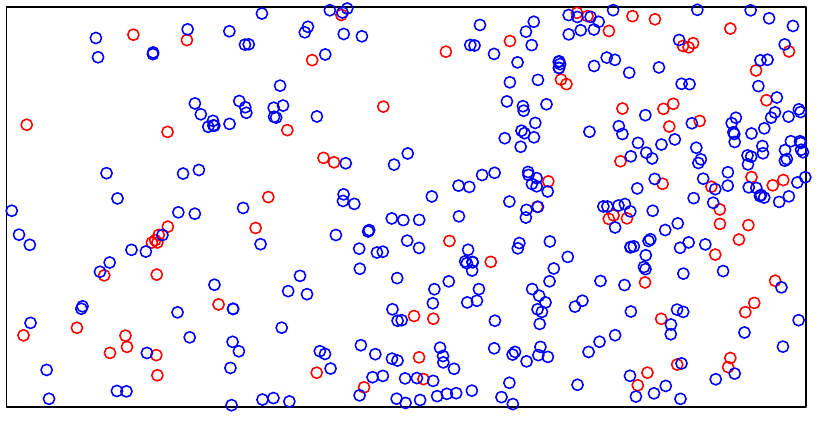}
  &\includegraphics[width=0.45\textwidth]{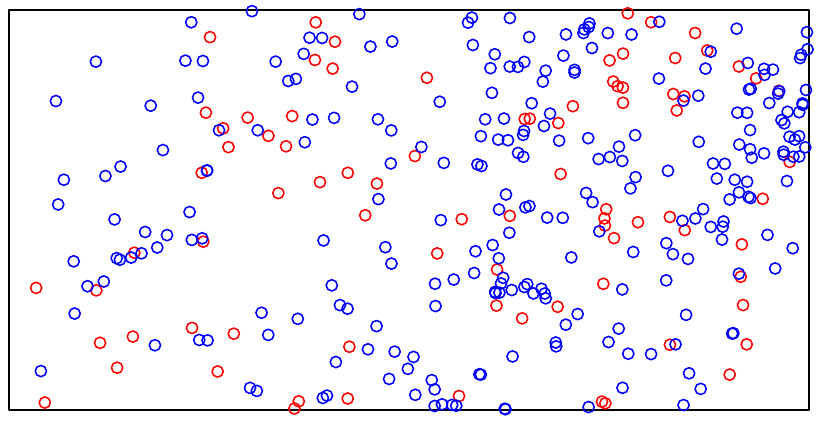}\\
2010 & 2015\\
\includegraphics[width=0.45\textwidth]{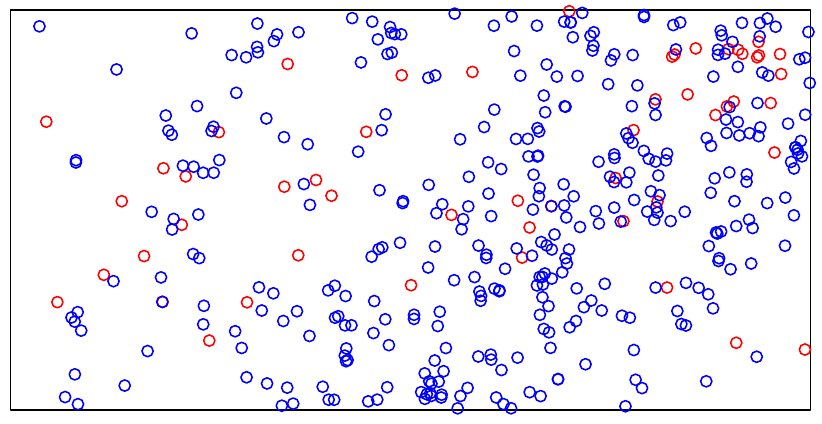}
  &\includegraphics[width=0.45\textwidth]{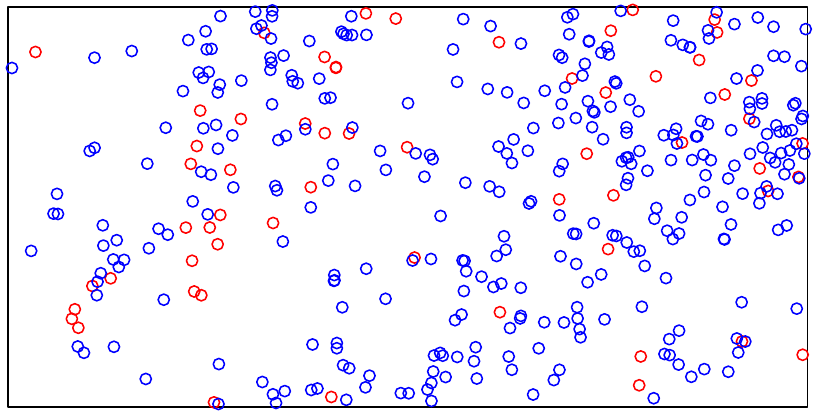}\\
\end{tabular}\vspace*{\bsl}
\caption{Spatial point patterns of locations of  {\em Capparis
    frondosa} trees in the eight censuses of the 50 ha plot in Barro
  Colorado Island (BCI), Panama. Top left plot: trees in first
  census. Remaining plots: recruits (red) and deaths (blue) relative
  to previous census.}
\label{fig:data}
\end{figure}

We consider tree census data from the 50 ha, $\widetilde W = [0,
1000\text{m}]\times[0, 500\text{m}]$, Barro Colorado Island (BCI) study plot \citep{condit2019complete}. The first
census was conducted in 1983  followed by  censuses
in 1985, 1990, 1995, 2000, 2005, 2010, and 2015. The censuses include
all trees with diameter of breast height $m\ge 10$mm. Figure~\ref{fig:data}
shows the locations of  \emph{Capparis frondosa} trees in the first
census and recruits and deaths for the remaining seven censuses. We here
ignore that the time-interval between the first and the second census
is smaller than for the remaining censuses. To some extent this is accounted for by the census dependent intercepts.
Table~\ref{tab:census} in the supplementary material summarizes the numbers of recruits and deaths in
each census. The population of {\em Capparis} trees seems to be
declining with a decreasing trend regarding number of recruits and increasing
trend regarding number of deaths. \rwrev{A detailed discussion of existing
BCI literature on recruitment and mortality is given in the supplementary Section~\ref{sec:existing}.}

We employ the log linear recruit intensity function
\eqref{eq:recruits_model}  with covariates copper (Cu),
potassium (K), phosphorus (P), pH, mineralized nitrogen (Nmin), elevation (dem), slope gradient (grad), convergence index
(convi), multi-resolution index of valley bottom flatness (mrvbf),
incoming mean annual solar radiation (solar), and topographic wetness
index (twi) available on a $5\times5\,\text{m}^2$ grid. For the
influence of existing trees we only distinguish between Capparis trees
($l=1$) and other trees ($l=2$). For the influence of Capparis on recruits
we use 
\eqref{eq:seedkernel} with $\psi_1=0.25$ and for the influence of
Capparis on deaths we use \eqref{eq:ci} with $\kappa_1=5$. For the
influence of other trees we compute influence functions of the form
\eqref{eq:ci} for all abundant species other than Capparis with more
than 500 trees and with $\kappa_{2}=5$. These influence functions are
averaged to get an influence function for other trees that is
used both for recruits and deaths. 

\begin{figure}[!htb]
  \centering
\begin{tabular}{ccc}
  \includegraphics[width=0.33\textwidth]{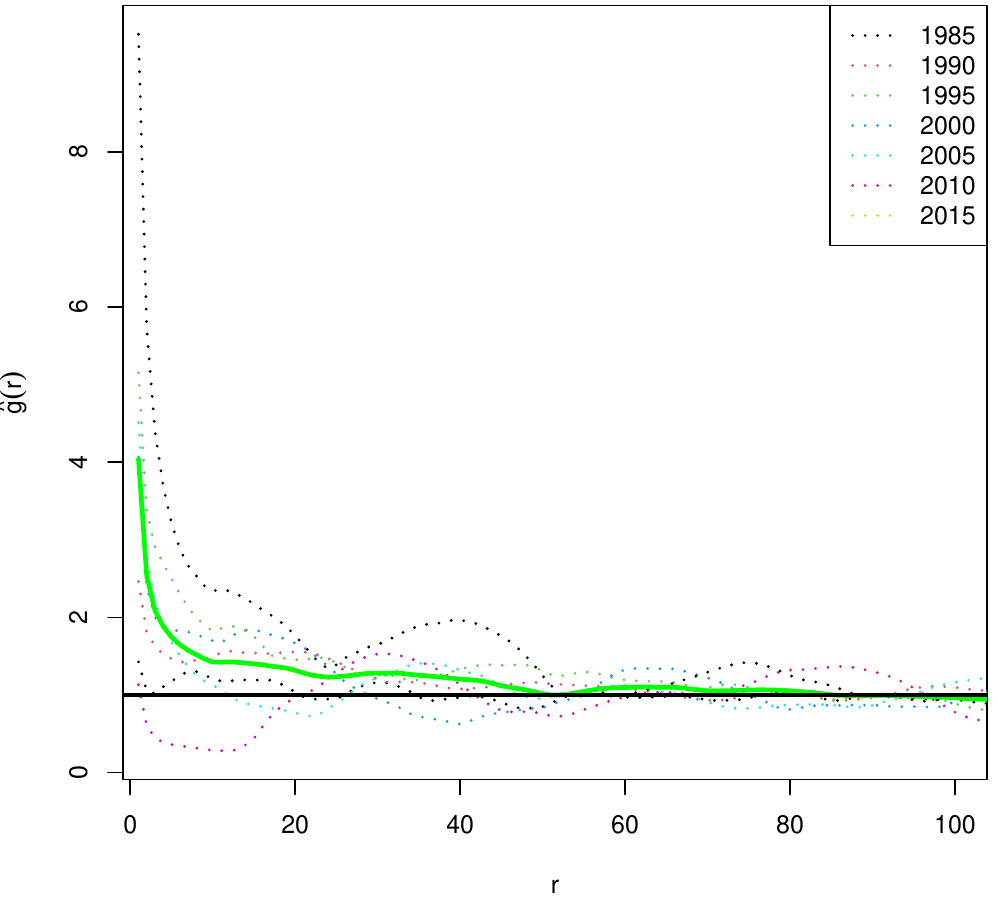} &
                                                                  \includegraphics[width=0.33\textwidth]{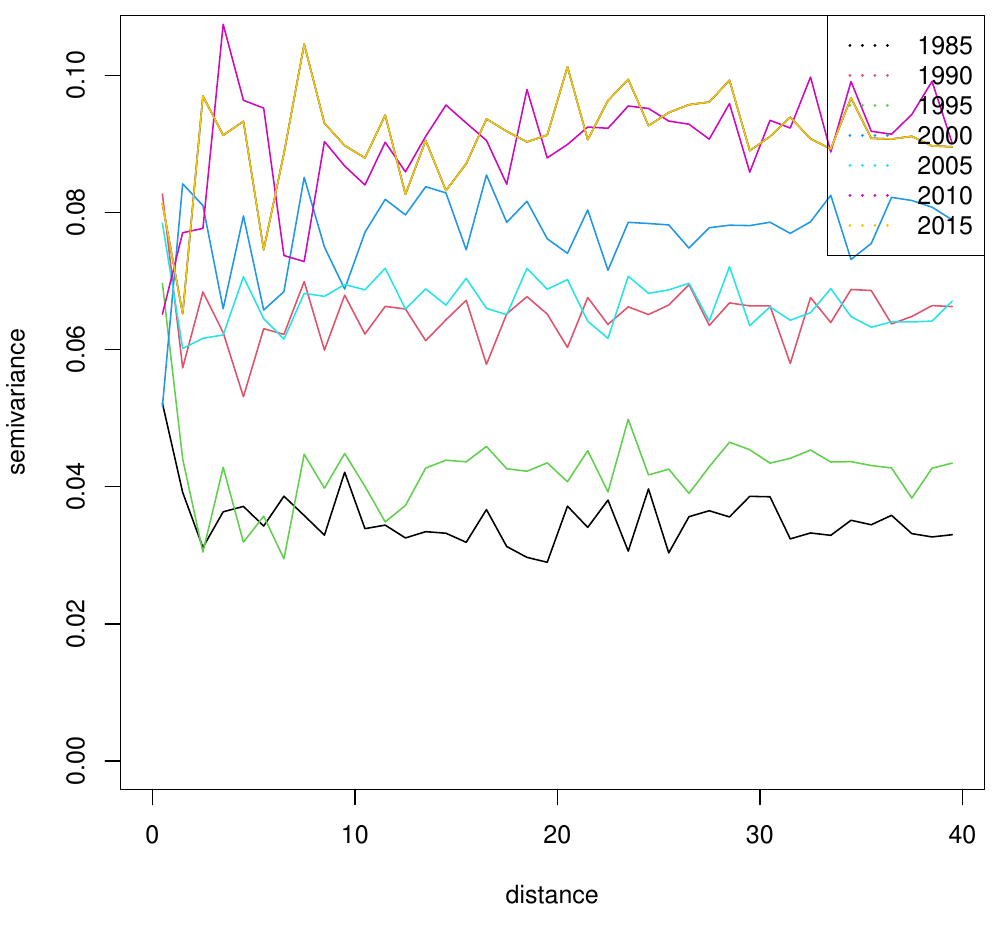} &
\includegraphics[width=0.33\textwidth]{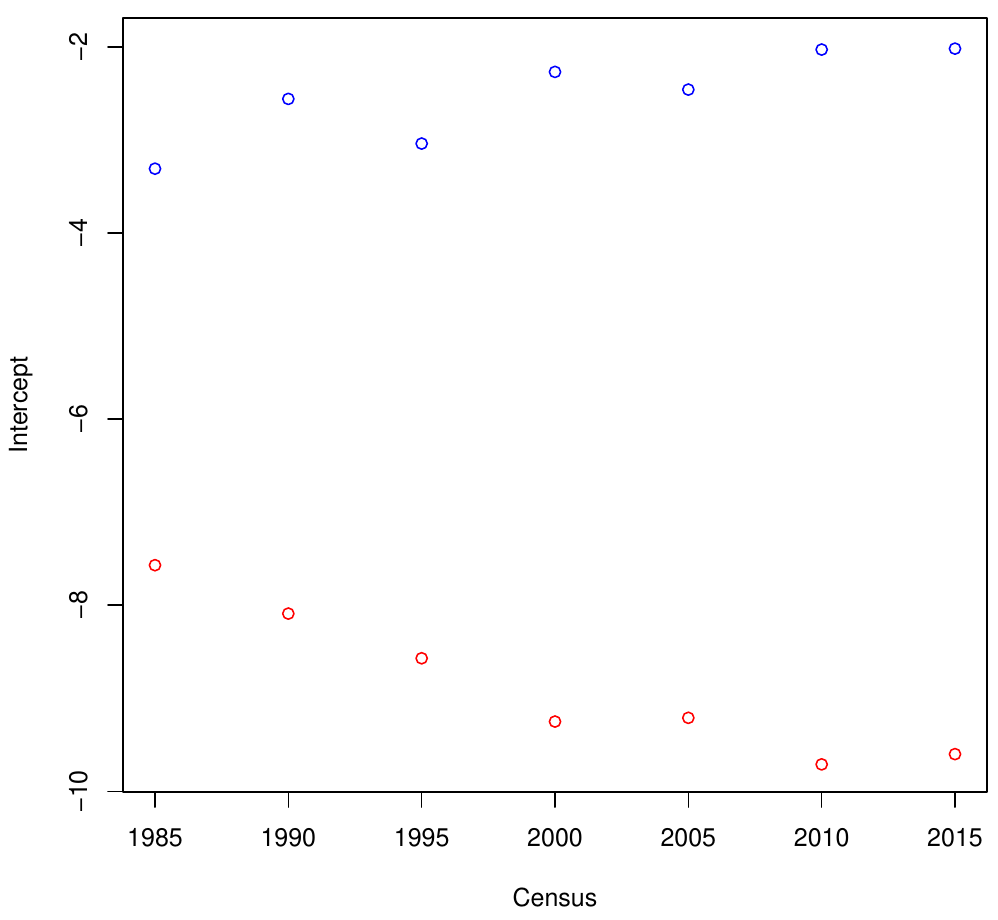}
\end{tabular}\vspace*{\bsl}
  \caption{Left: estimated pair correlation functions for each
    census. Middle: variograms for death indicators  for each census. Right:
    time dependent intercepts for recruits (blue) and deaths (red).}\label{fig:estim_pcf}
\end{figure}

The left plot in Figure~\ref{fig:estim_pcf}
shows non-parametric estimates of pair correlation functions for each
point pattern of recruits. Despite the variability
between the estimates, all estimates seem to stabilize around 1 after
a distance of 55m which we use as the truncation distance for variance
estimation. The middle plot shows variograms for the death events. For the deaths we also truncation distance 55m. There does not appear to be strong spatial
correlation between death events and indeed our estimated standard
deviations for the death regression parameter estimates are only slightly
larger than those (not shown) obtained assuming conditionally
independent death events.  The right plot shows the intercepts for recruits that
decrease over time and the intercepts for deaths that increase over
time. The composite likelihood estimates of the remaining
regression parameters for respectively recruits and deaths are given in
Table~\ref{tab:betapars} together with $p$-values and INLA results (see below).
\begin{table}[!htb]
\centering
\caption{Recruits (left) and death (right) composite likelihood parameter estimates, INLA posterior means, $p$-values based on asymptotic normality, and indication of whether 0 is inside INLA 95\% posterior credibility interval.}\label{tab:betapars}
\begin{tabular}{l|rrrr|rrrr|}
                & estm.\ & post.\ mean  & $p$ & 0 inside & estm.\ & post.\ mean & $p$ & 0 inside\\
\hline
convi         & -0.01  & -0.01  & 0.00 &no & -3e-3  &  0.00 &  0.09 & yes\\
Cu            & 0.03   & 0.05   & 0.33 &yes&  0.10  &  0.11 &  0.00 & no\\
dem           & 0.02   & 0.02   & 0.06 &yes&  4e-3  &  0.01 &  0.54 & yes\\
grad          & -2.79  & -3.25  & 0.03 &no &  0.76  &  0.63 &  0.32 &yes\\
K             & 0.00   &  0.00  & 0.85 &yes& -2e-3  &  0.00 &  0.17 & yes\\
mrvbf         & -0.07  & -0.03  & 0.24 &yes&  0.02  &  0.00 &  0.64 &yes\\
Nmin          & 0.03   & 0.02   & 0.00 &no &  1e-3  &  0.00 &  0.88 &yes\\
P             & -2e-3  & 0.01   & 0.96 &yes& -0.03  & -0.03 &  0.32 & yes\\
pH            & -0.02  &-0.01   & 0.90 &yes&  0.55  &  0.60 &  0.00 & no\\
solar         & -5e-3  & 0.00   & 0.00 &no &  3e-3  &  0.00 &  0.01 & no\\
twi           & 0.01   &-0.03   & 0.84 &yes&  0.03  &  0.03 &  0.39 & yes\\
Infl.\ others & 1e-3   & 0.01   & 0.93 &yes& -0.01  & -0.00 &  0.21& yes\\
Infl.\ Capparis & 0.64 & 0.29   & 0.00 &no &  2e-4  &  0.32 &  0.65 & no
\end{tabular}
\end{table}
We do not attempt a formal investigation of the significance of the
various covariates. However, based on the $p$-values, there is some
evidence that recruit intensity is negatively associated with the
covariates convi, grad, and solar and positively associated with Nmin and,
not surprisingly, presence of existing Capparis trees. Probability of
death appears to be positively associated with high level of Cu, pH and solar. However, no dependence on existing trees is detected.

\rwrev{
  We also analyzed the BCI data set using INLA  as described in
  Section~\ref{sec:inla} with a discretization into $5\times 5$m cells
  for the recruits. The INLA posterior means for the regression
  parameters are quite similar to our estimates and this also holds
  for conclusions regarding significance based on whether zero is
  contained in the 95\% credibility intervals or not (except for
  Capparis influence on deaths). However, our method is much faster
  with computing times in seconds 53.2 seconds (recruits) and 12.6
  seconds (deaths) compared to 14.6 minutes and 34.9 minutes for INLA.}



\section{Discussion}

Our methodology is essentially free of assumptions regarding the second-order
properties of the space-time point process but a risk of misspecification for the intensity  and death
probability regression models remains. This does not necessarily render
regression parameter estimates meaningless. As in
\cite{choiruddin:coeurjolly:waagepetersen:21} one might define `least
wrong' regression models that minimize a composite
likelihood Kullback-Leibler distance to the true intensity and death
probability models. The composite likelihood parameter estimates then estimate the corresponding `least wrong' regression parameter
values.

We have focused on estimation of regression parameters. However, scale
parameters in the models for influence of existing trees need to be
determined too. If suitable values of these parameters can not be identified from biological insight, one
might include these parameters in the composite
likelihood estimation. However, computations are more cumbersome
and it may be necessary to restrict maximization to a discrete set of
candidate parameter values similar to the approach for irregular
parameters of Markov point processes in the \texttt{spatstat} package \citep{baddeley2015spatial}.

We have not provided a theoretically well founded method
for truncation distance selection for the variance matrix estimators. Our
simulation study, however, indicates robustness to the choice of truncation distance. As in the data example, plots of estimates of pair correlation functions and variograms may give some idea of suitable truncation distances.




\section*{Acknowledgements}

\rwrev{We thank the associate editor and two reviewers for constructive and helpful comments that led to substantial improvements.} Francisco Cuevas-Pacheco was supported by Agencia Nacional de Investigaci{\'o}n y Desarrollo, Proyecto ANID/FONDECYT/INICIACION 11240330. \rwrev{Rasmus Waagepetersen was supported by grants VIL57389, Villum Fonden, and NNF23OC0084252, Novo Nordisk Foundation.}


%
\section*{Supporting Information}

Supporting material for the simulation study and the BCI data section is available with
this paper at the Biometrics website on Wiley Online
Library. The code for the simulation study and data example is available at 
\verb+https://github.com/FcoCuevas87/CL_ST-PointProcesses+

\section*{Data availability}
\rwrev{The BCI data are available at} \verb+https://datadryad.org/stash/dataset/doi:10.15146/5xcp-0d46+.

\bibliographystyle{apalike}
\bibliography{Bib/library2}

\section{Supplementary material}

\subsection{Background on models fitted with \texttt{INLA}}\label{sec:inla}

The INLA package implements latent Gaussian field models for space and
space-time data \citep{rue2009approximate,lindgren2011explicit}. In
case of point pattern data, these are converted to count data by
considering counts of points within $L \ge 1$ quadratic cells with
center points $u_1,\ldots,u_L$ of a specified grid covering the
observation window.  Let $N_{ik}$ denote the number of recruits in
the $i$th grid cell at time $k$ and let
$\xi_{\oo,k}=\{\xi_{\oo,k}(u)\}_{u \in \R^2}$ denote Gaussian random
fields for recruits and deaths, $\oo=\b,\dd$. For recruit counts we
assume that counts at time $k$ are conditionally independent given
$\H_{k-1}$ and $\xi_{\b,k}$ with 
\[
  N_{ik} | \H_{k-1},\xi_{\b,k} \sim \text{Poisson}(\lambda_{ik}), \quad i=1,\ldots, L, k=1,\ldots,K,
\]
where
\begin{equation*}\label{eq:recruits_model}
  \lambda_{ik} = \exp\left[\beta_{\b 0,k}  +\bZ_{k-1}^\T(u_i)\bbeta_{\mathrm{b}}
      +   \bc_{k-1}(u_i)^\T \bgamma_\b + \xi_{\b,k}(u_i) \right].
  \end{equation*}
The deaths are assumed to be conditionally independent given
$\H_{k-1}$ and $\xi_{\dd,k}$ with death probabilities
\begin{equation*}\label{eq:inladeathprob}
  p_k(u|\H_{k-1},\xi_{\dd,k}) =   \frac{\exp\left[ \eta_k(u|\H_{k-1},\xi_{\dd,k}) \right ] }{1 + \exp\left[ \eta_k(u|\H_{k-1},\xi_{\dd,k}) \right ]},
\end{equation*}
where
\begin{equation*}\label{eq:deathseta*}
  \eta_k(u|\H_{k-1},\xi_{\dd,k}) =  \beta_{\dd 0,k}+\bZ_{k-1}^\T(u) \bbeta_{\mathrm{d}} +  \bd_{k-1}(x)^\T \bgamma_{\dd} + \xi_{\mathrm{d}, k}(u).
\end{equation*}

For $\mathrm{o}=\b, \mathrm{d}$, we assume that $\xi_{\mathrm{o}, k}$, $k=1,\ldots,K$ form a zero-mean space-time Gaussian process with space-time separable covariance function
\[
  \Cov[\xi_{\mathrm{o}, k}(u), \xi_{\mathrm{o},k+t}(u+h)] = \sigma^{2}_{\mathrm{o}} c_{\mathrm{o},1}(t) c_{\mathrm{o},2}(\|h\|), \quad u,h\in\RR^{2}, k,t\in\ZZ,
\]
where $\sigma^{2}_{\mathrm{o}}$ is the variance,
\[
   c_{\mathrm{o},1}(t) = \varrho_{\mathrm{o}}^{t}, \quad t\in\ZZ,
\]
is the temporal first-order autoregressive correlation function with
$-1\leq \varrho_{\mathrm{o}} \leq 1$, and
\[
 c_{\mathrm{o}, 2}(r) = \frac{2^{1-\nu}}{\Gamma(\nu)} ( r /\xi_{\mathrm{o}})^{\nu} K_{\nu}( r / \xi_{\mathrm{o}}), \quad r\geq 0,
\]
is the spatial Matern correlation function with the correlation scale parameter $\xi_{\mathrm{o}} > 0$ and shape parameter $\nu=1$, and where $K_{\nu}(\,\cdot\,)$ denotes the modified Bessel function of the second kind.

In INLA, computationally efficient representations of the Gaussian fields $\xi_{\oo,k}$, $\oo=\b,\dd$ are obtained using stochastic partial differential equation (SPDE) representations and associated efficient numerical methods \citep{lindgren2011explicit}. Inference is implemented in the Bayesian framework using the integrated nested Laplace approximation \citep{rue2009approximate}. We use INLA default settings for the Gaussian field numerical methods and default priors for the regression and covariance parameters.

One issue with using INLA for point process data is the need for choosing a grid for the discretization of the point process data into count data. Results will converge as increasingly fine grids are used \citep{waagepetersen2004convergence} but in practice the results will be sensitive to the choice of grid. Note further that the marginal death probability is obtained by integrating out the random effect $\xi_{\dd,k}(u)$ in \eqref{eq:inladeathprob} which does not result in a logistic probability marginally.

\subsection{Supplementary figures and tables for simulation study.}
\begin{figure}
  \centering
  \begin{tabular}{c}
    \includegraphics[width=0.8\textwidth]{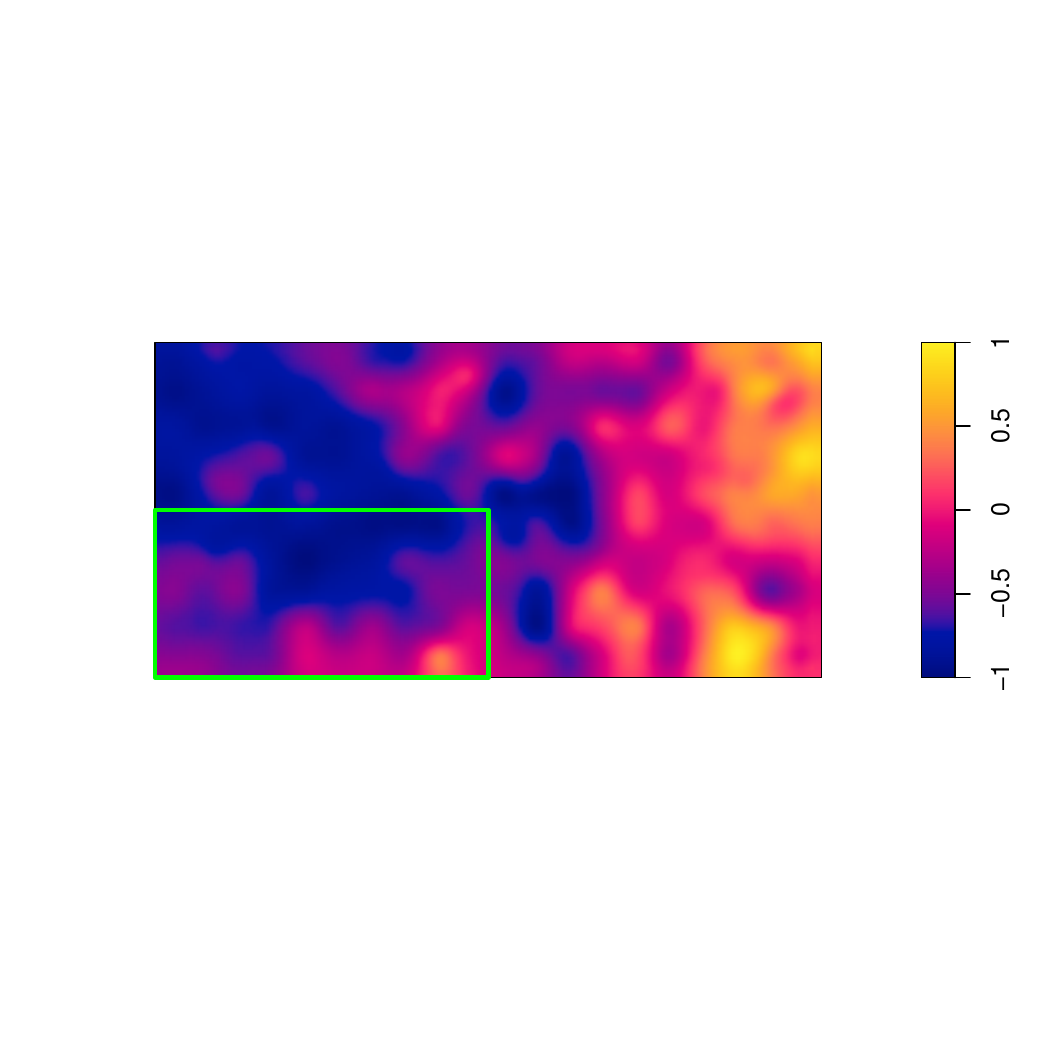}\\
    \includegraphics[width=0.8\textwidth]{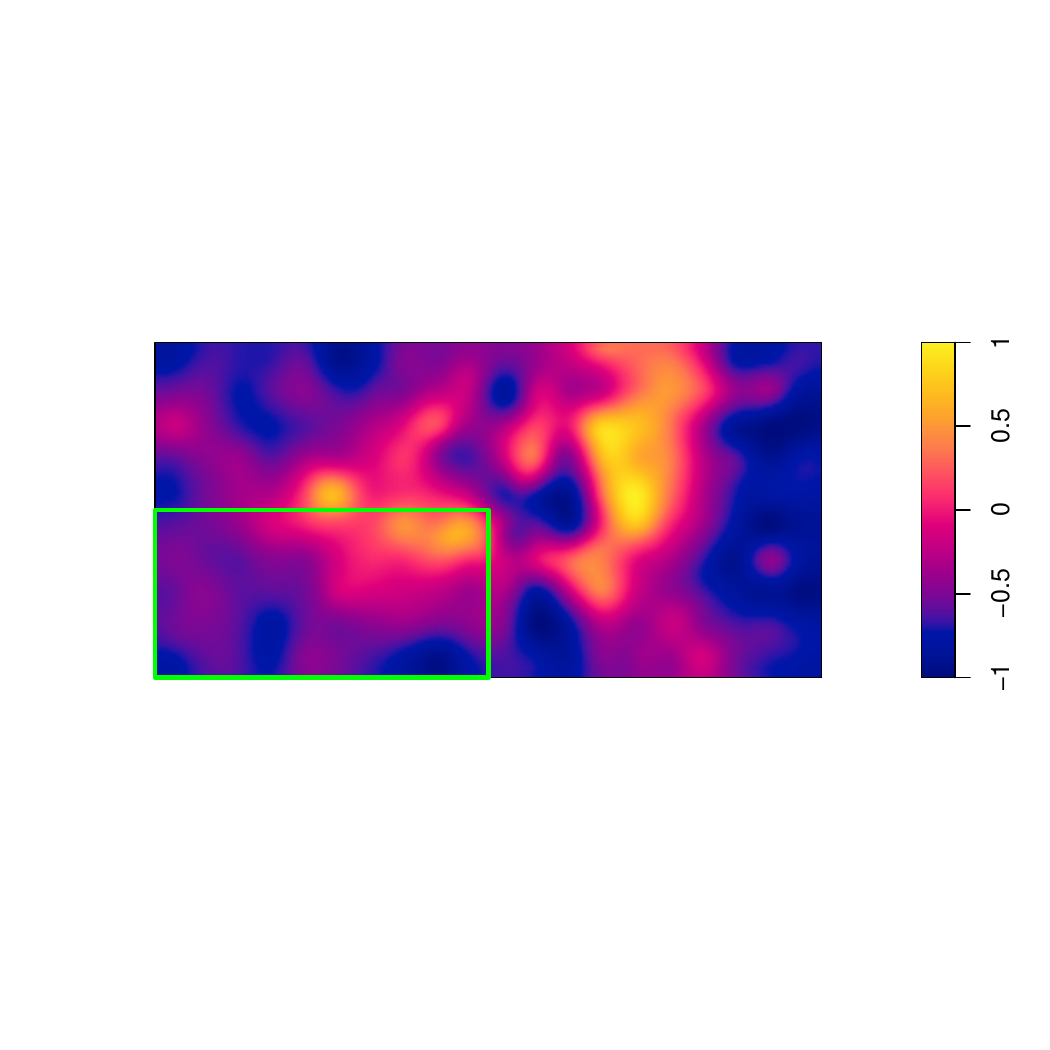}\\[\bsl]
    \end{tabular}
      \caption{Covariates $Z^{(1)}$ (top) and $Z^{(2)}$
        (bottom) on $\widetilde W_2$. Green rectangle shows
        $\widetilde W_1$.}\label{fig:covariates}
\end{figure}

\begin{figure}
\centering
\begin{tabular}{ccc}
        \adjustbox{valign=m,vspace=1pt}{\includegraphics[width=0.25\linewidth]{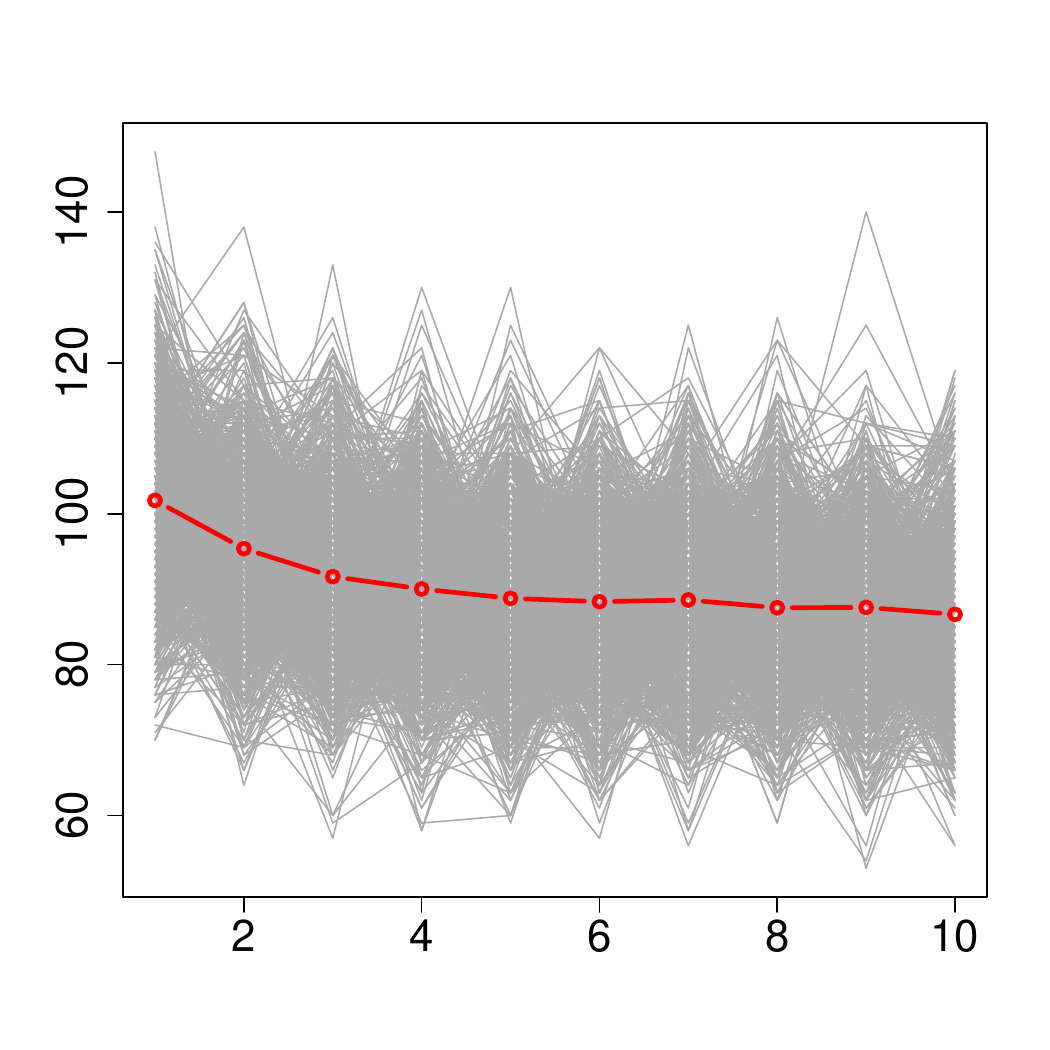}}
      & \adjustbox{valign=m,vspace=1pt}{\includegraphics[width=0.25\linewidth]{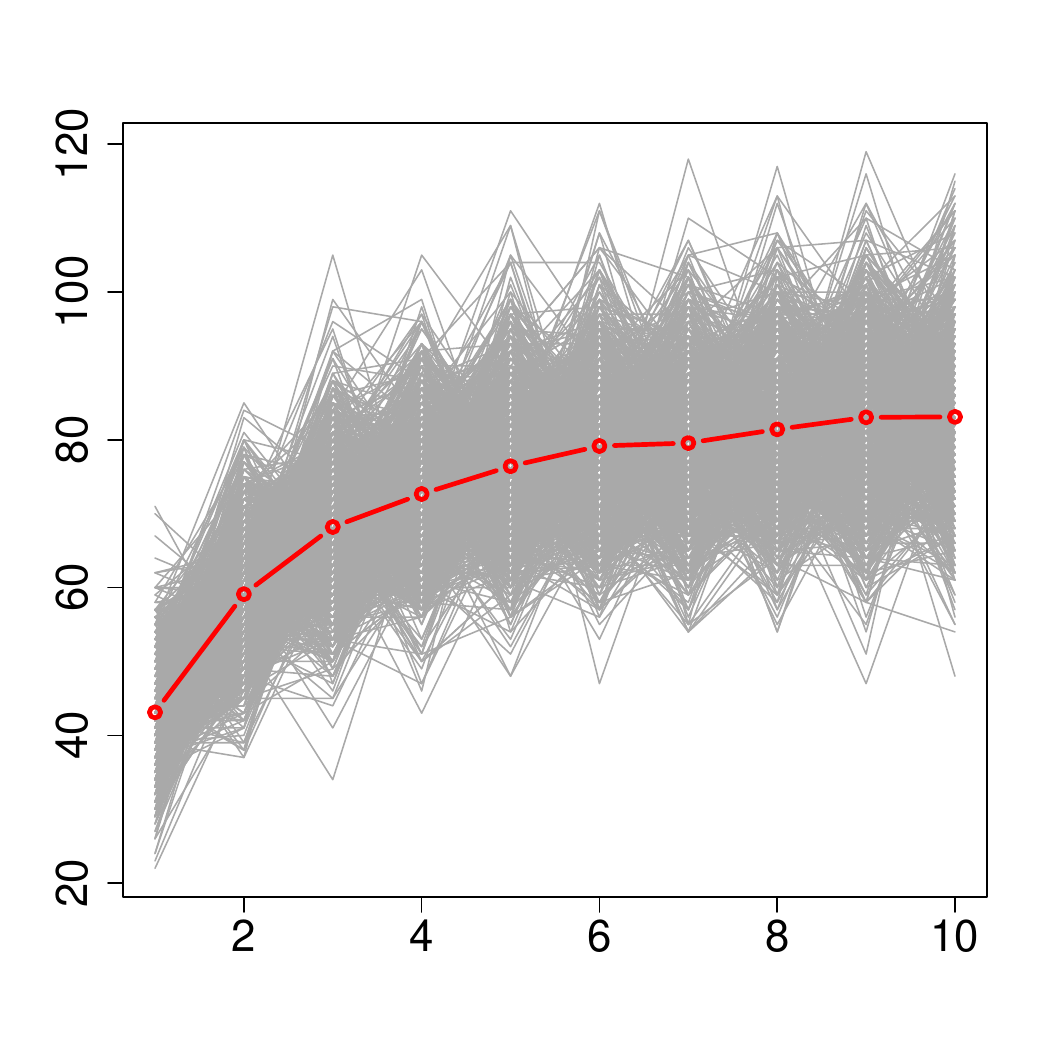}}
      & \adjustbox{valign=m,vspace=1pt}{\includegraphics[width=0.25\linewidth]{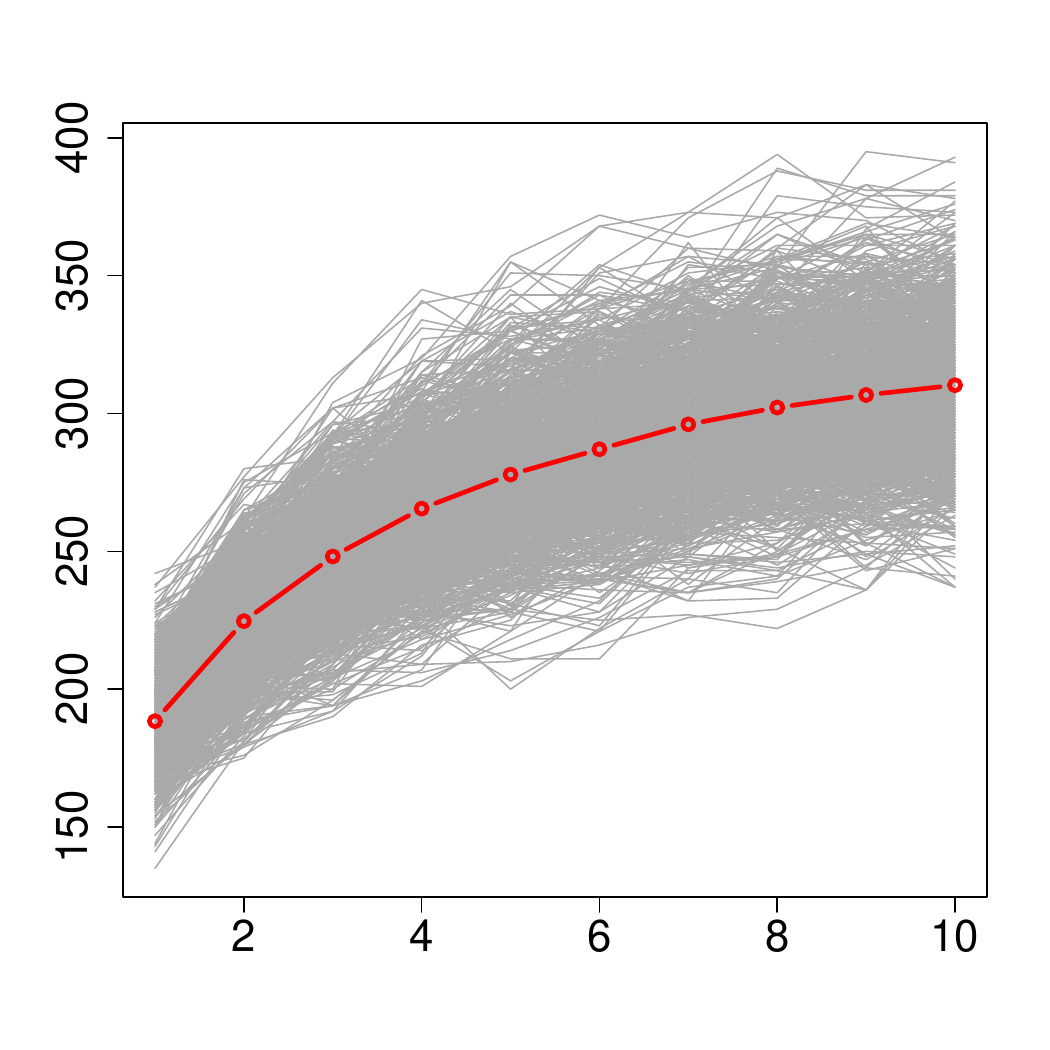}}\\
        \adjustbox{valign=m,vspace=1pt}{\includegraphics[width=0.25\linewidth]{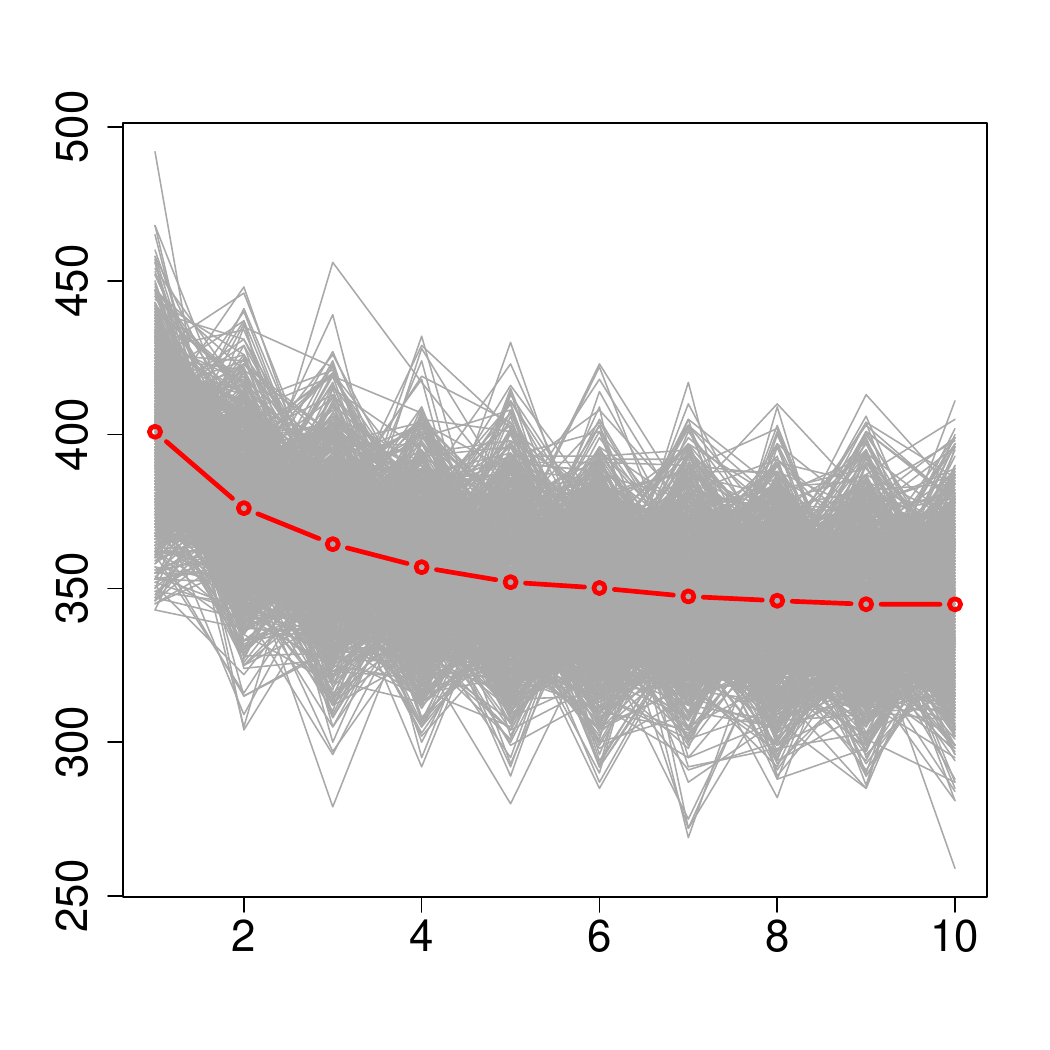}}
&       \adjustbox{valign=m,vspace=1pt}{\includegraphics[width=0.25\linewidth]{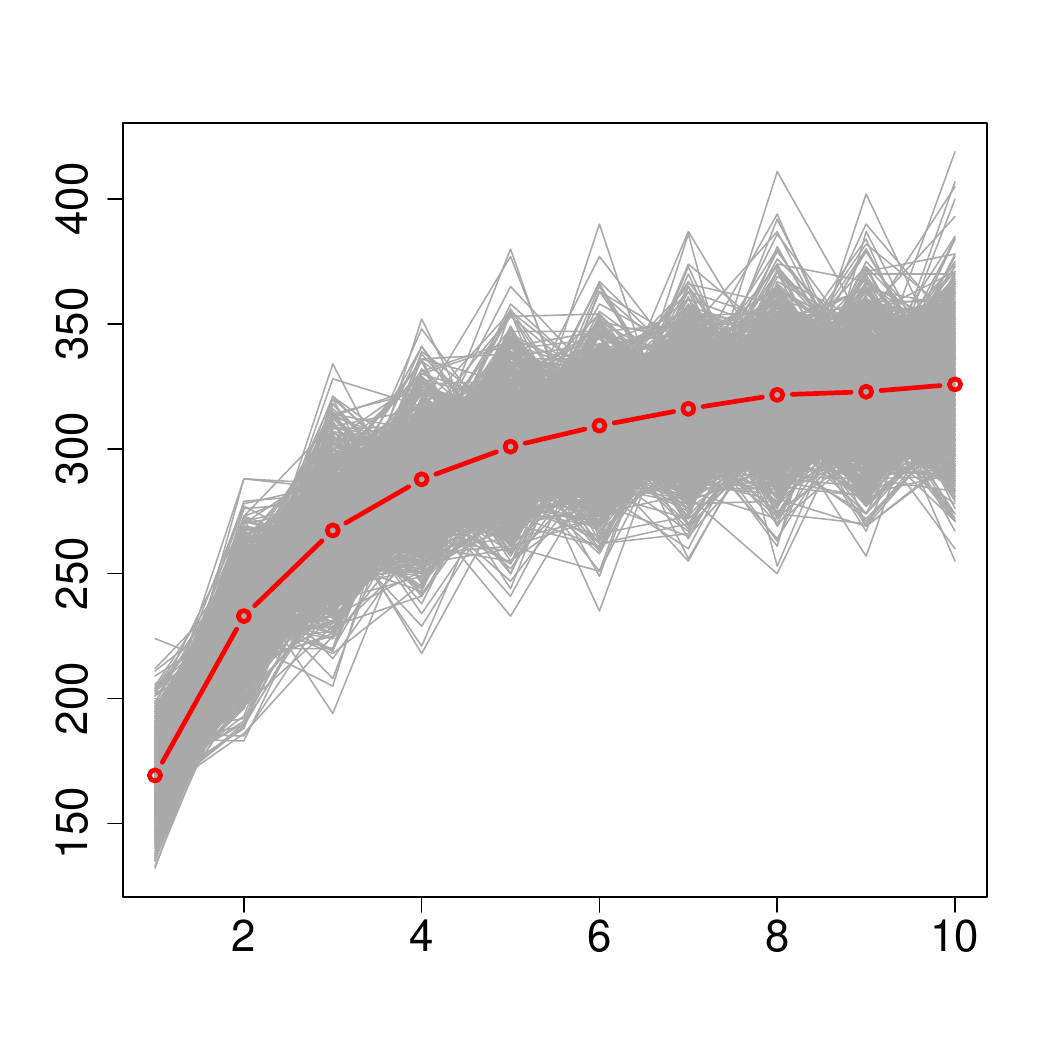}}
&       \adjustbox{valign=m,vspace=1pt}{\includegraphics[width=0.25\linewidth]{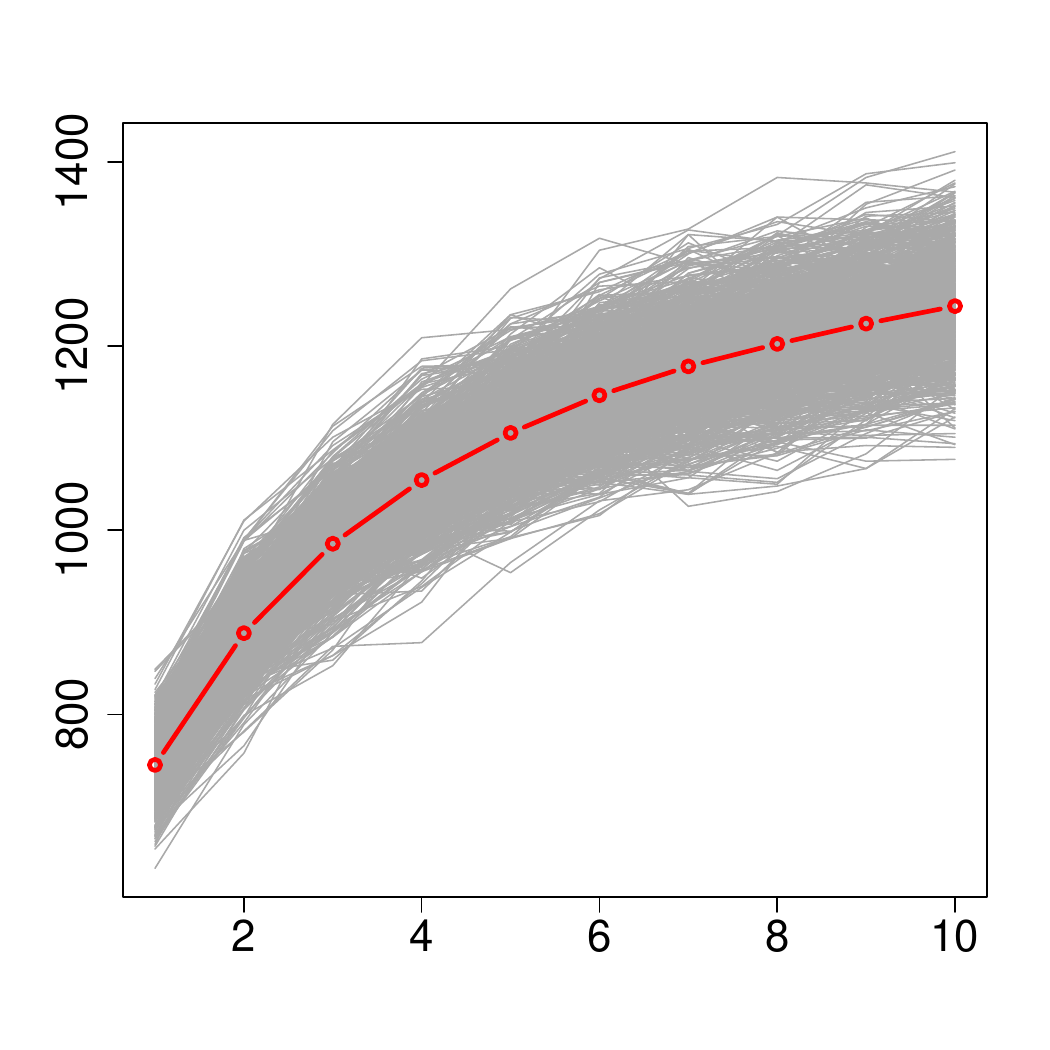}}
  \end{tabular}
  \caption{Gray lines show for each simulation numbers of recruits (left), numbers of deaths (middle), and the total amount of trees (right) per time
    step for $\widetilde W_{1}$ (upper row) and $\widetilde W_{2}$ (lower row). The red
    curves show the averages of the numbers over the 1000 simulations.}\label{fig:counting_values}
\end{figure}

Figure~\ref{fig:covariates} shows the covariates used for the
simulation study. These are realizations of Gaussian fields with
Mat{\' e}rn covariance functions with parameters
 $(\sigma_{1}, \xi_{1}, \nu_{1}) = (1/3, 28, 0.50)$ and  $(\sigma_{2},
 \xi_{2}, \nu_{2}) = (1/3, 16, 1.75)$ respectively. Figure \ref{fig:counting_values}
shows the evolution of the numbers of recruits and deaths for the
windows $\widetilde W_{1}$ and $\widetilde W_{2}$.

\begin{figure}
\centering
\begin{tabular}{ccccc}
   $\beta_{0\b}$ & $\beta_{1\b}$ & $\beta_{2\b}$ & $\gm_{1\b}$ & $\gm_{2\b}$ \\
    \includegraphics[width=0.2\linewidth]{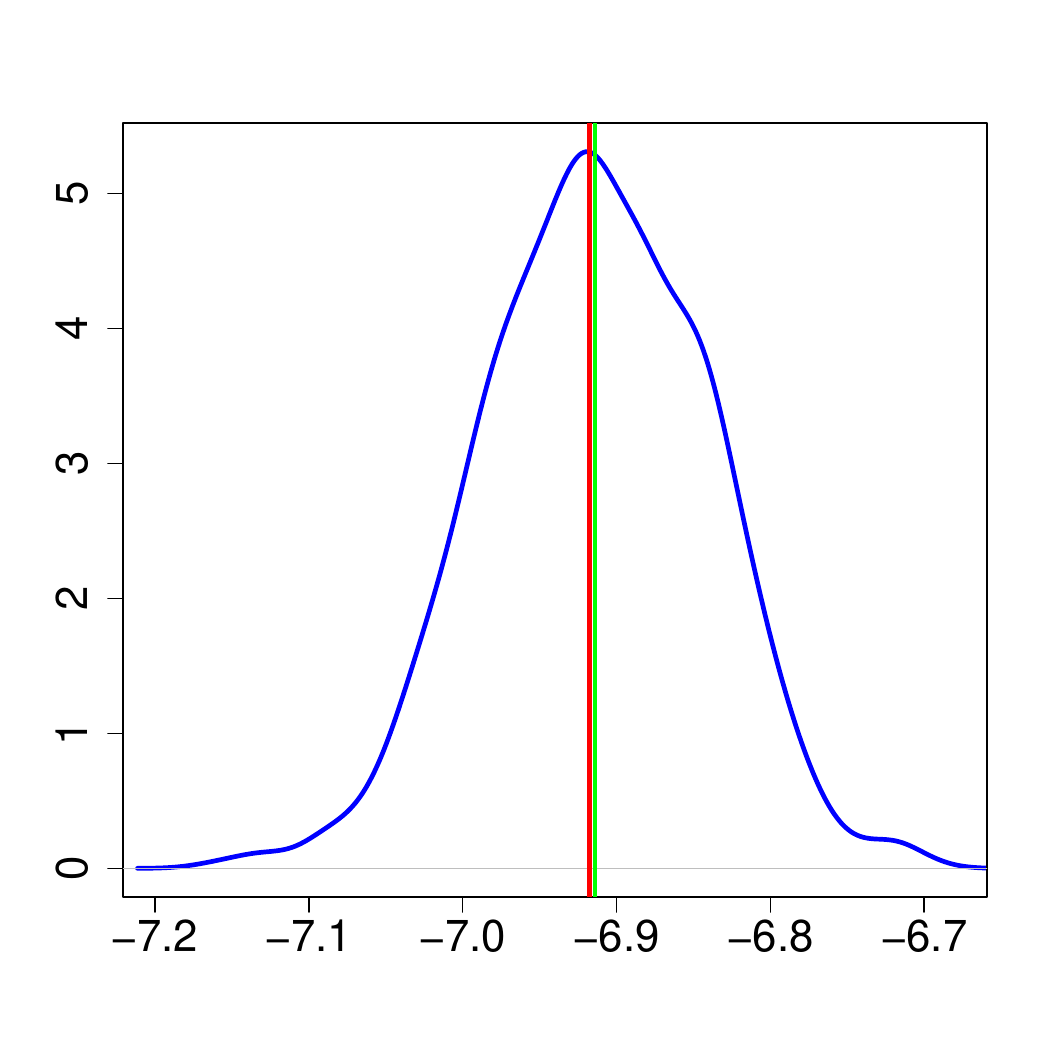} &
    \includegraphics[width=0.2\linewidth]{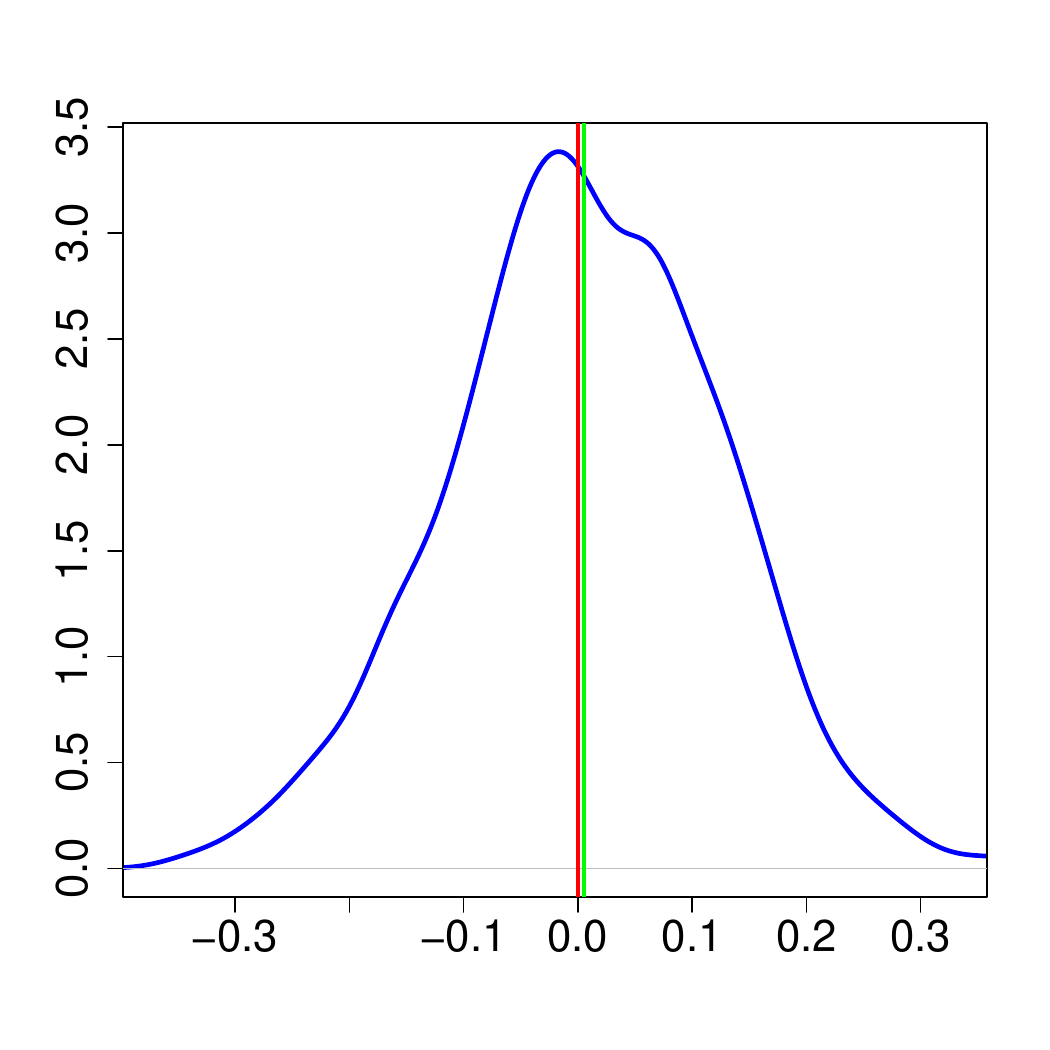} &
    \includegraphics[width=0.2\linewidth]{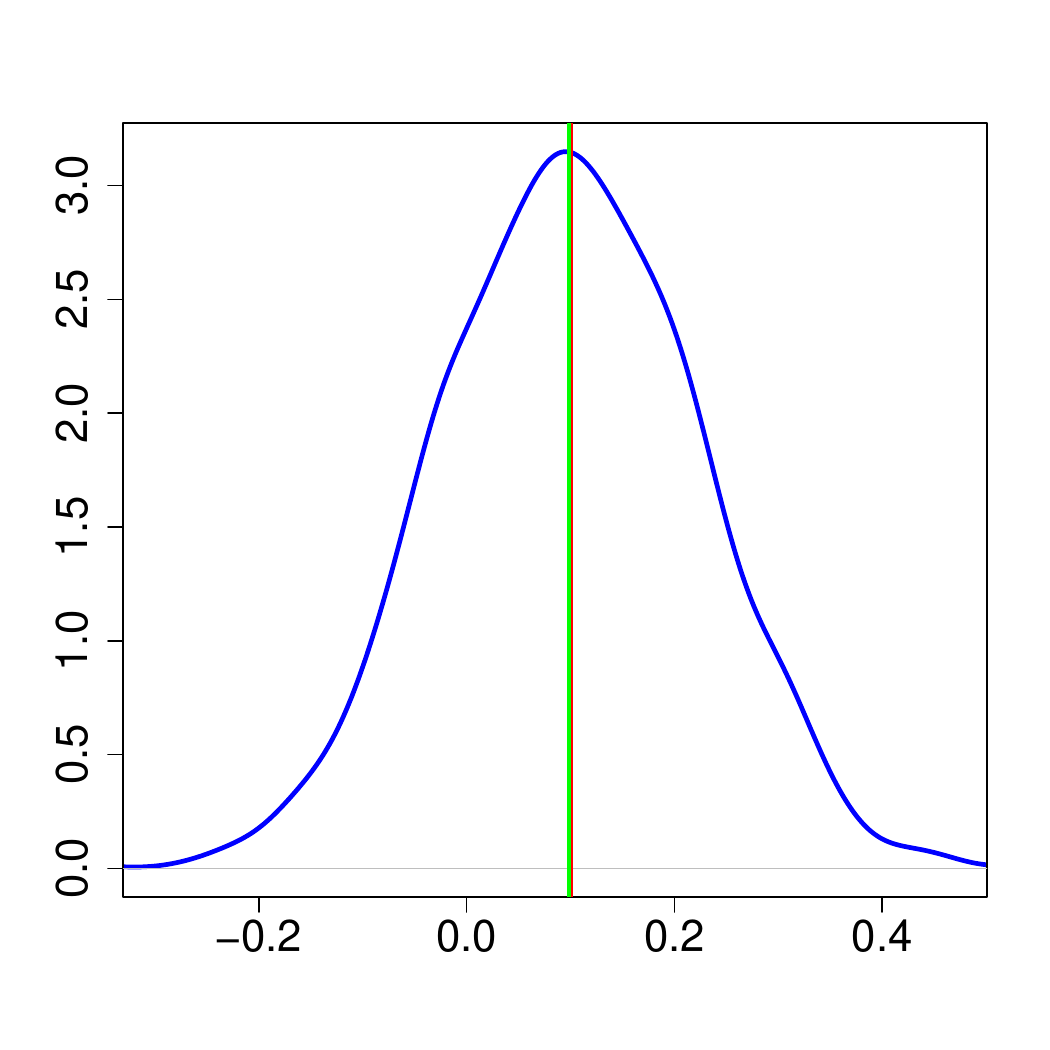} &
    \includegraphics[width=0.2\linewidth]{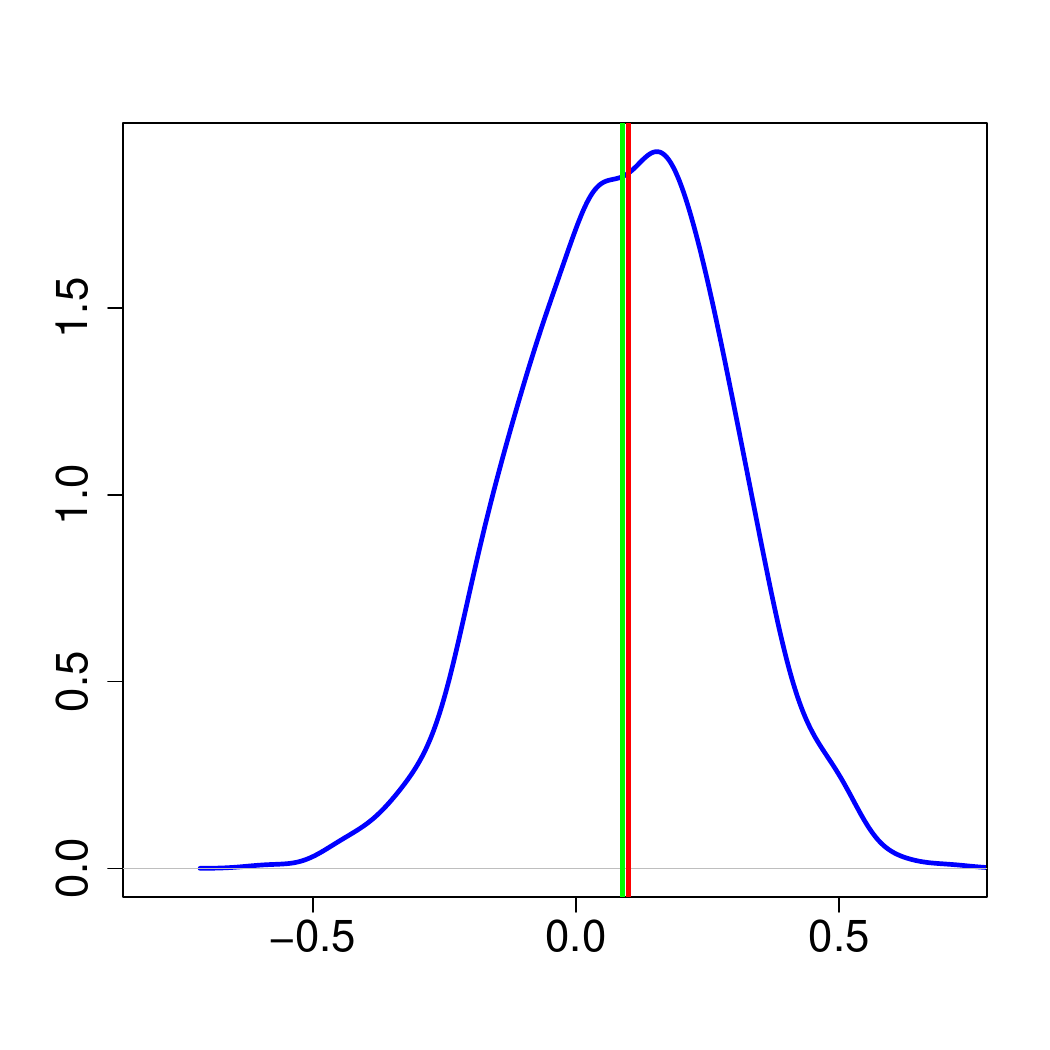} & 
    \includegraphics[width=0.2\linewidth]{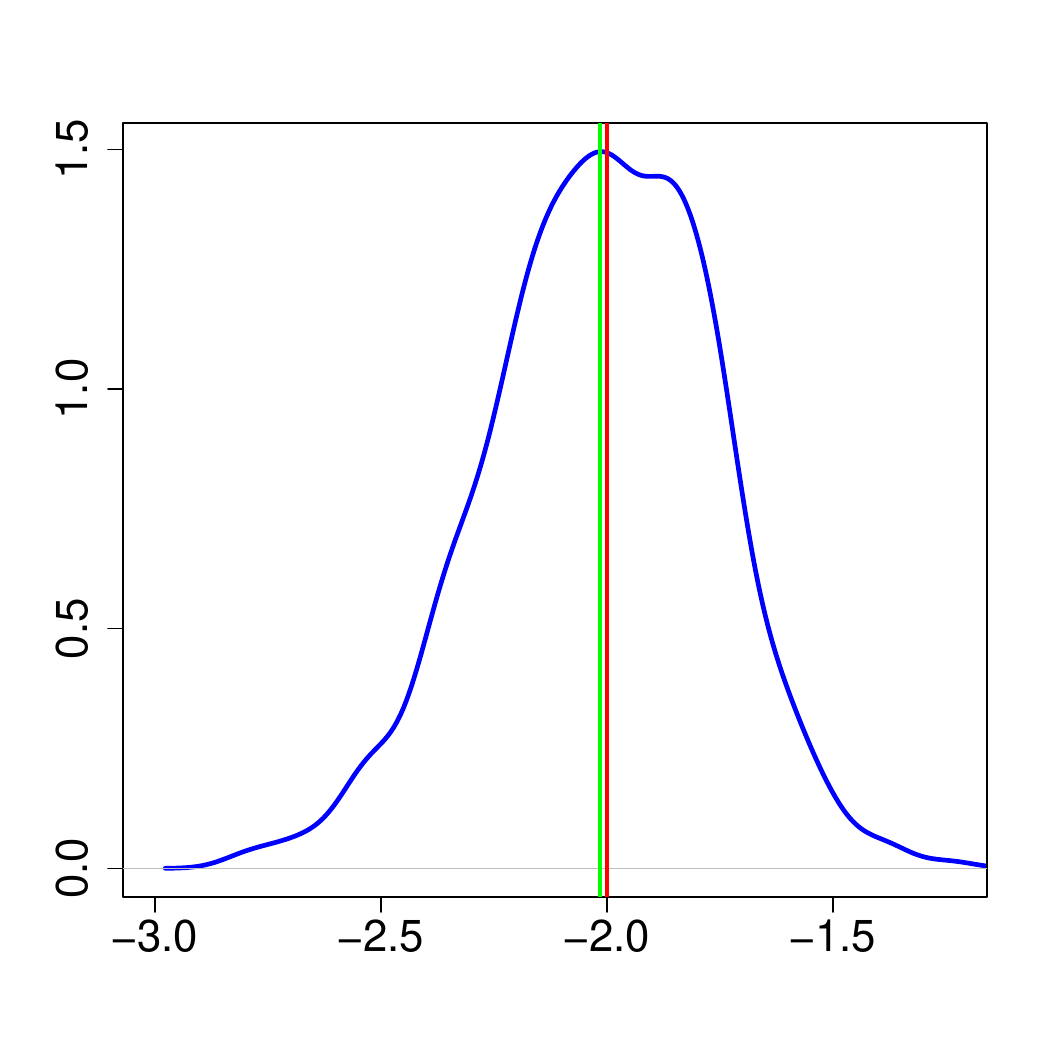} \\
    \includegraphics[width=0.2\linewidth]{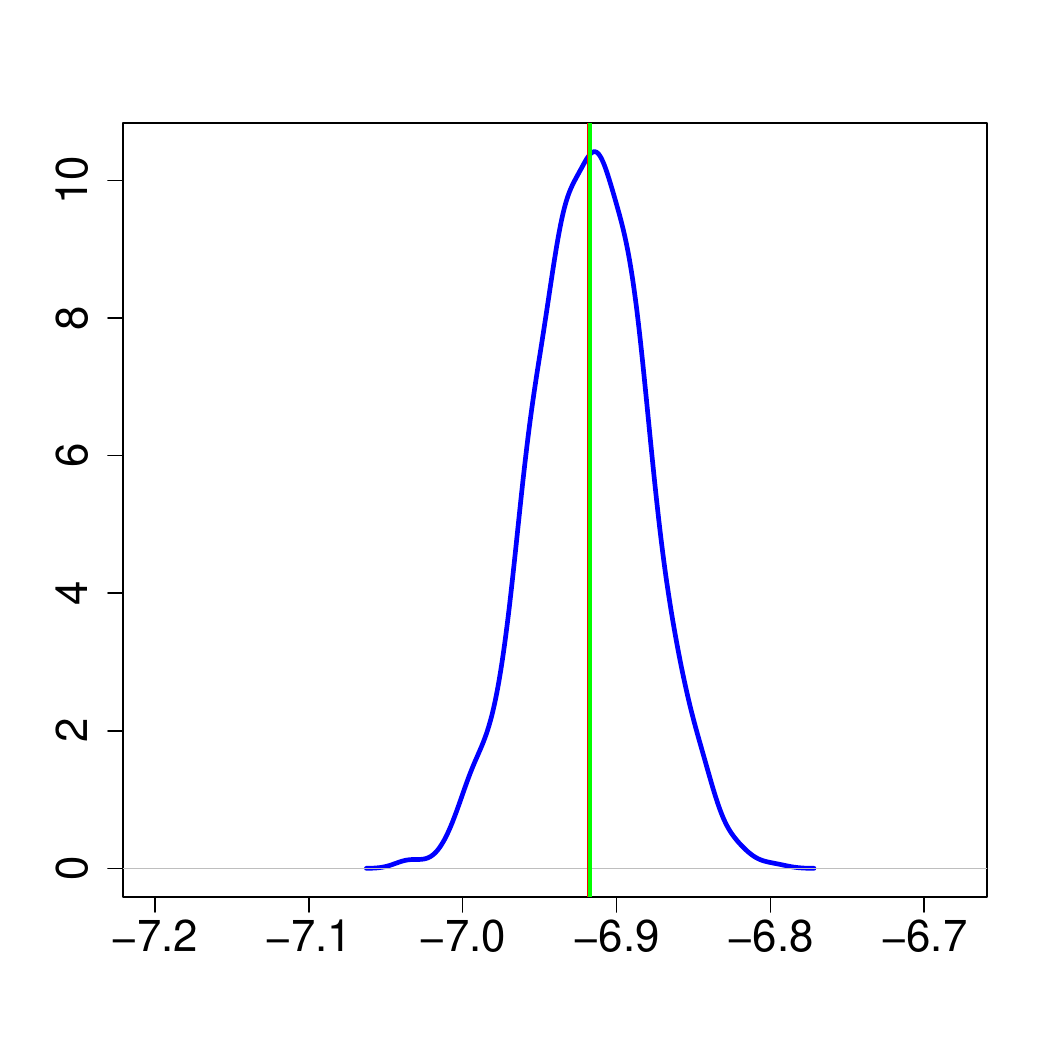} &
    \includegraphics[width=0.2\linewidth]{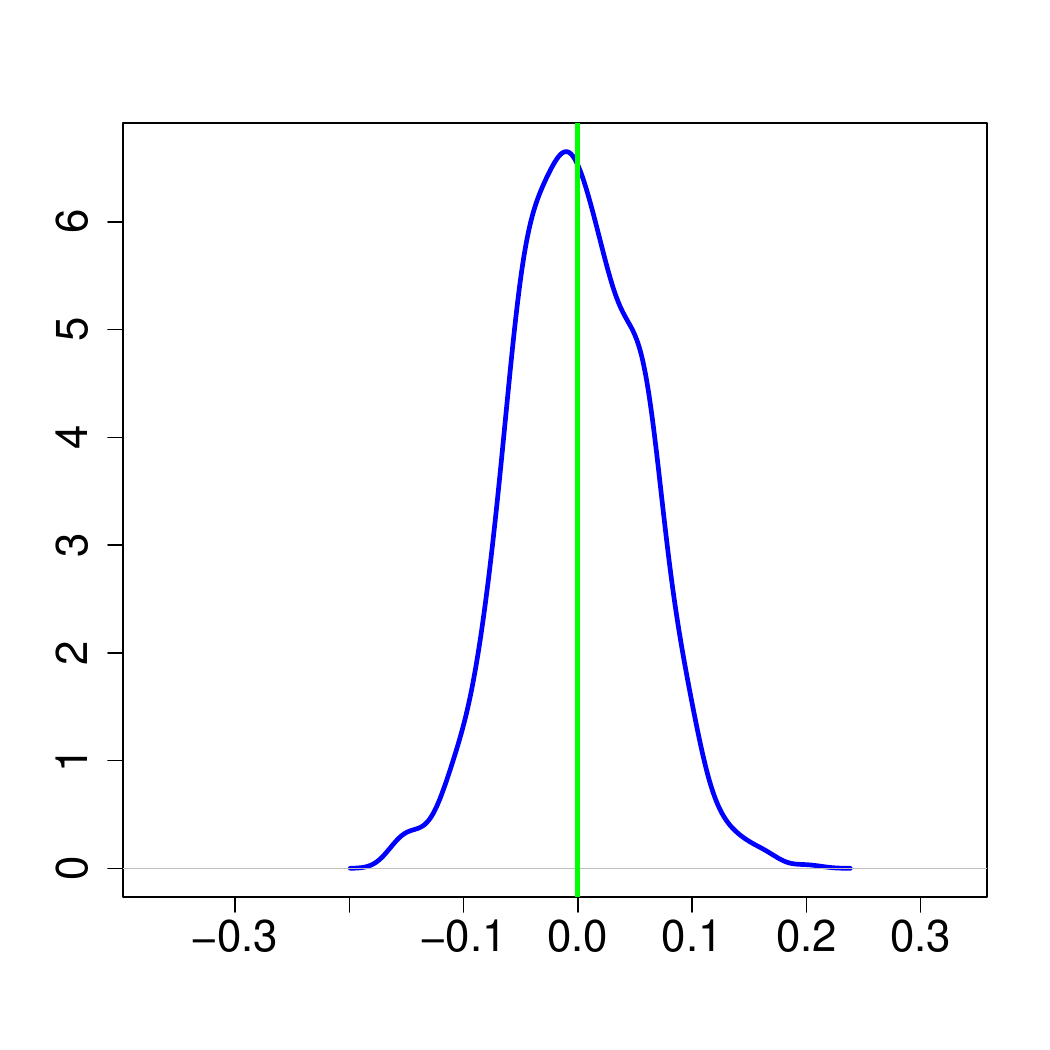} &
    \includegraphics[width=0.2\linewidth]{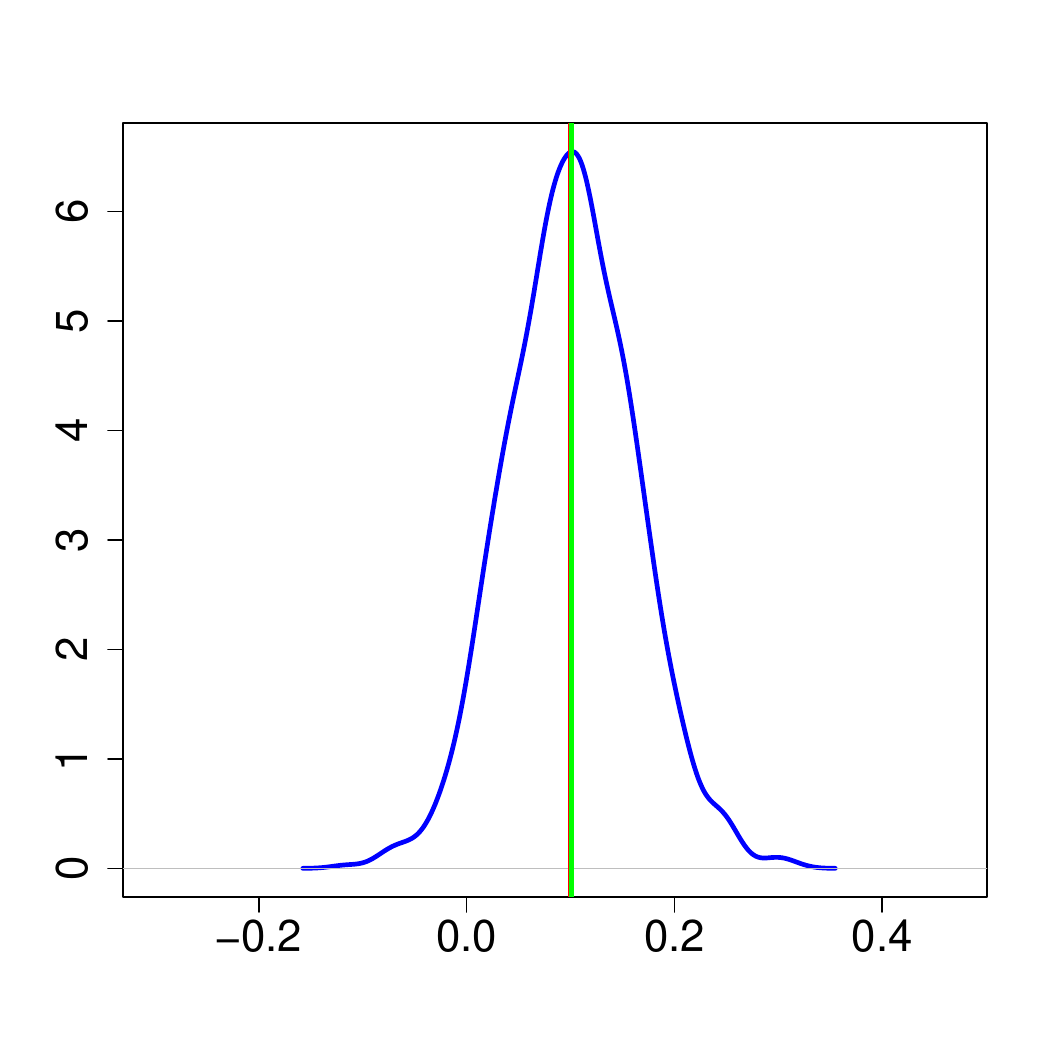} &
    \includegraphics[width=0.2\linewidth]{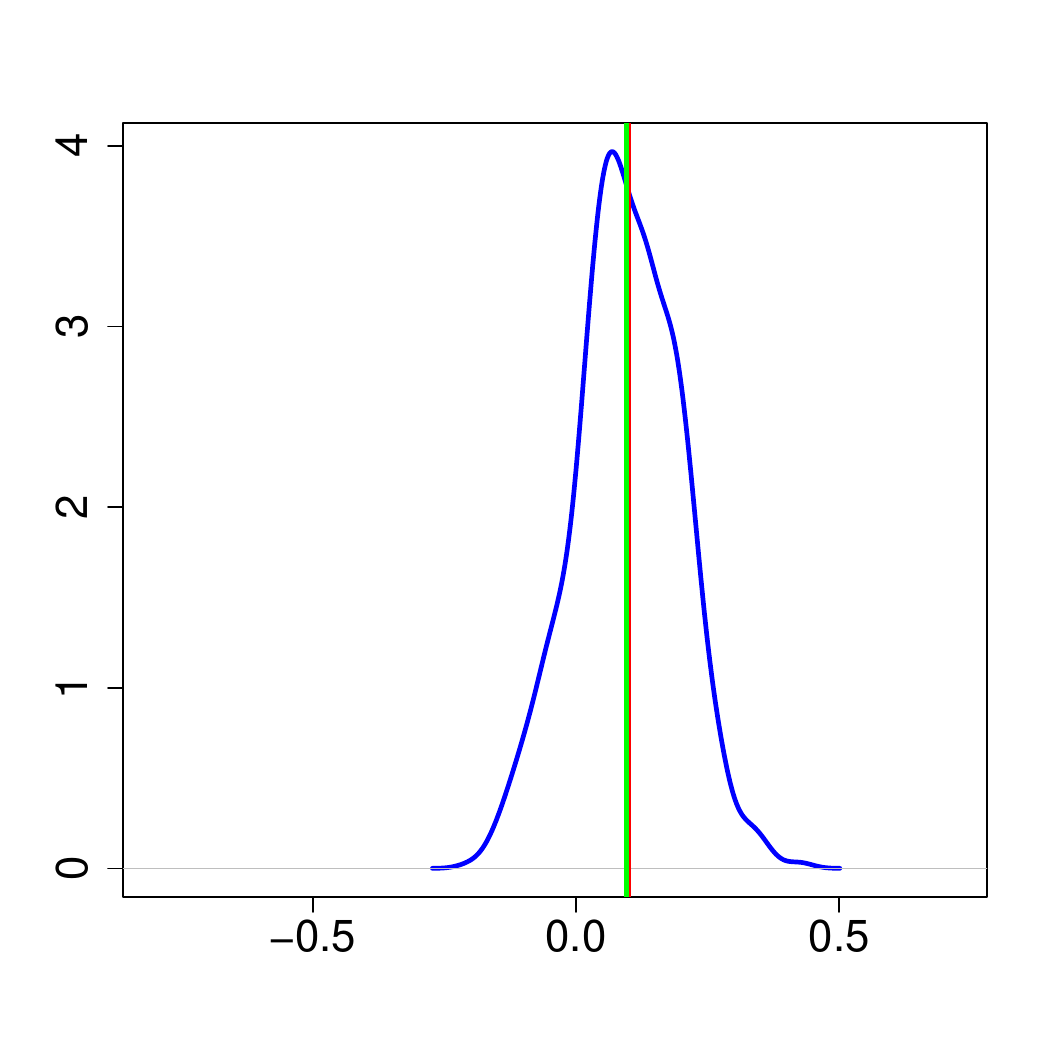} & 
    \includegraphics[width=0.2\linewidth]{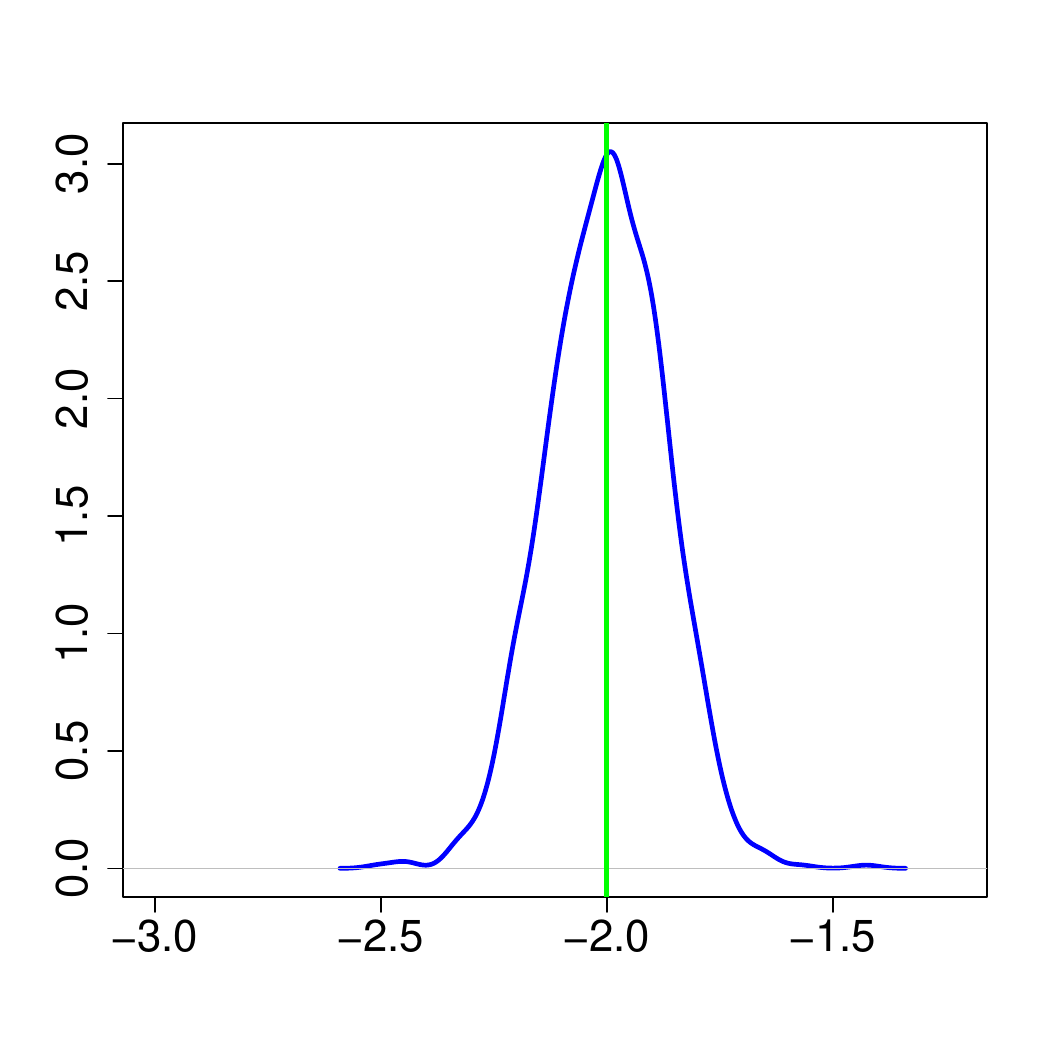}\\[\bsl]
\end{tabular}
\caption{Kernel density estimates of simulated parameter estimates for recruits
  intensity for $\widetilde W_1$ (upper row) and $\widetilde W_2$ (lower row). Red line shows the true parameter value and green line the mean of the simulated parameter estimates.}\label{fig:recruitsestimates}
\end{figure}

\begin{figure}
\centering
\begin{tabular}{ccccc}
   $\beta_{0\dd}$ & $\beta_{1\dd}$ & $\beta_{2\dd}$ & $\gm_{1\dd}$ & $\gm_{2\dd}$ \\
    \includegraphics[width=0.2\linewidth]{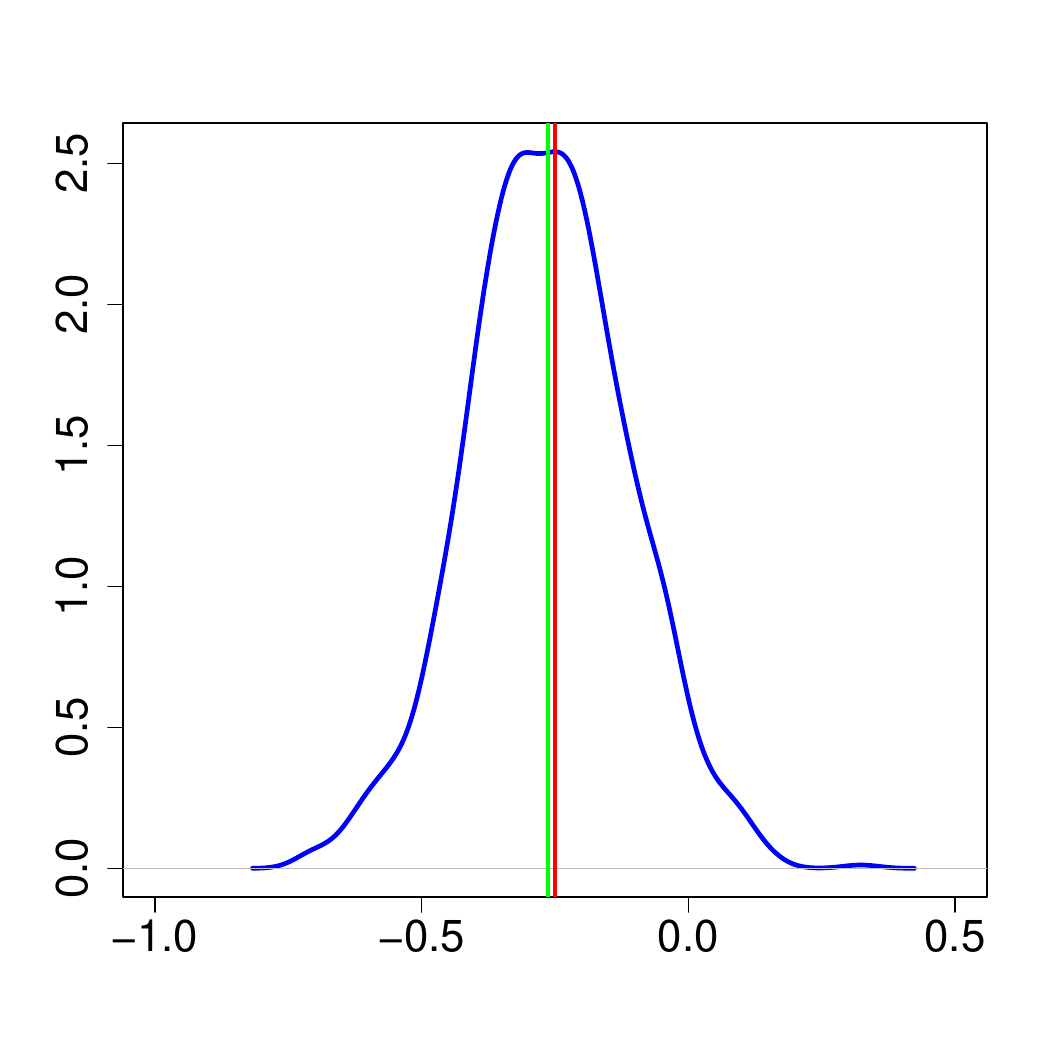} &
    \includegraphics[width=0.2\linewidth]{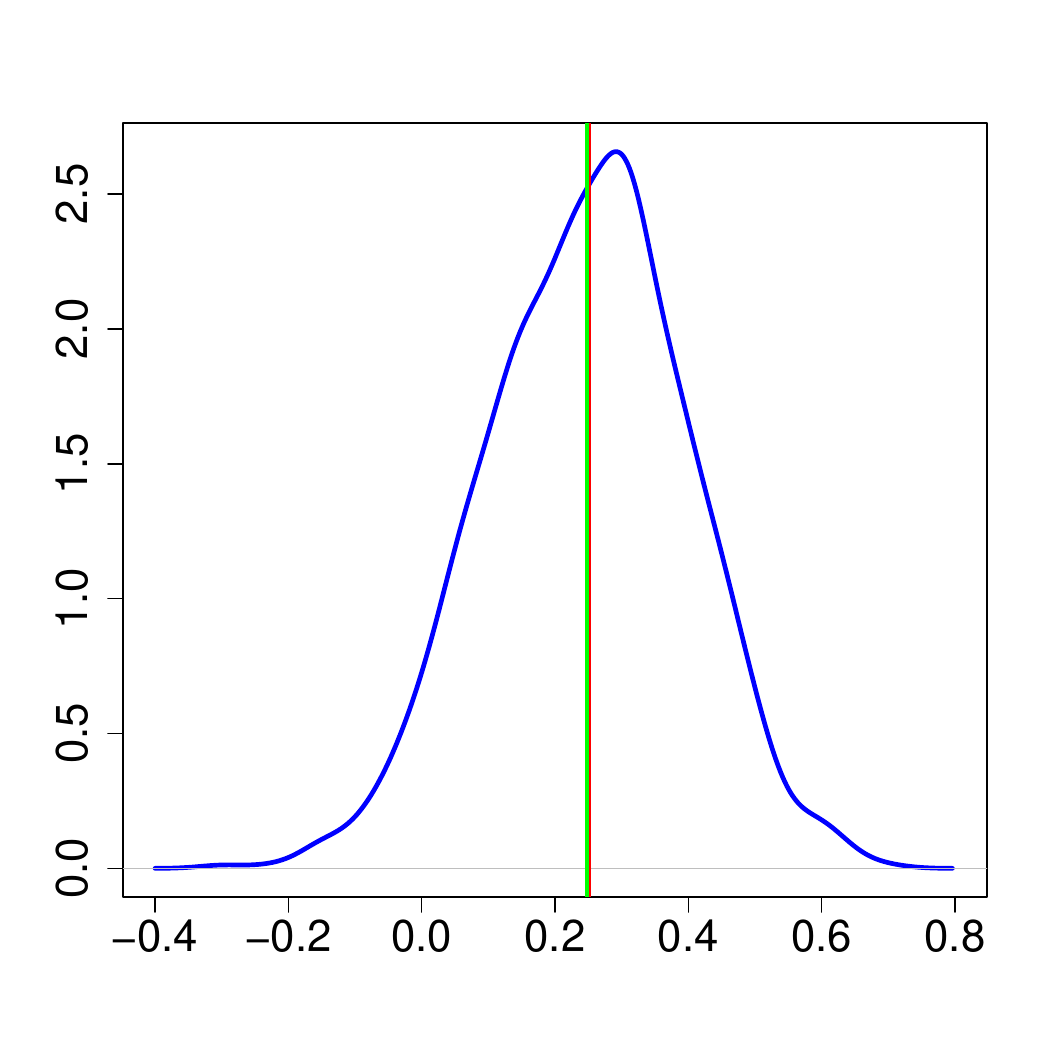} &
    \includegraphics[width=0.2\linewidth]{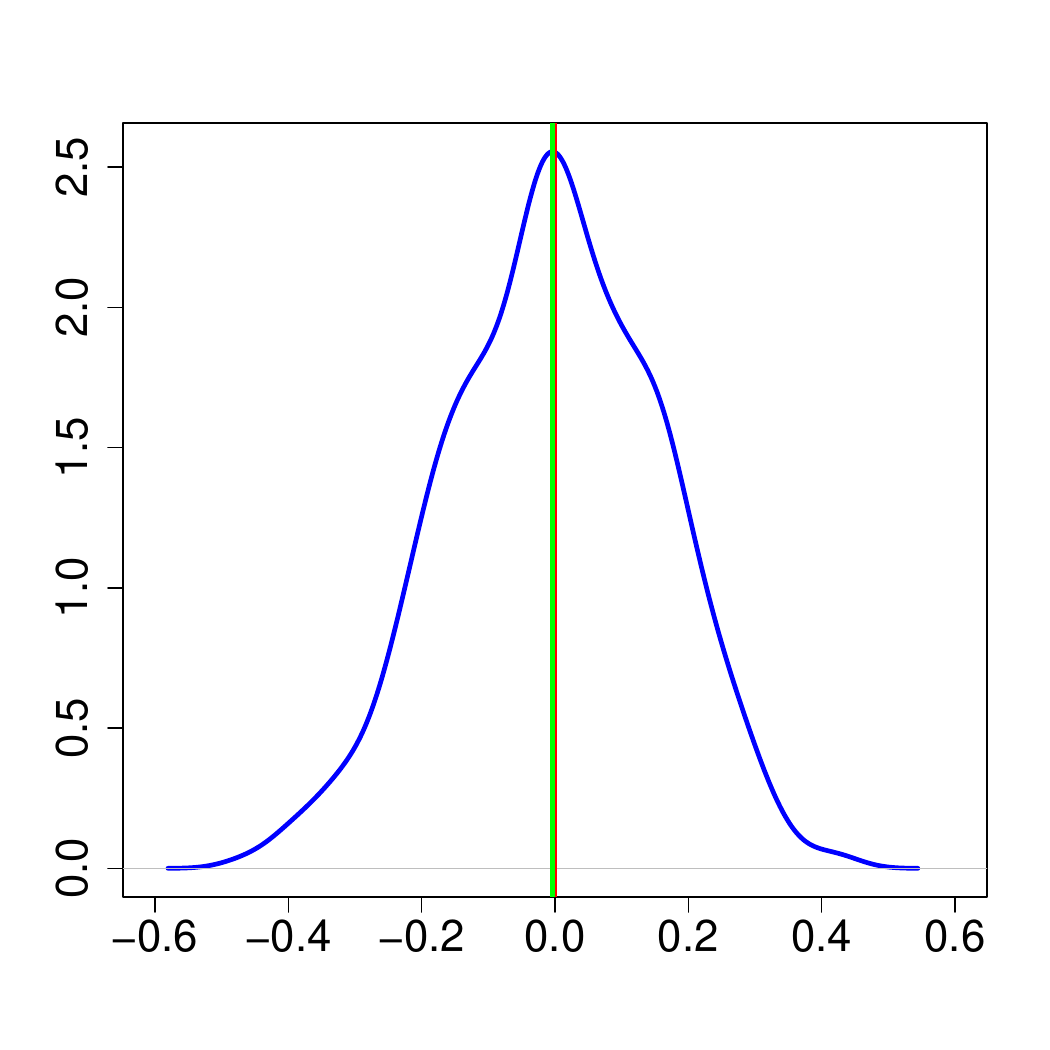} &
    \includegraphics[width=0.2\linewidth]{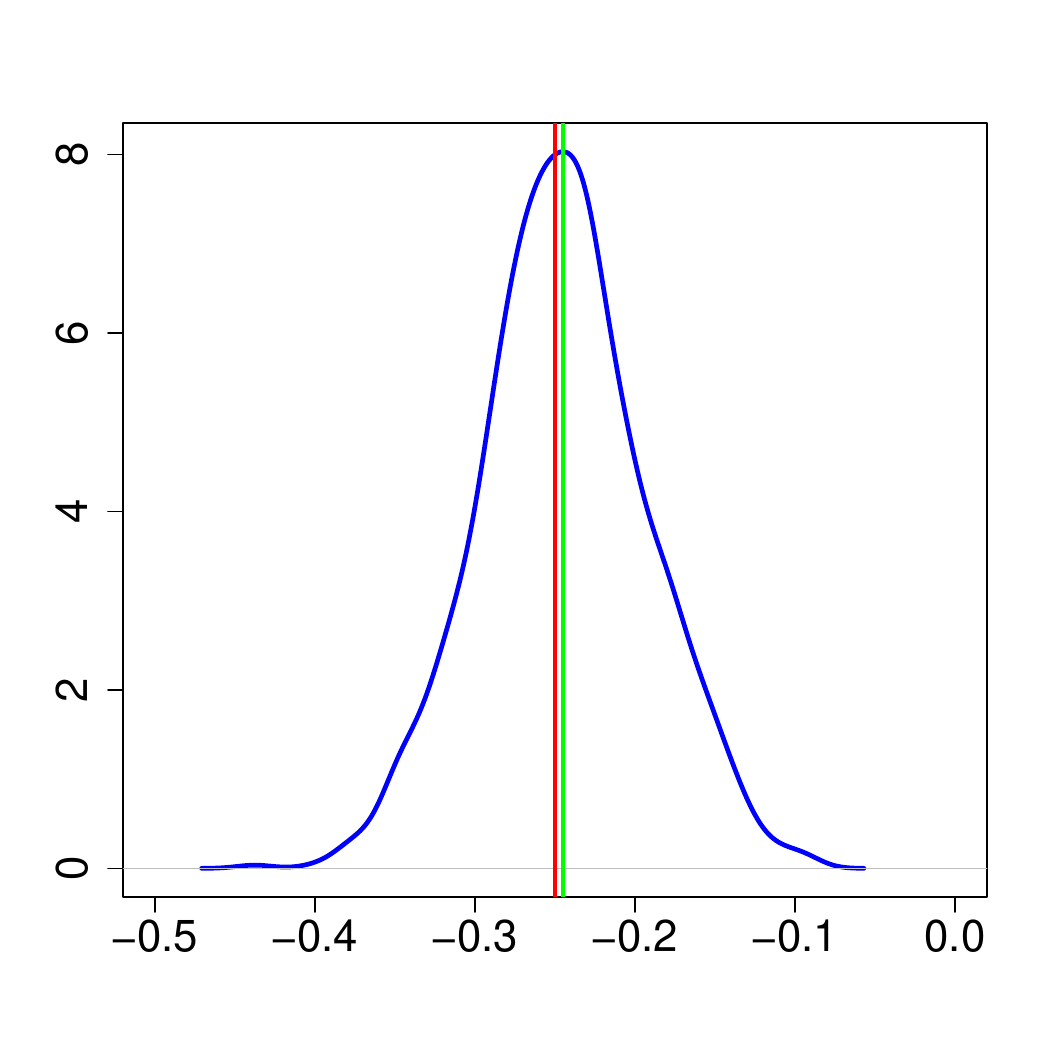} &
    \includegraphics[width=0.2\linewidth]{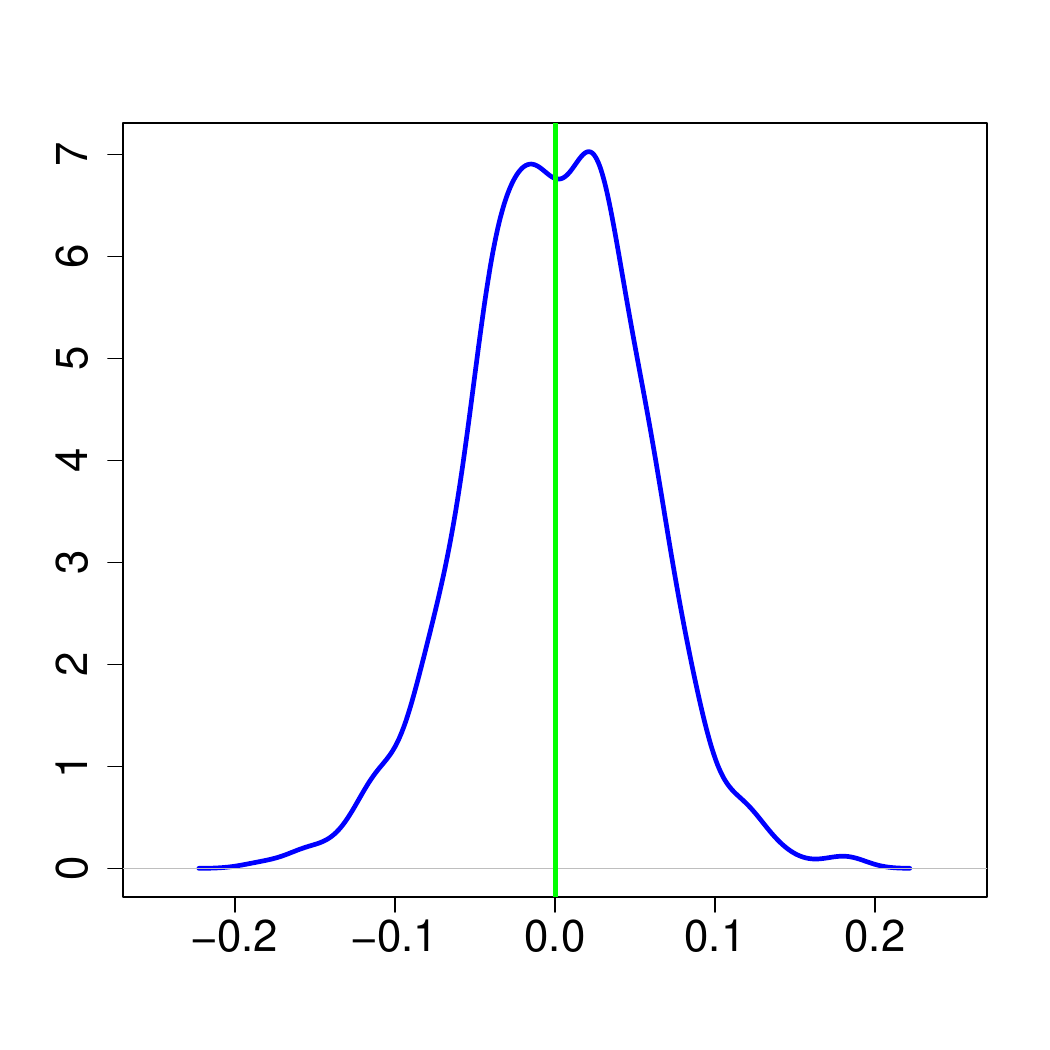}\\
    \includegraphics[width=0.2\linewidth]{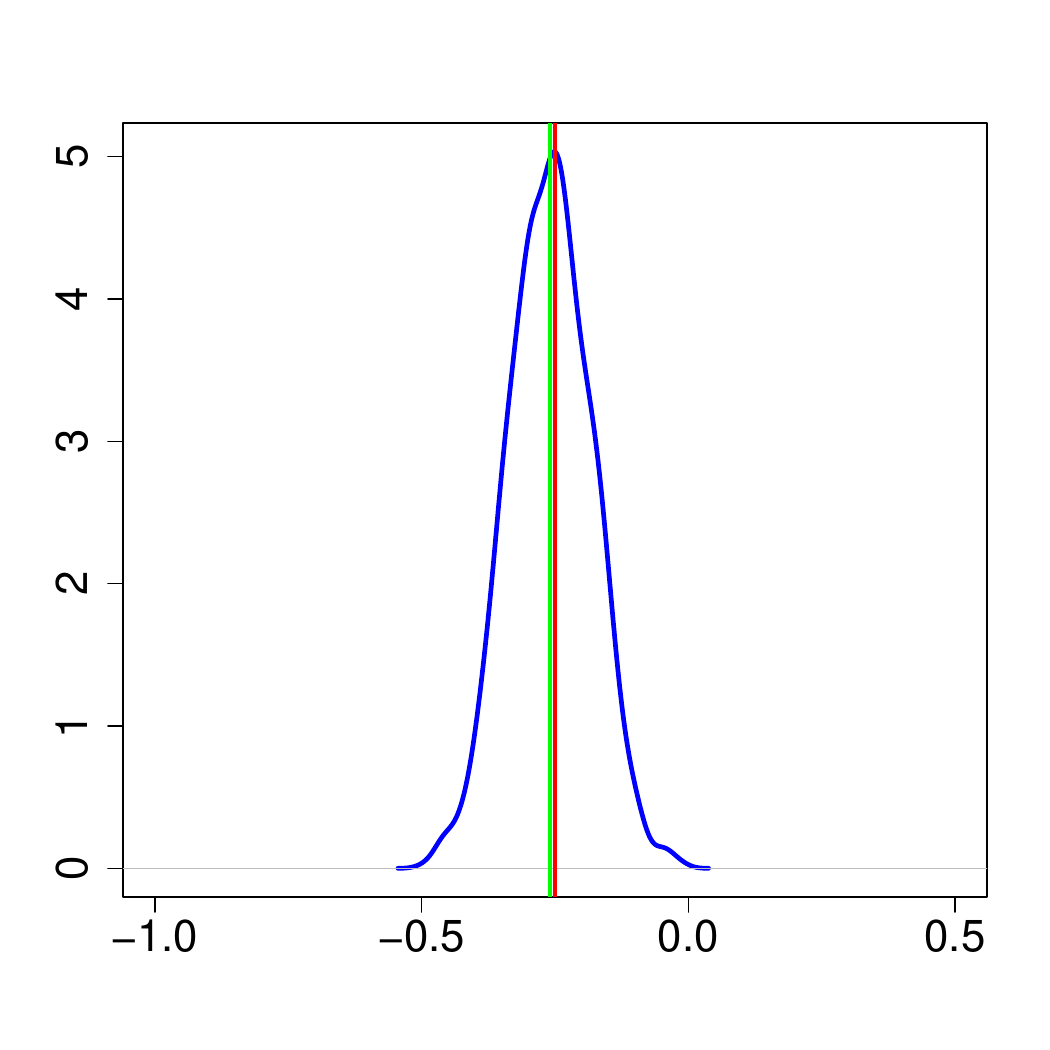} &
    \includegraphics[width=0.2\linewidth]{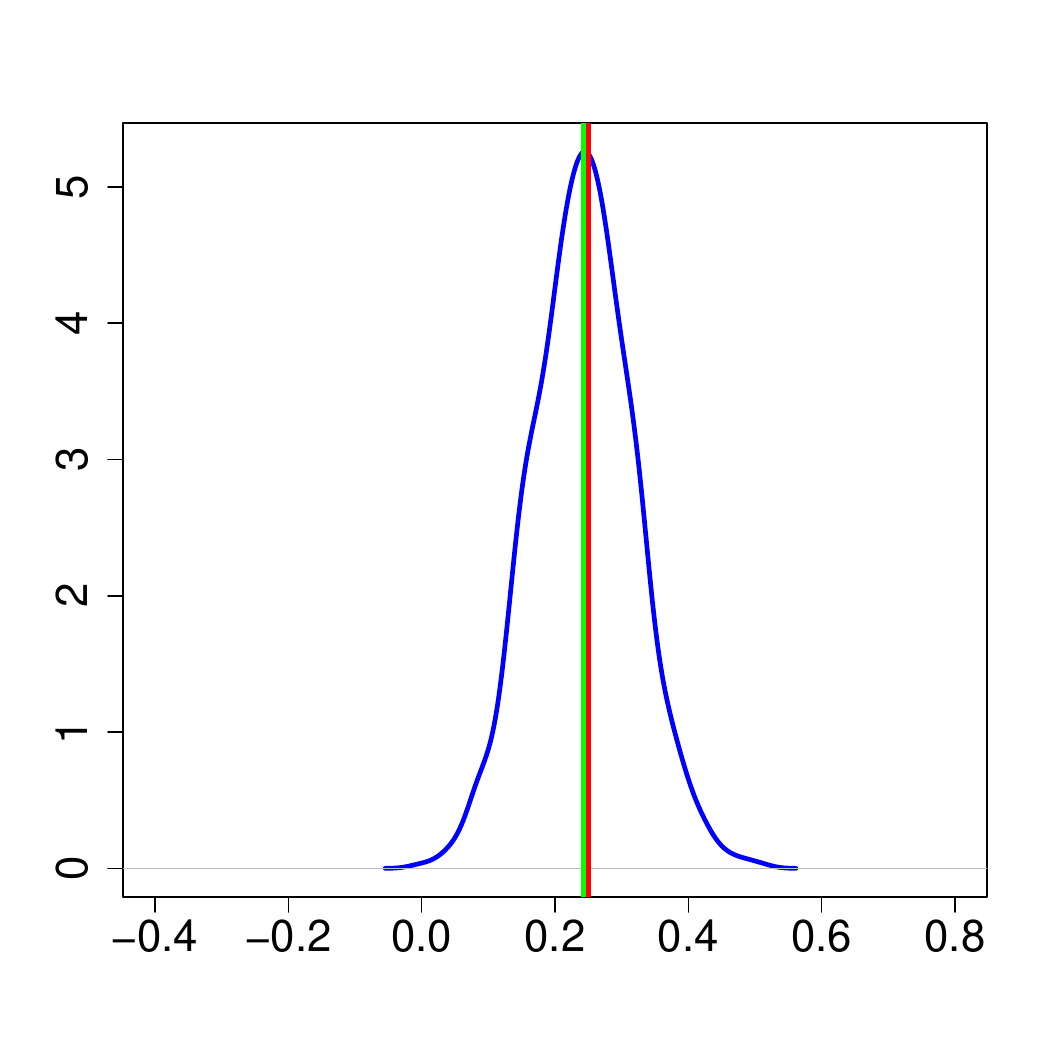} &
    \includegraphics[width=0.2\linewidth]{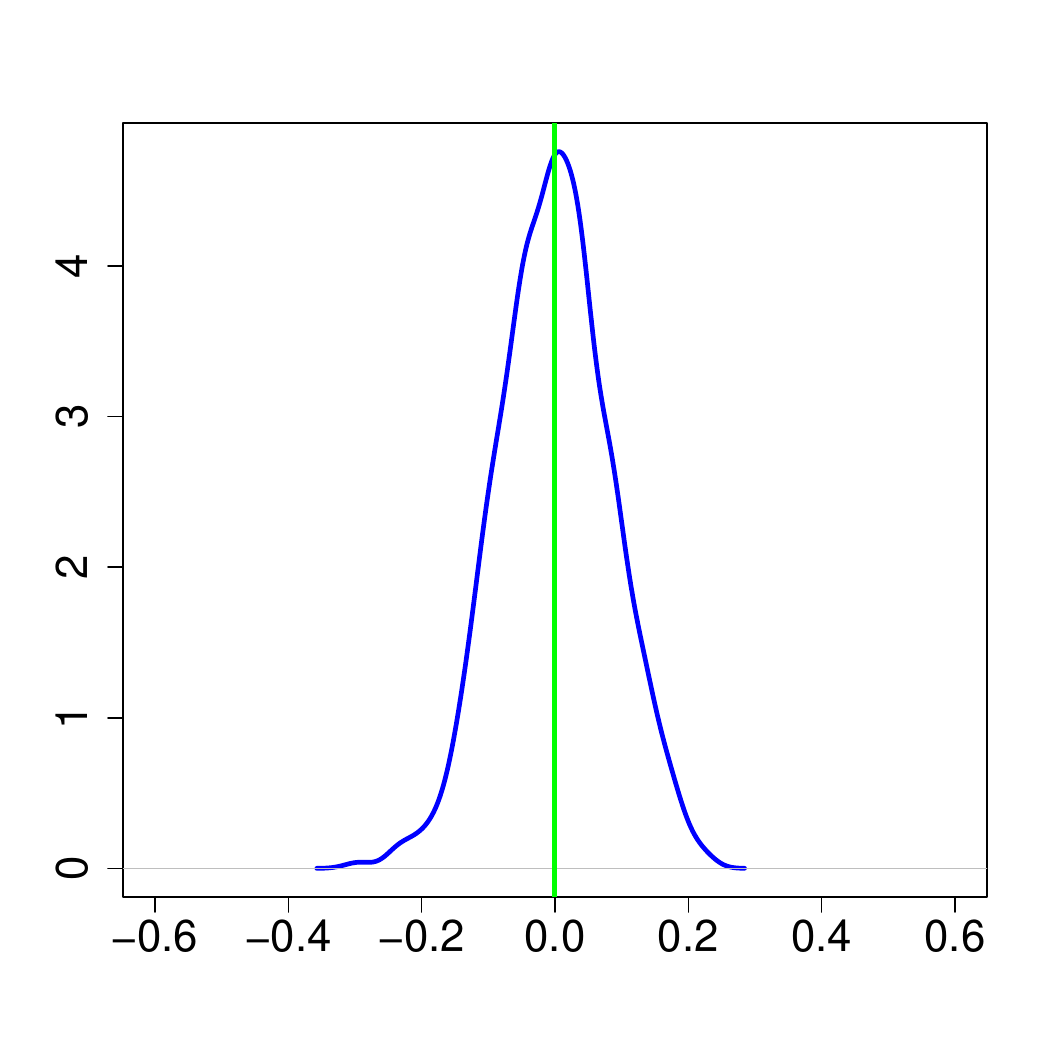} &
    \includegraphics[width=0.2\linewidth]{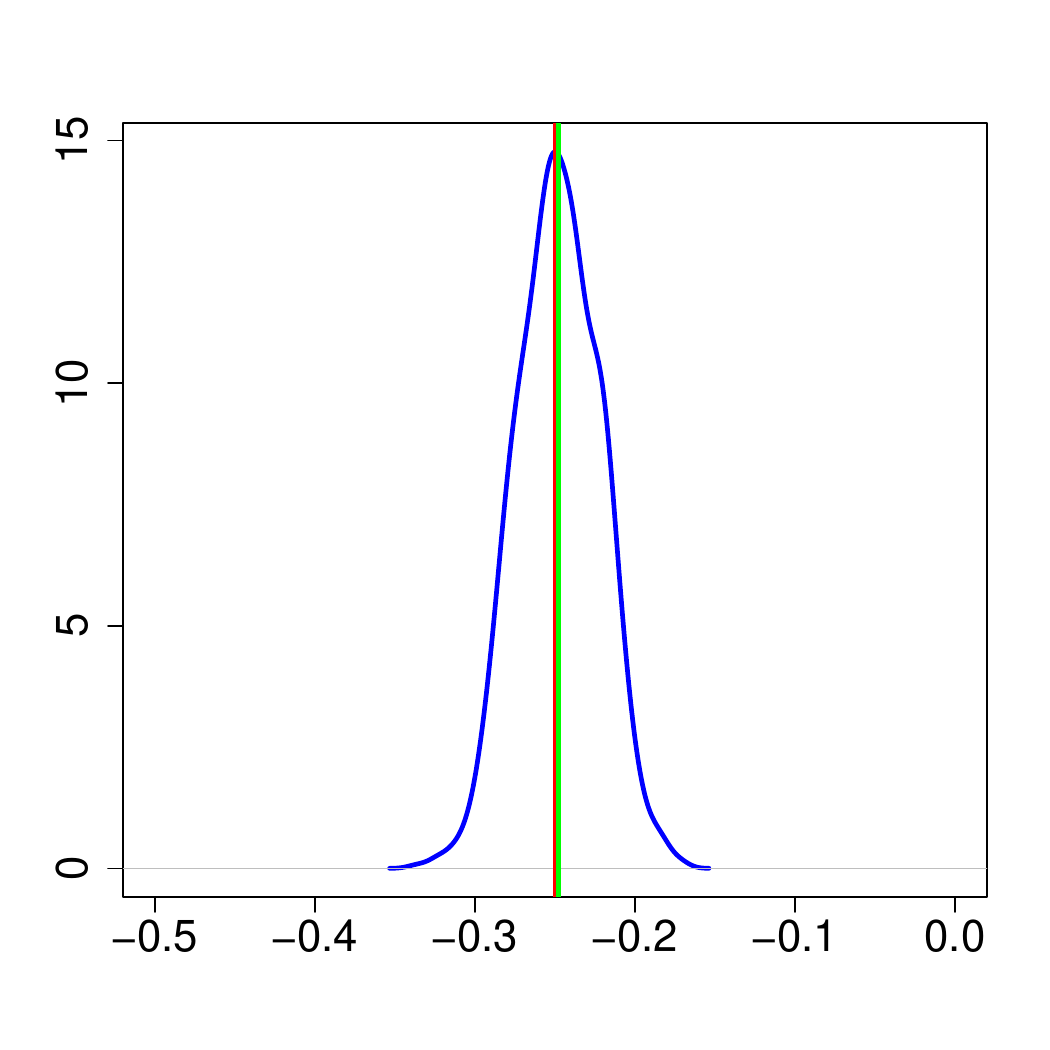} &
    \includegraphics[width=0.2\linewidth]{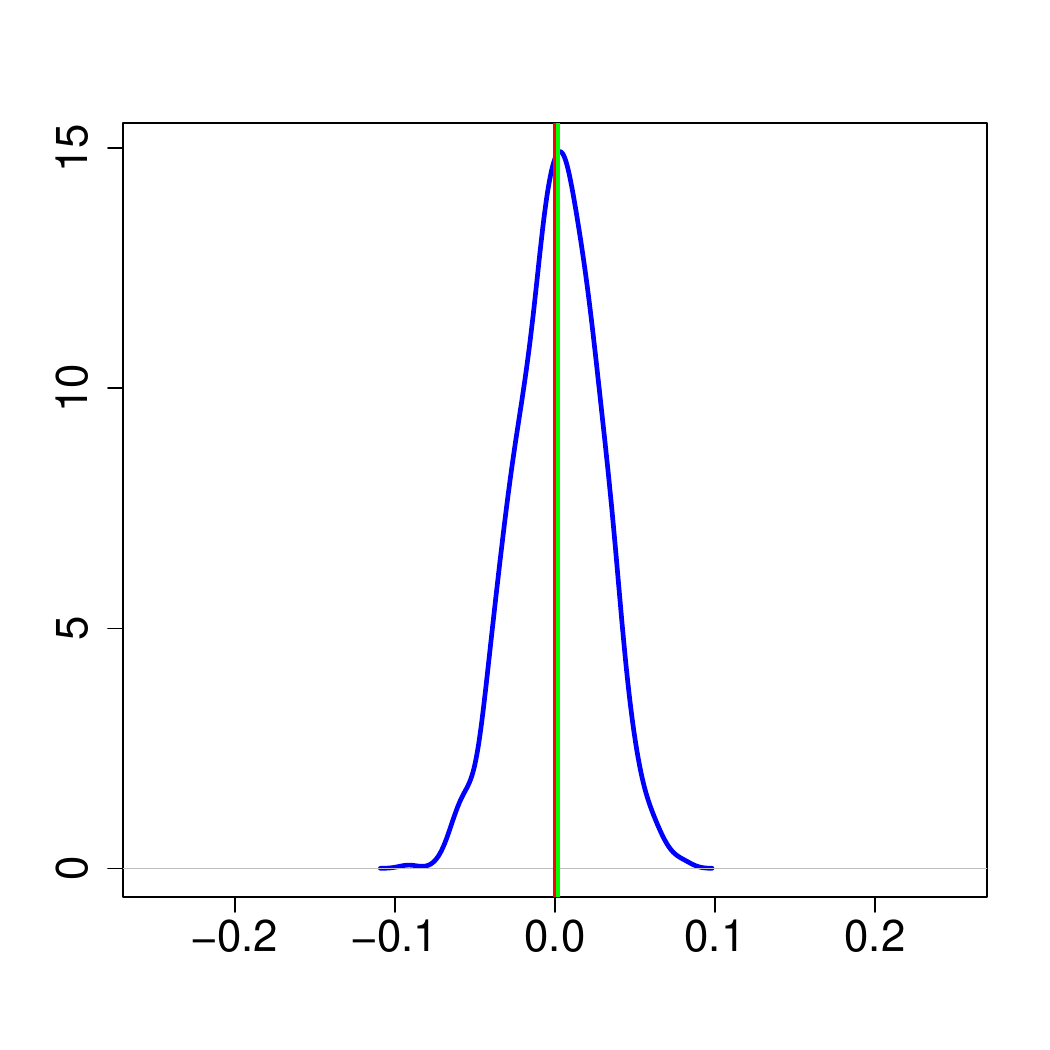}\\[\bsl]
\end{tabular}
\caption{Kernel density estimates of simulated parameter estimates for death
  probabilities for $\widetilde W_1$ (upper row) and $\widetilde W_2$ (lower row). Red line shows the true parameter value and green line the mean of the simulated parameter estimates.}\label{fig:deathestimates}
\end{figure}

Figures~\ref{fig:recruitsestimates} and \ref{fig:deathestimates} show
kernel density estimates of the $1000$ simulated recruit and death regression parameter estimates.

\begin{figure}
\centering
\begin{tabular}{ccccc}
     $\beta_{0\dd}$ &$\beta_{1\dd}$ & $\beta_{2\dd}$ & $\gm_{1\dd}$ & $\gm_{2\dd}$ \\
        \adjustbox{valign=m,vspace=1pt}{\includegraphics[width=0.2\linewidth]{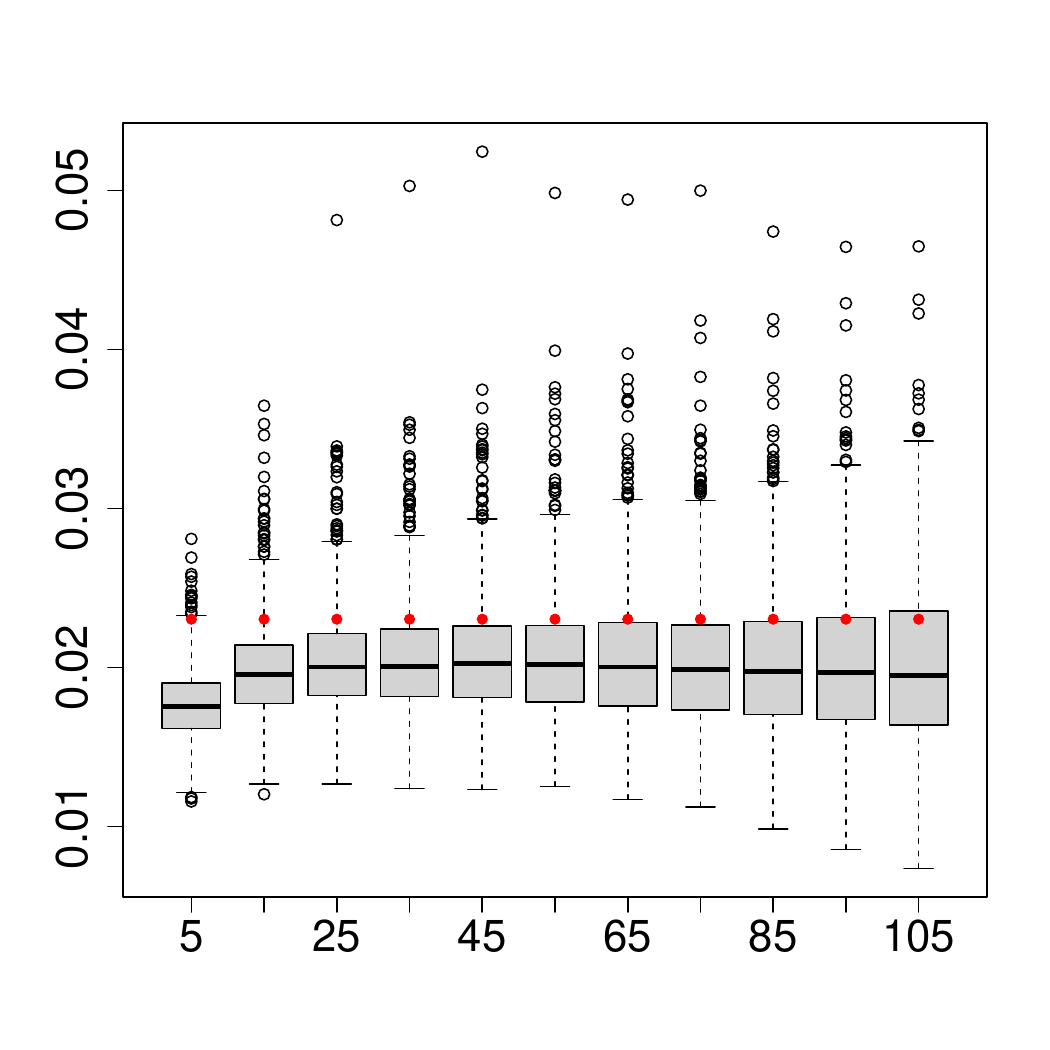}} & \adjustbox{valign=m,vspace=1pt}{\includegraphics[width=0.2\linewidth]{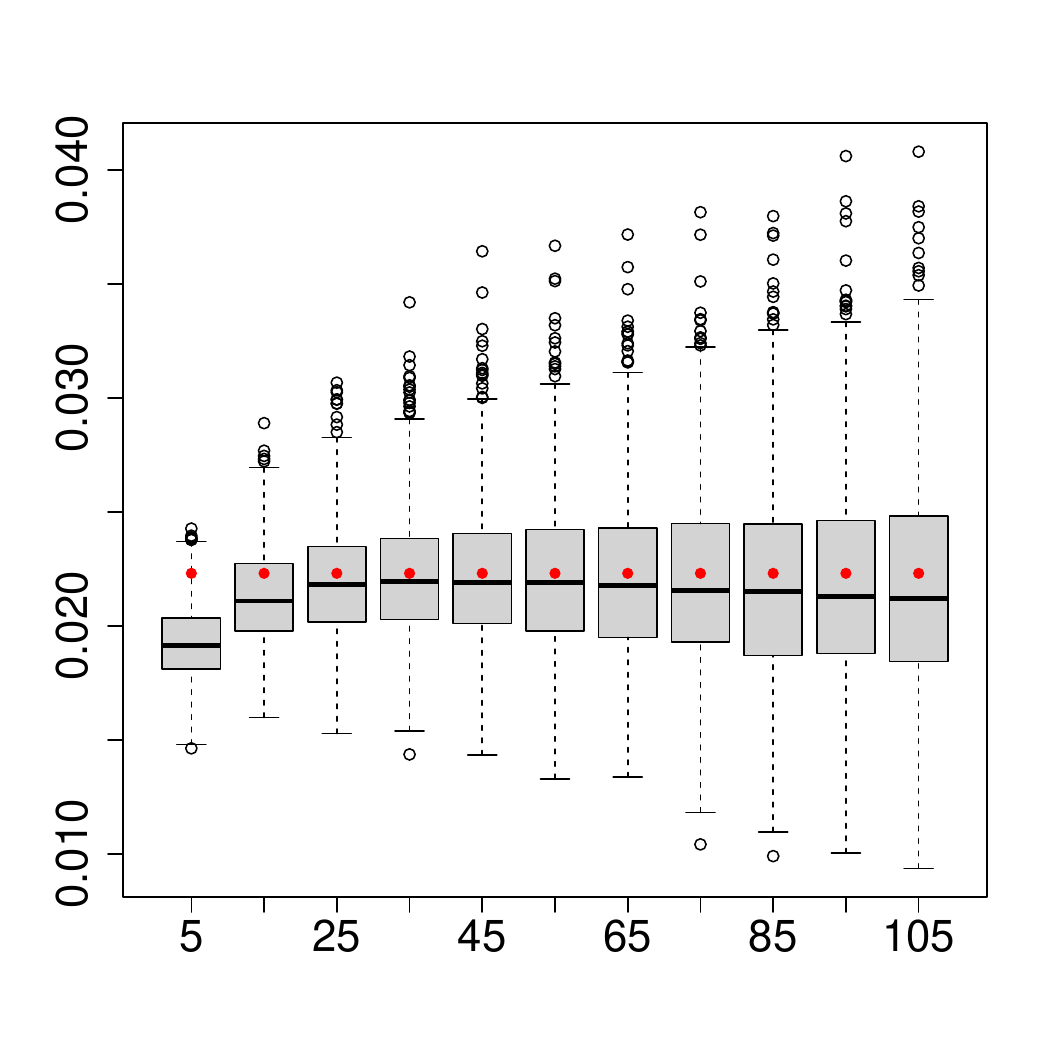}} &
        \adjustbox{valign=m,vspace=1pt}{\includegraphics[width=0.2\linewidth]{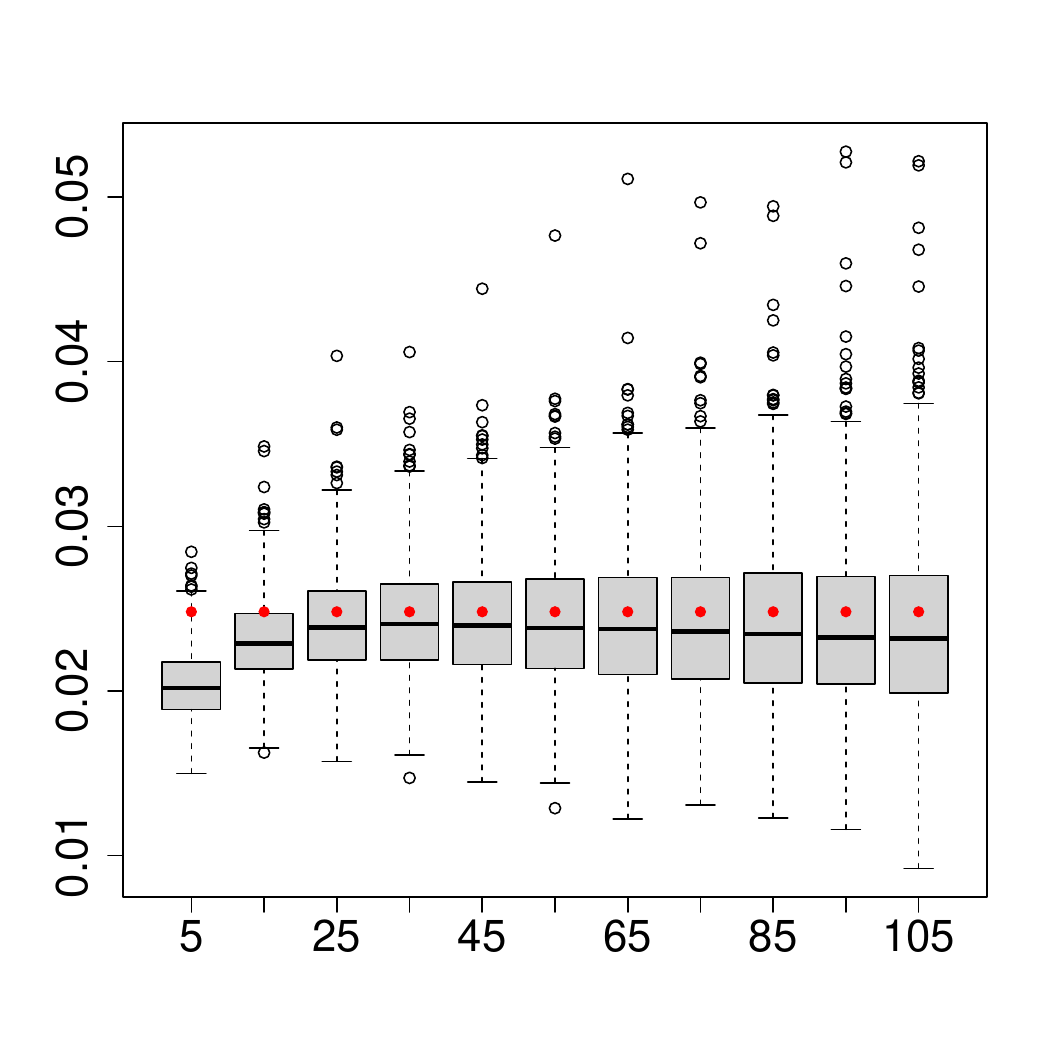}} &
        \adjustbox{valign=m,vspace=1pt}{\includegraphics[width=0.2\linewidth]{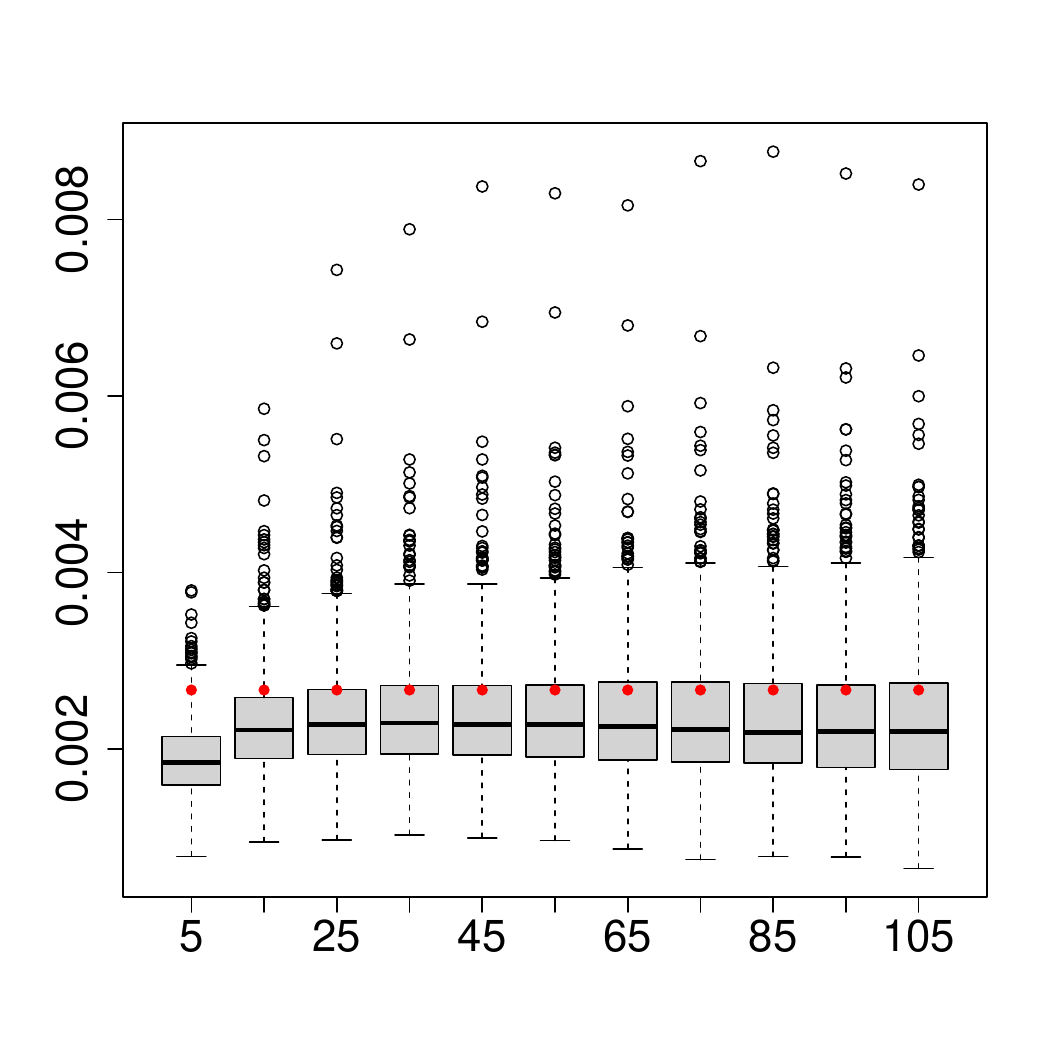}} & 
        \adjustbox{valign=m,vspace=1pt}{\includegraphics[width=0.2\linewidth]{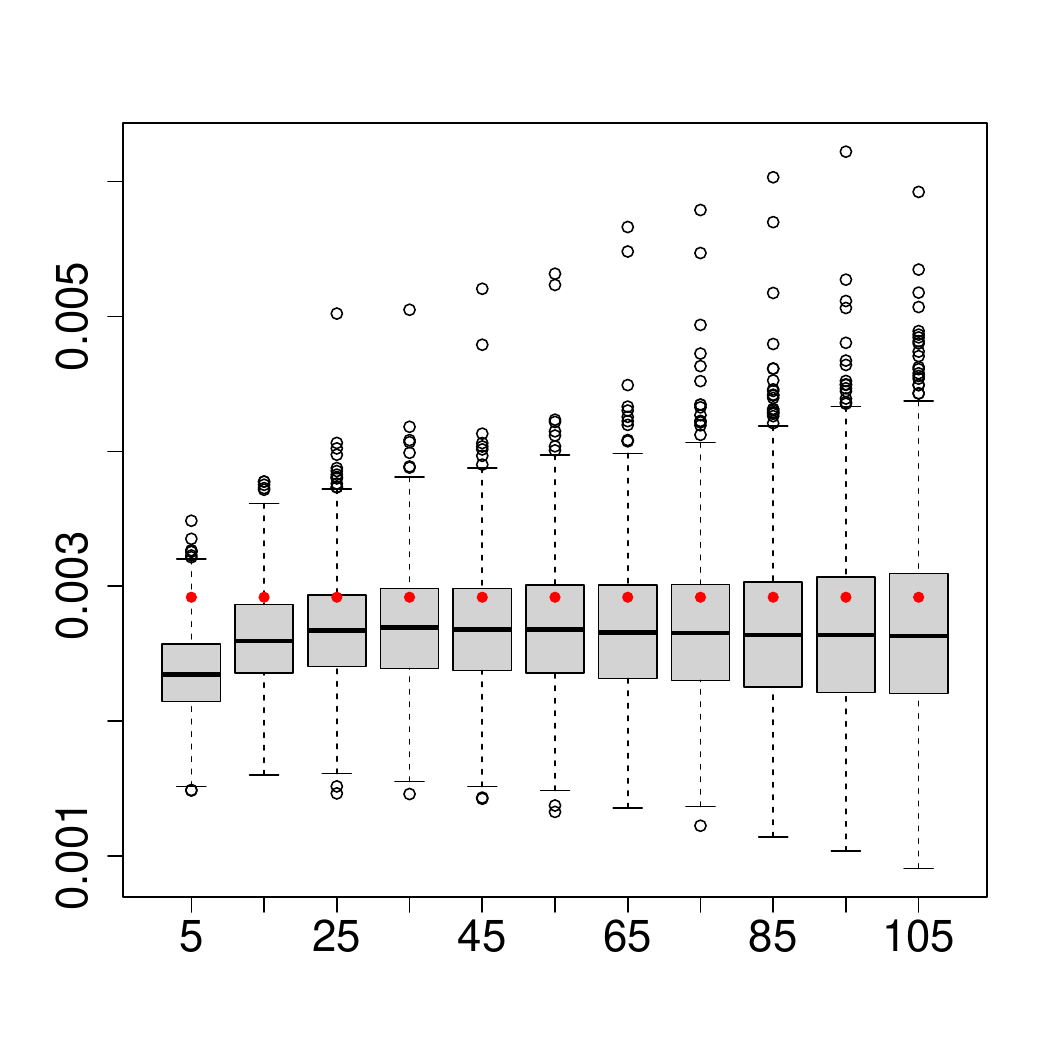}}\\
        \adjustbox{valign=m,vspace=1pt}{\includegraphics[width=0.2\linewidth]{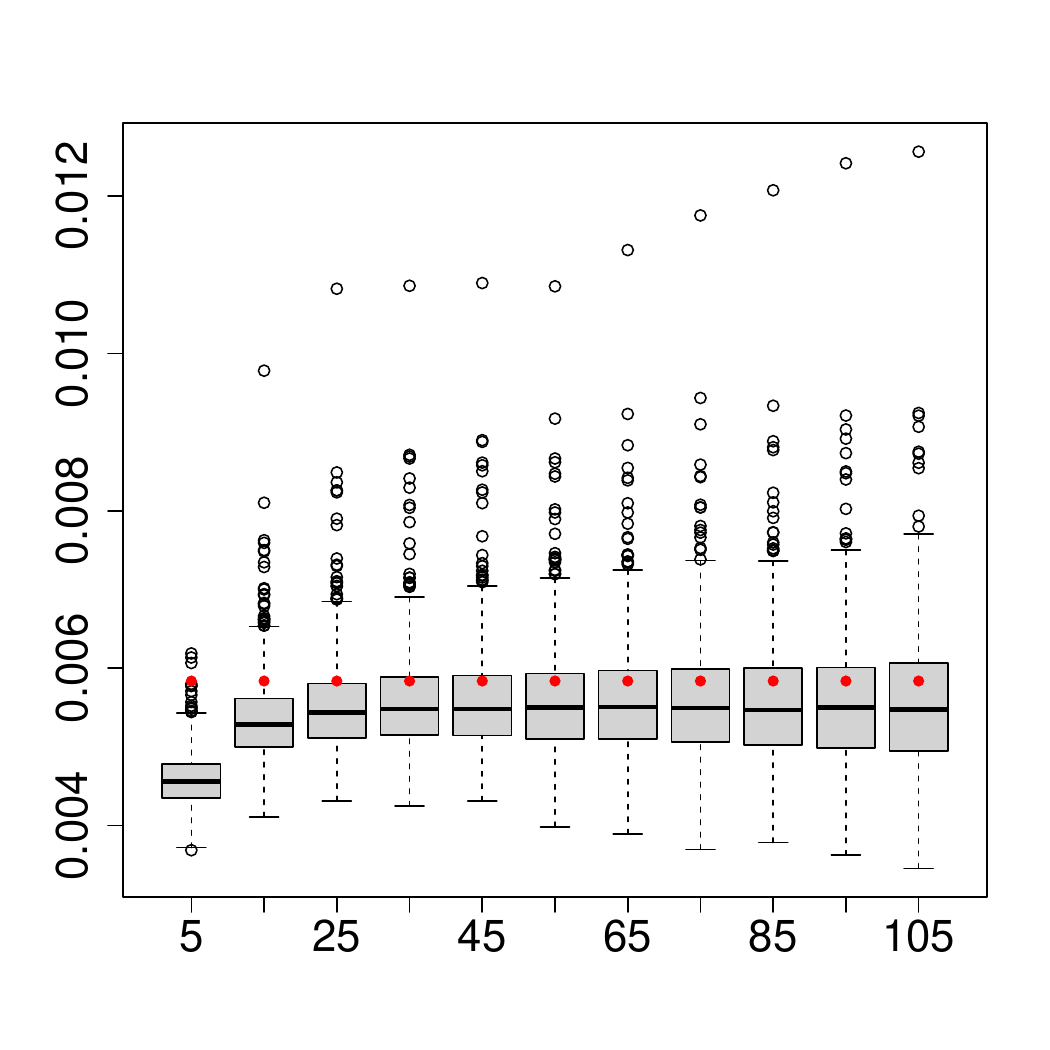}} &        \adjustbox{valign=m,vspace=1pt}{\includegraphics[width=0.2\linewidth]{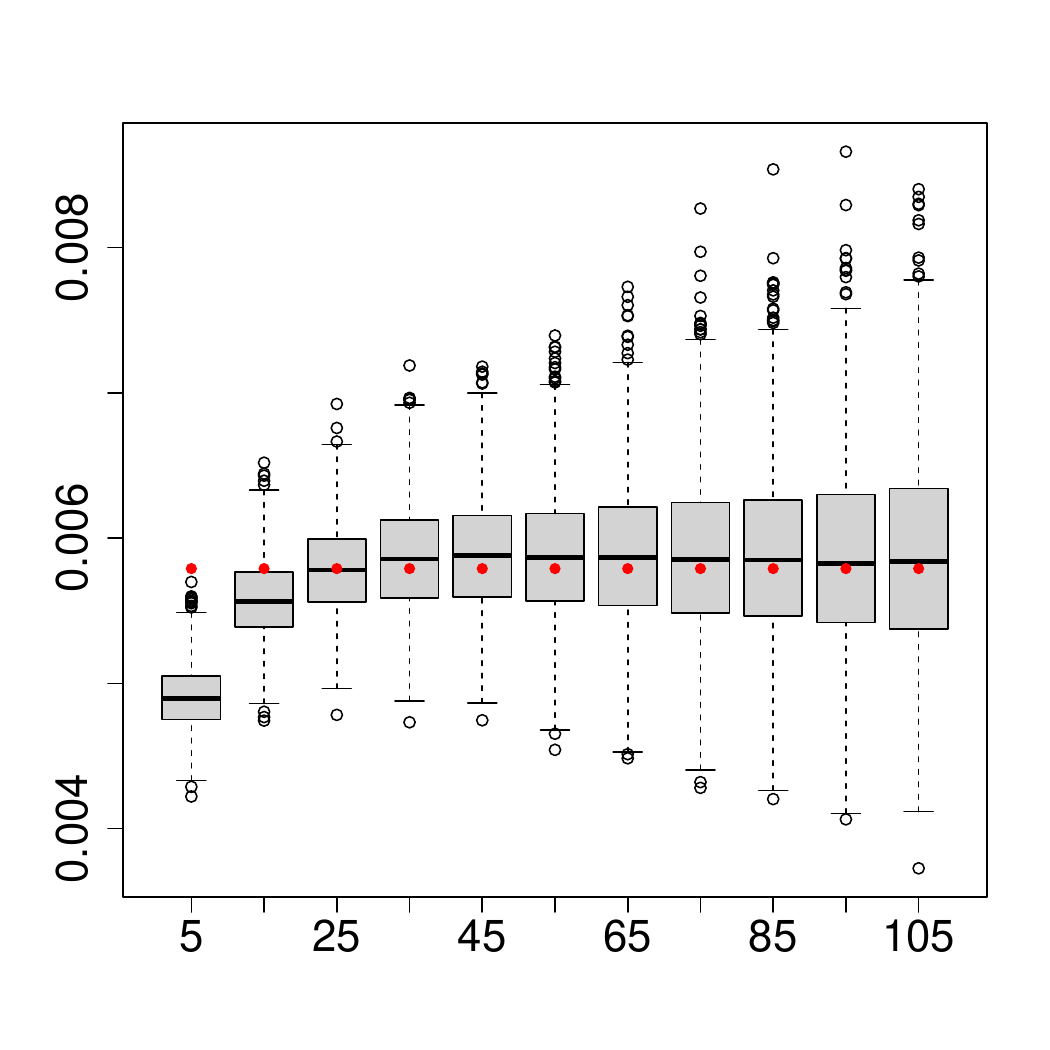}} &
        \adjustbox{valign=m,vspace=1pt}{\includegraphics[width=0.2\linewidth]{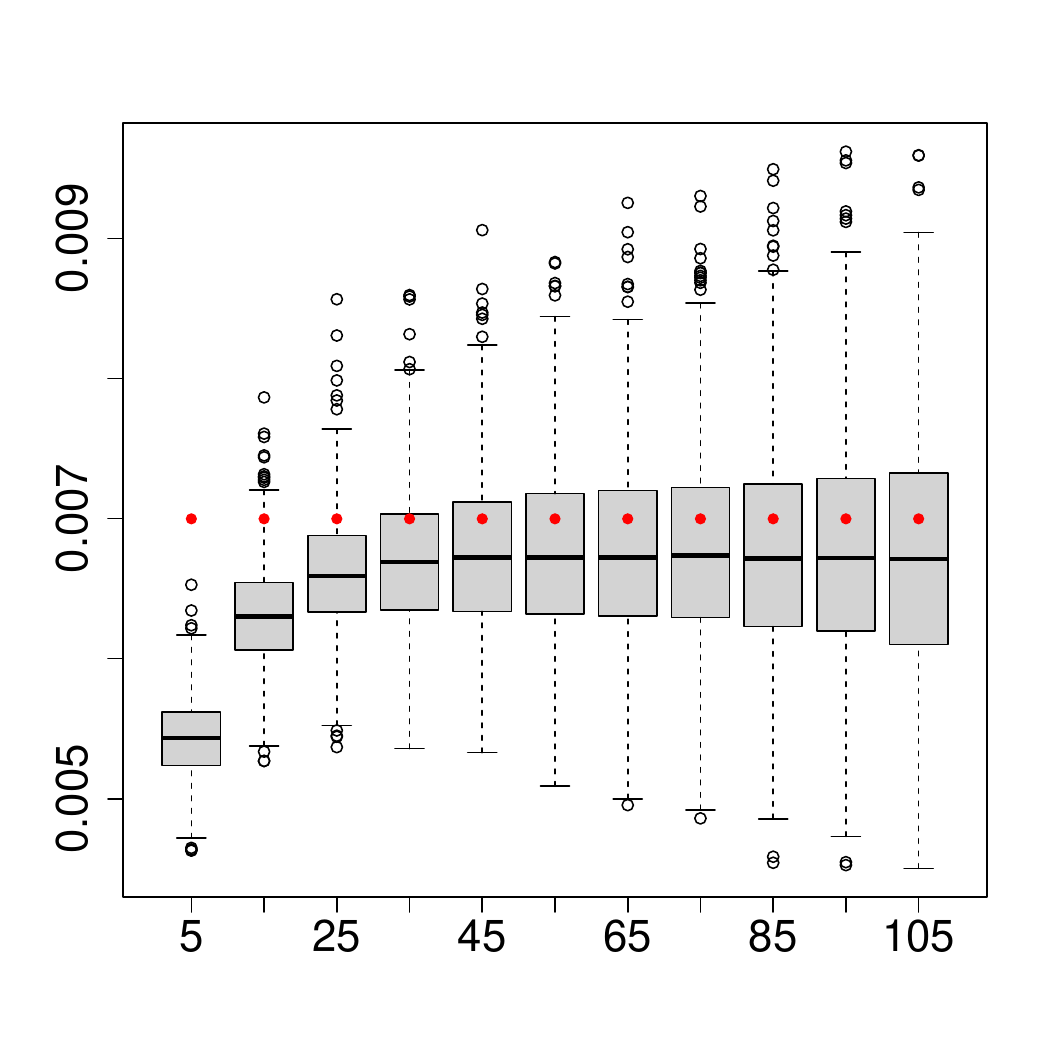}} &
        \adjustbox{valign=m,vspace=1pt}{\includegraphics[width=0.2\linewidth]{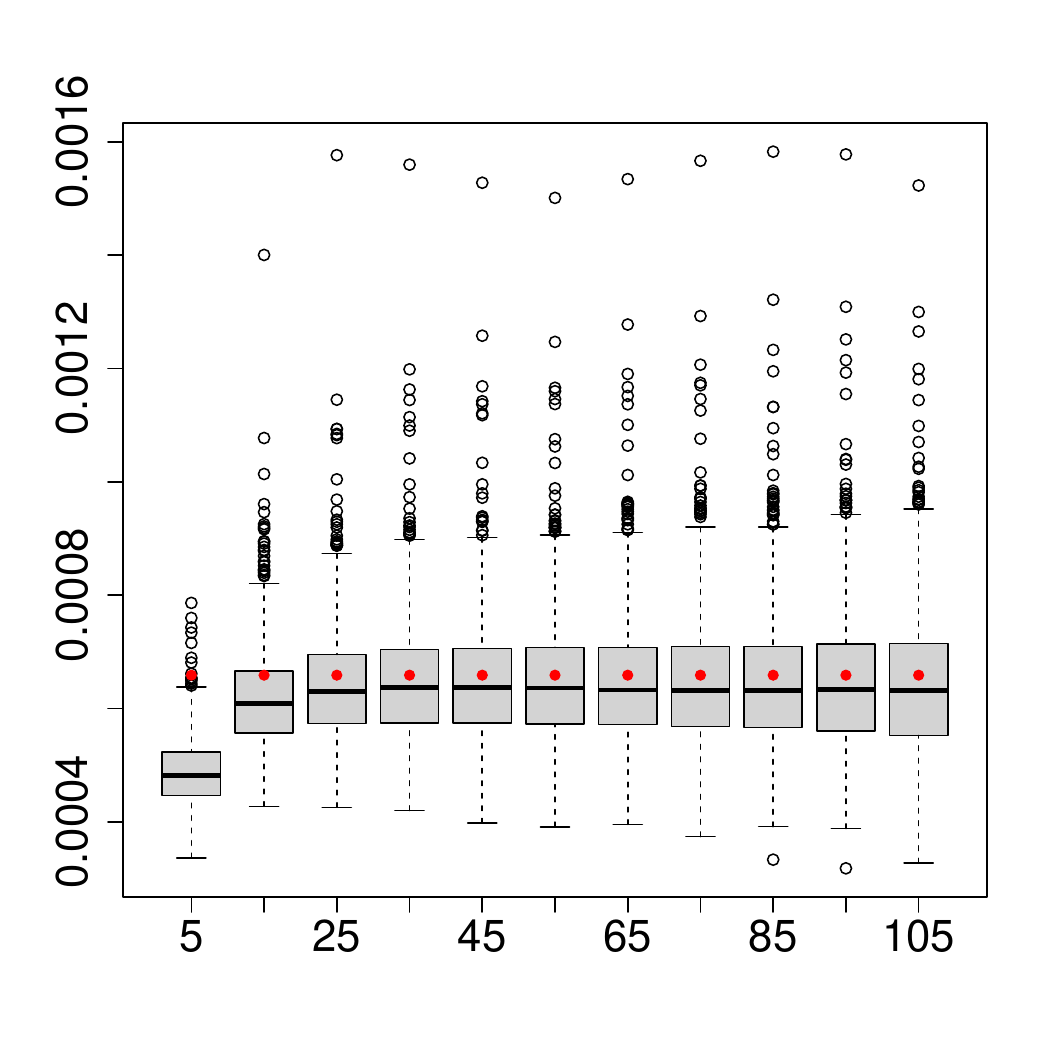}} &
        \adjustbox{valign=m,vspace=1pt}{\includegraphics[width=0.2\linewidth]{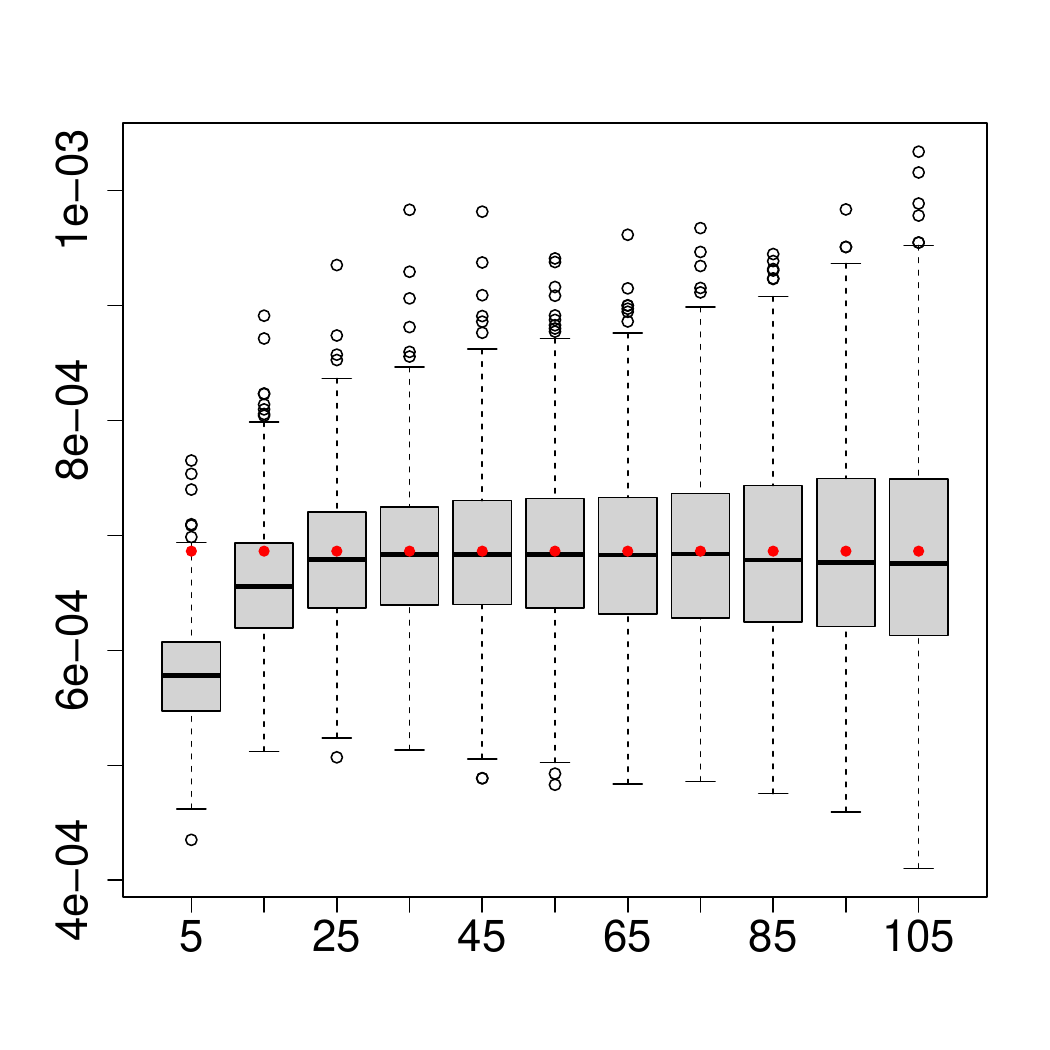}}\\[\bsl]
  \end{tabular}
  \caption{Boxplots of the estimated variances for the estimated death
    parameters using window $\widetilde W_{1}$ (upper row) and $\widetilde W_{2}$ (lower row) for
    different truncation distances. The red dots show the empirical variance of the simulated parameter estimates.}\label{fig:estvardeaths}
\end{figure}

Figure~\ref{fig:estvardeaths} shows boxplots of
estimates of the variances of
death parameter estimates
for different choices of truncation distances equally spaced between 5 and
155m.

\subsubsection{Posterior means estimates using  INLA}\label{sec:posteriorresultssimulation}

\begin{figure}
\centering
\begin{tabular}{ccccc}
     $\beta_{0\b}$ &$\beta_{1\b}$ & $\beta_{2\b}$ & $\gm_{1\b}$ & $\gm_{2\b}$ \\
        \adjustbox{valign=m,vspace=1pt}{\includegraphics[width=0.2\linewidth]{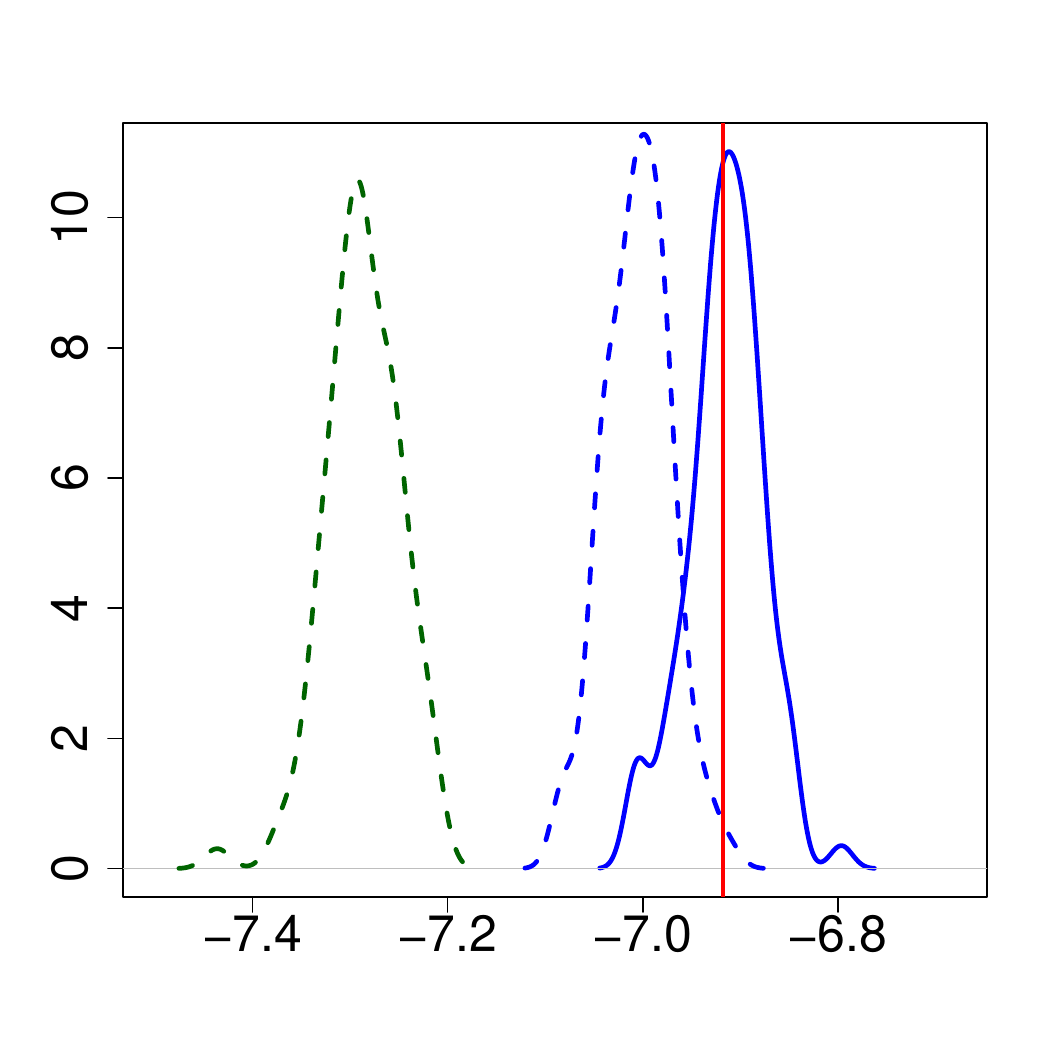}} & \adjustbox{valign=m,vspace=1pt}{\includegraphics[width=0.2\linewidth]{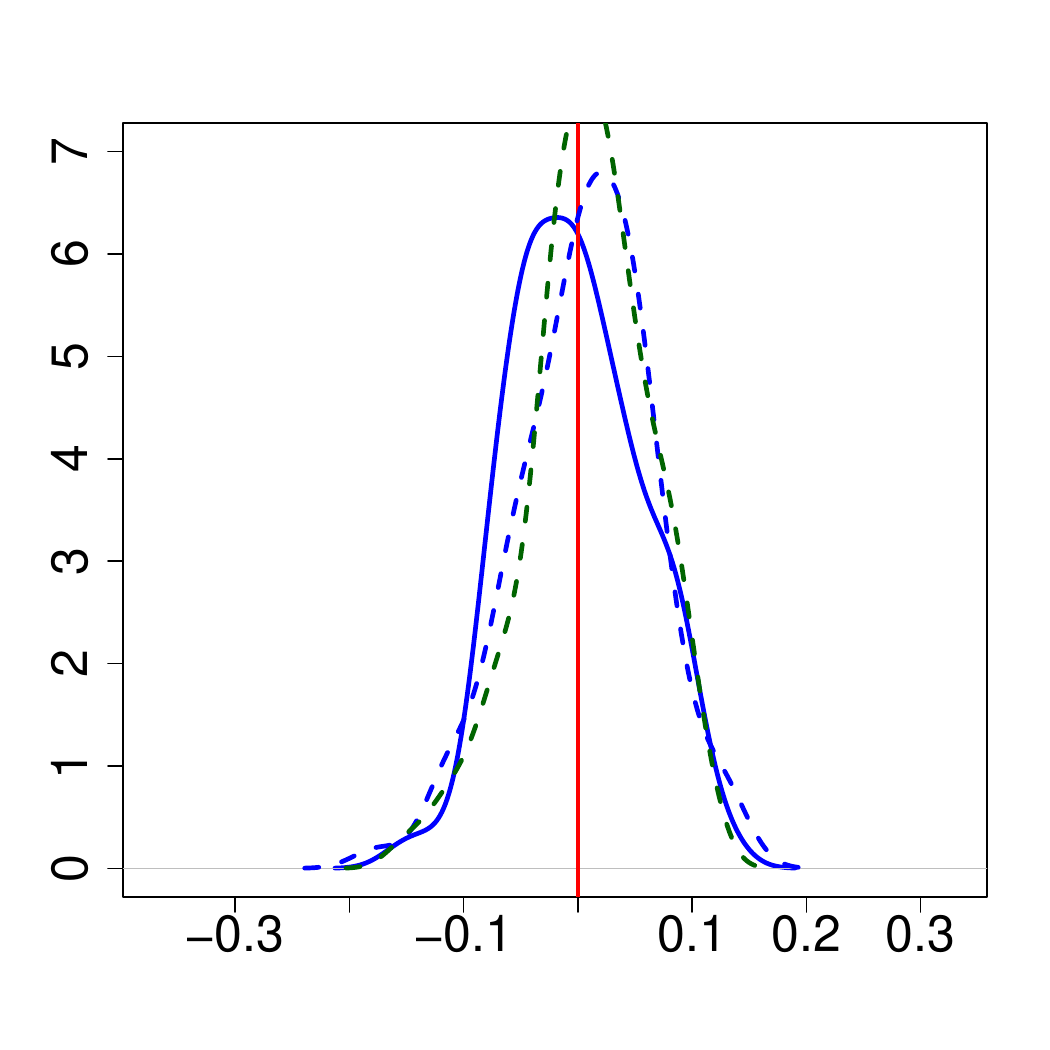}} &
        \adjustbox{valign=m,vspace=1pt}{\includegraphics[width=0.2\linewidth]{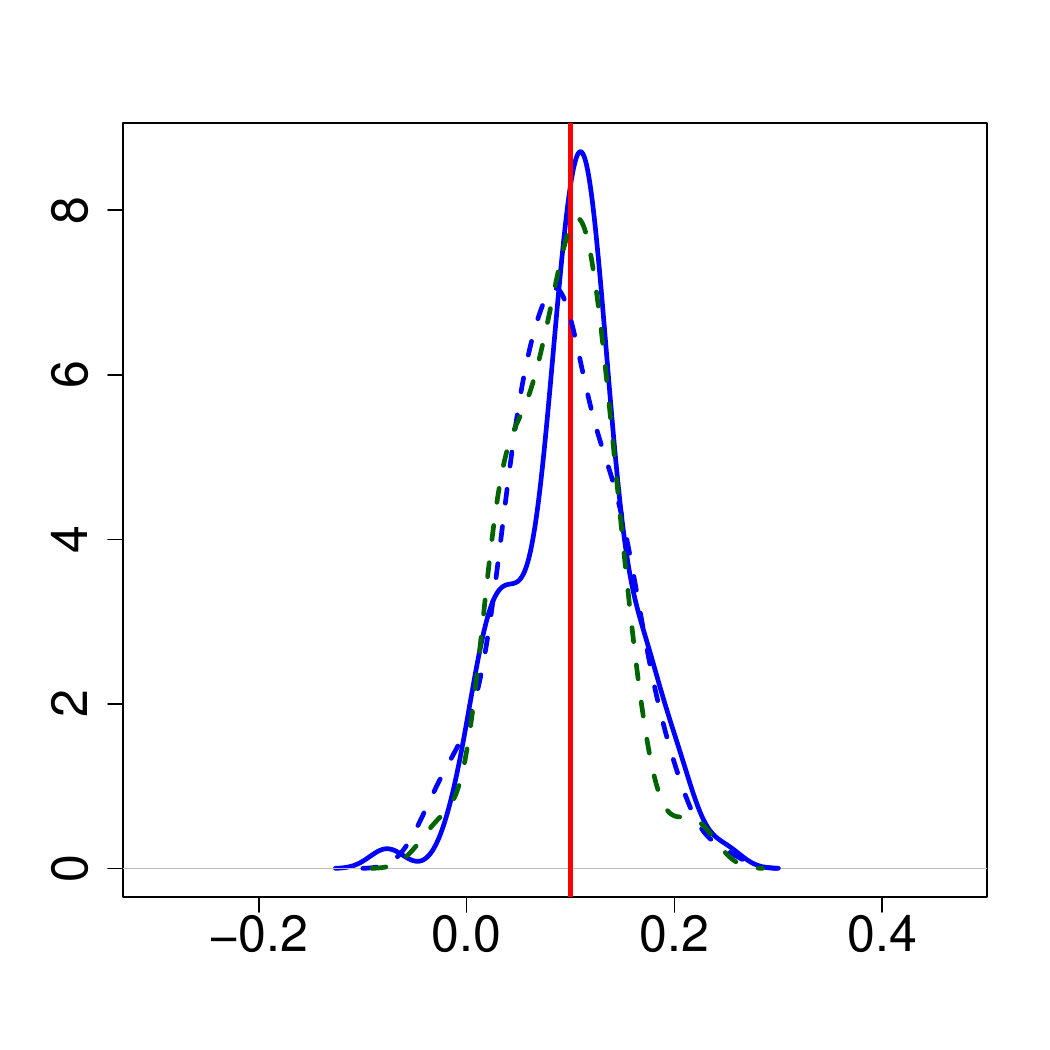}} &
        \adjustbox{valign=m,vspace=1pt}{\includegraphics[width=0.2\linewidth]{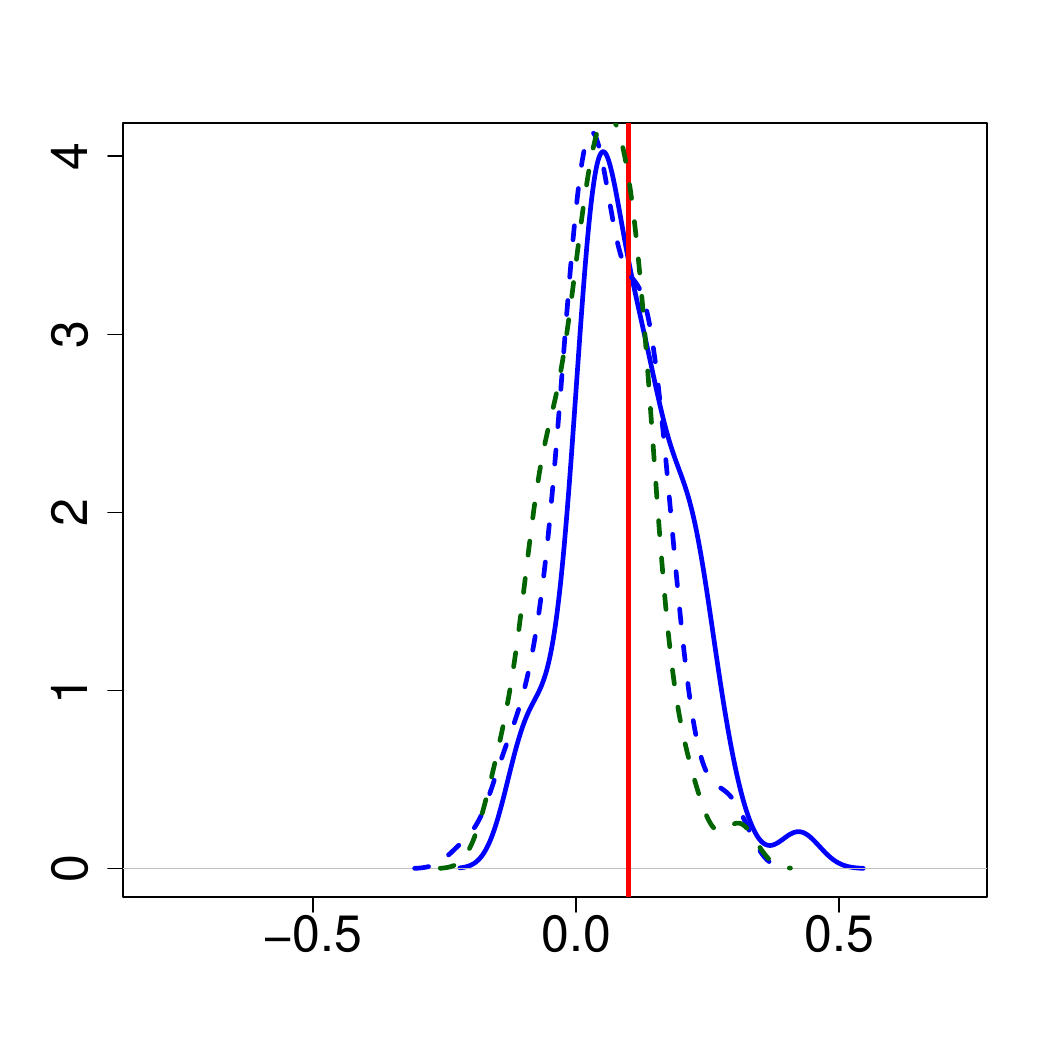}} & 
                                                                                                                         \adjustbox{valign=m,vspace=1pt}{\includegraphics[width=0.2\linewidth]{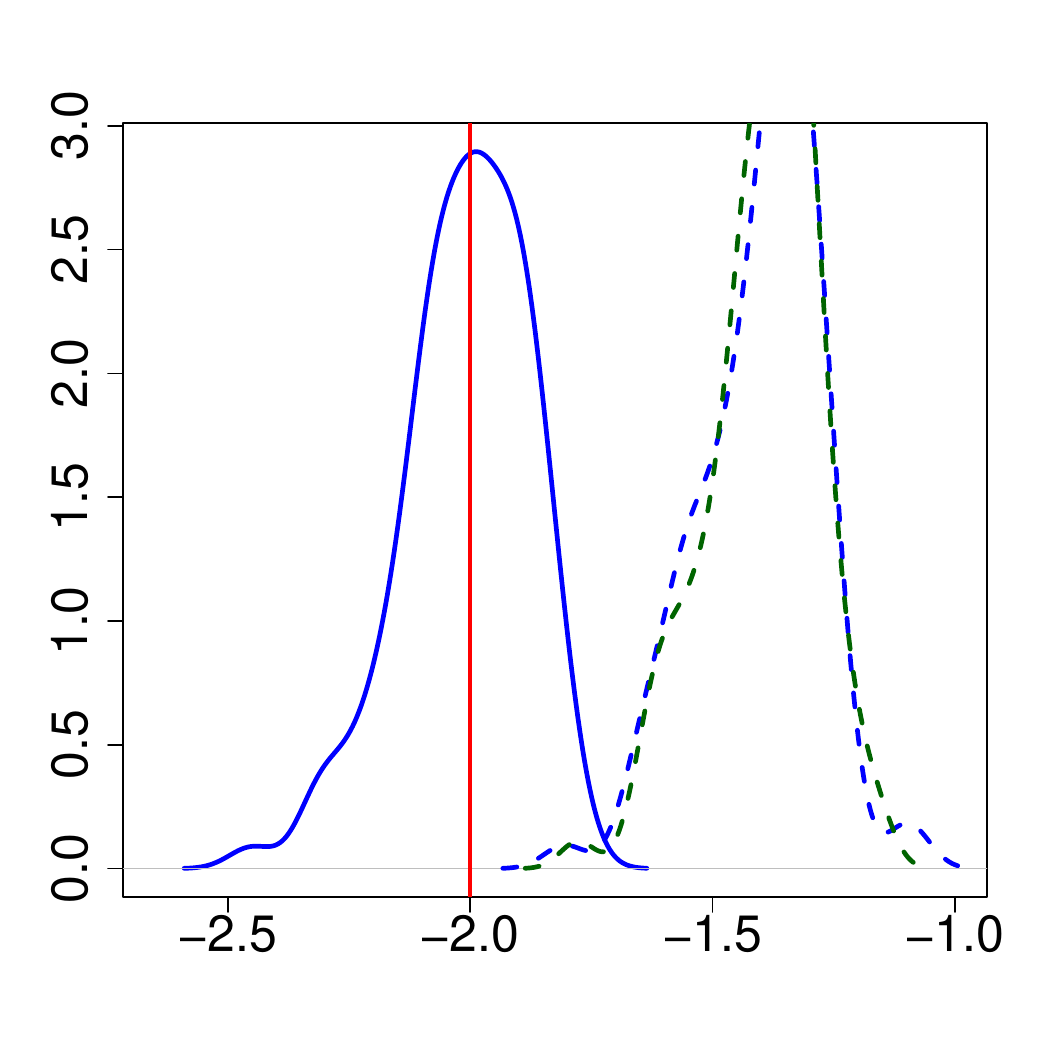}}\\
  $\beta_{0\dd}$ &$\beta_{1\dd}$ & $\beta_{2\dd}$ & $\gm_{1\dd}$ & $\gm_{2\dd}$ \\
        \adjustbox{valign=m,vspace=1pt}{\includegraphics[width=0.2\linewidth]{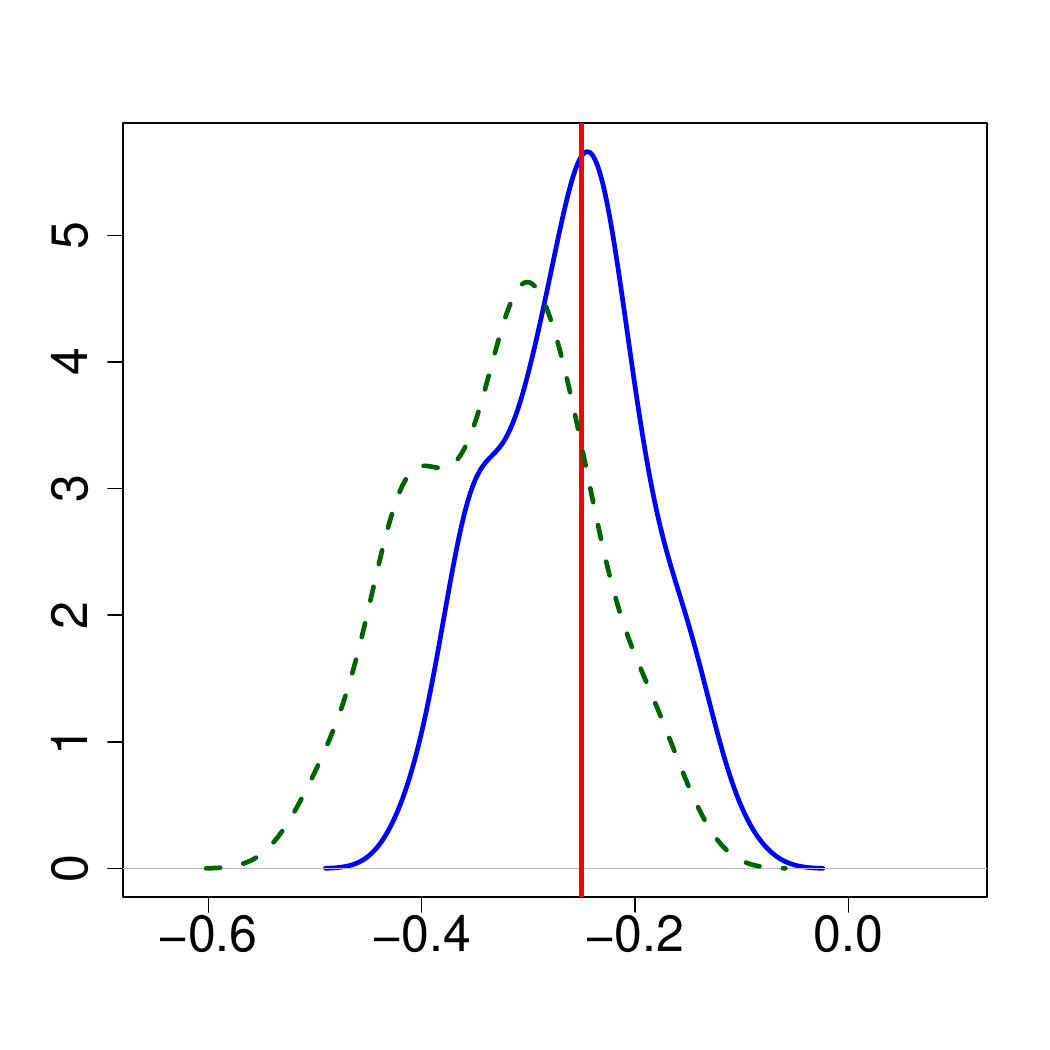}} &        \adjustbox{valign=m,vspace=1pt}{\includegraphics[width=0.2\linewidth]{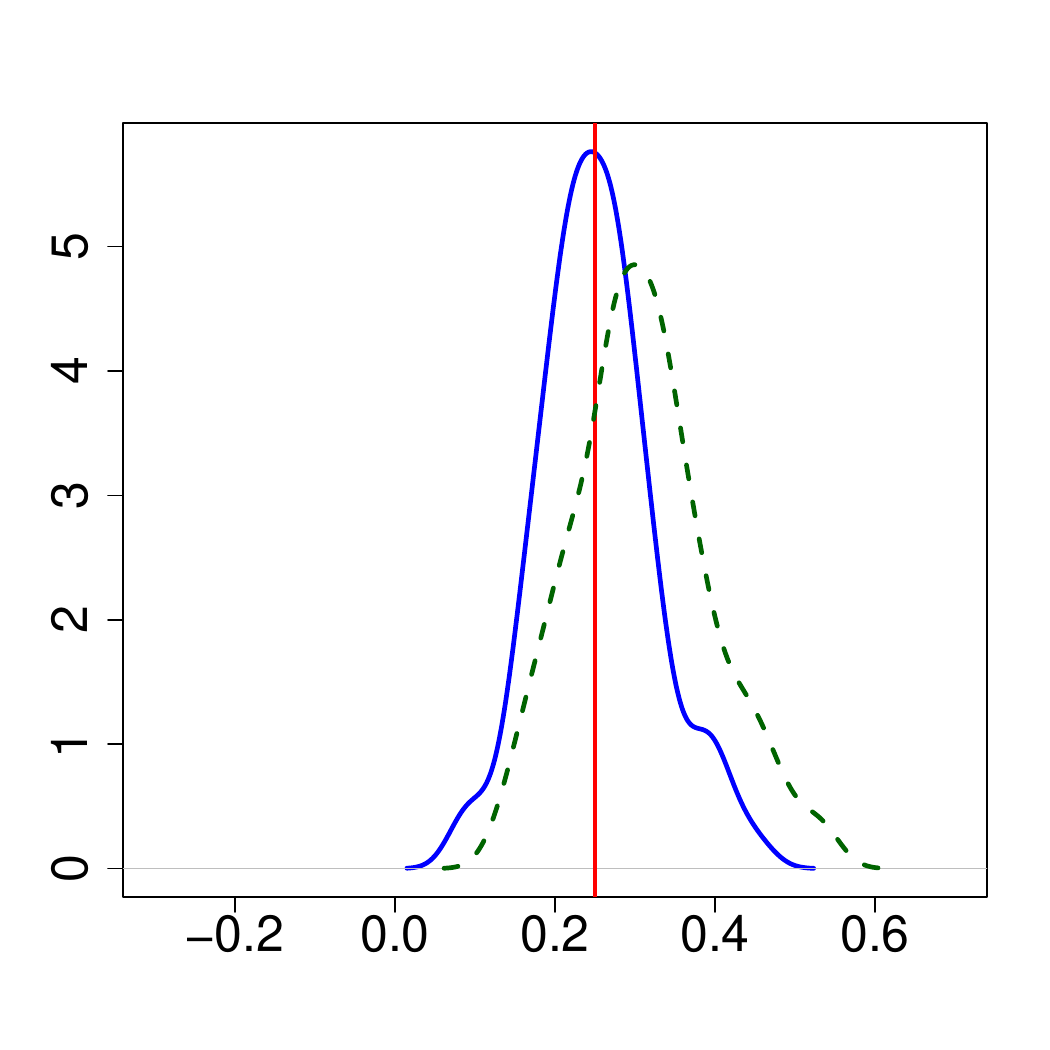}} &
        \adjustbox{valign=m,vspace=1pt}{\includegraphics[width=0.2\linewidth]{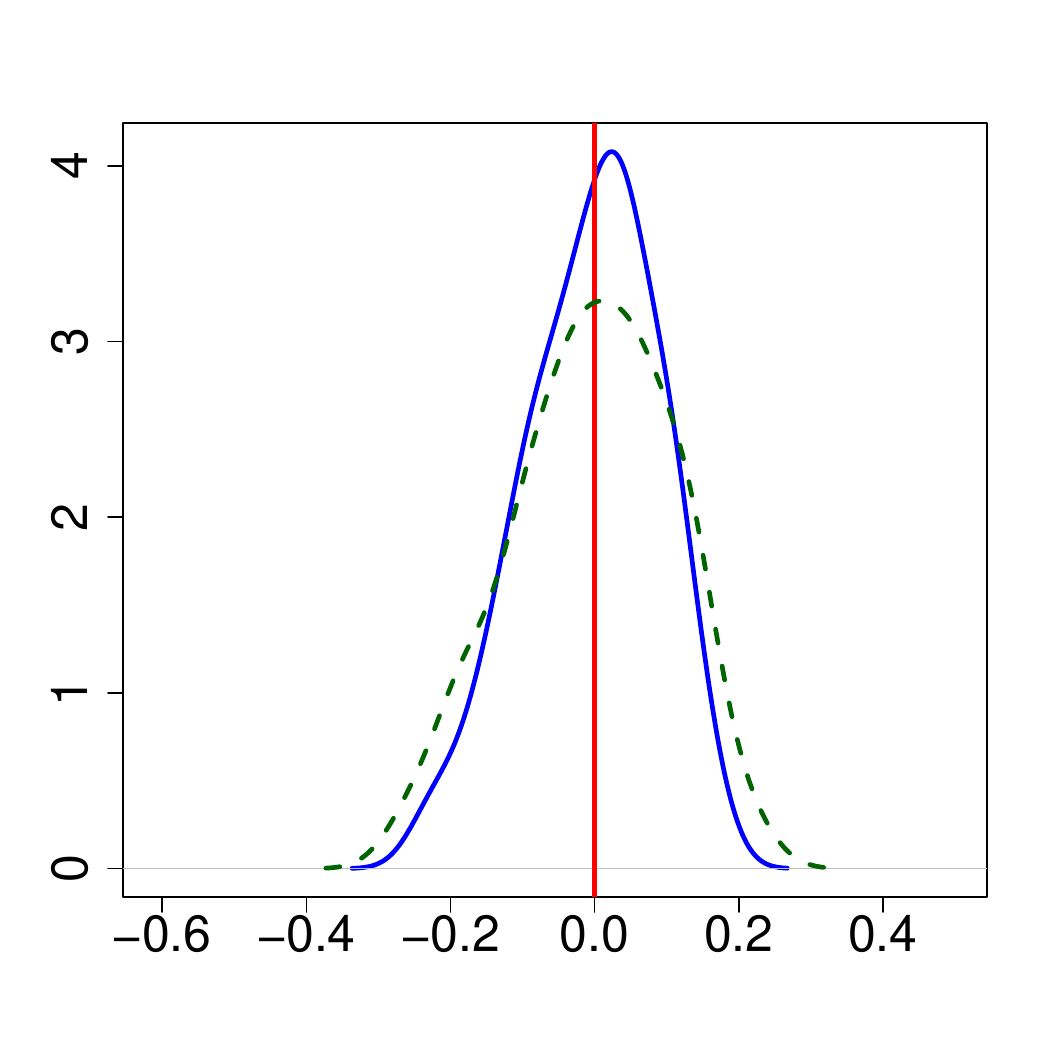}} &
        \adjustbox{valign=m,vspace=1pt}{\includegraphics[width=0.2\linewidth]{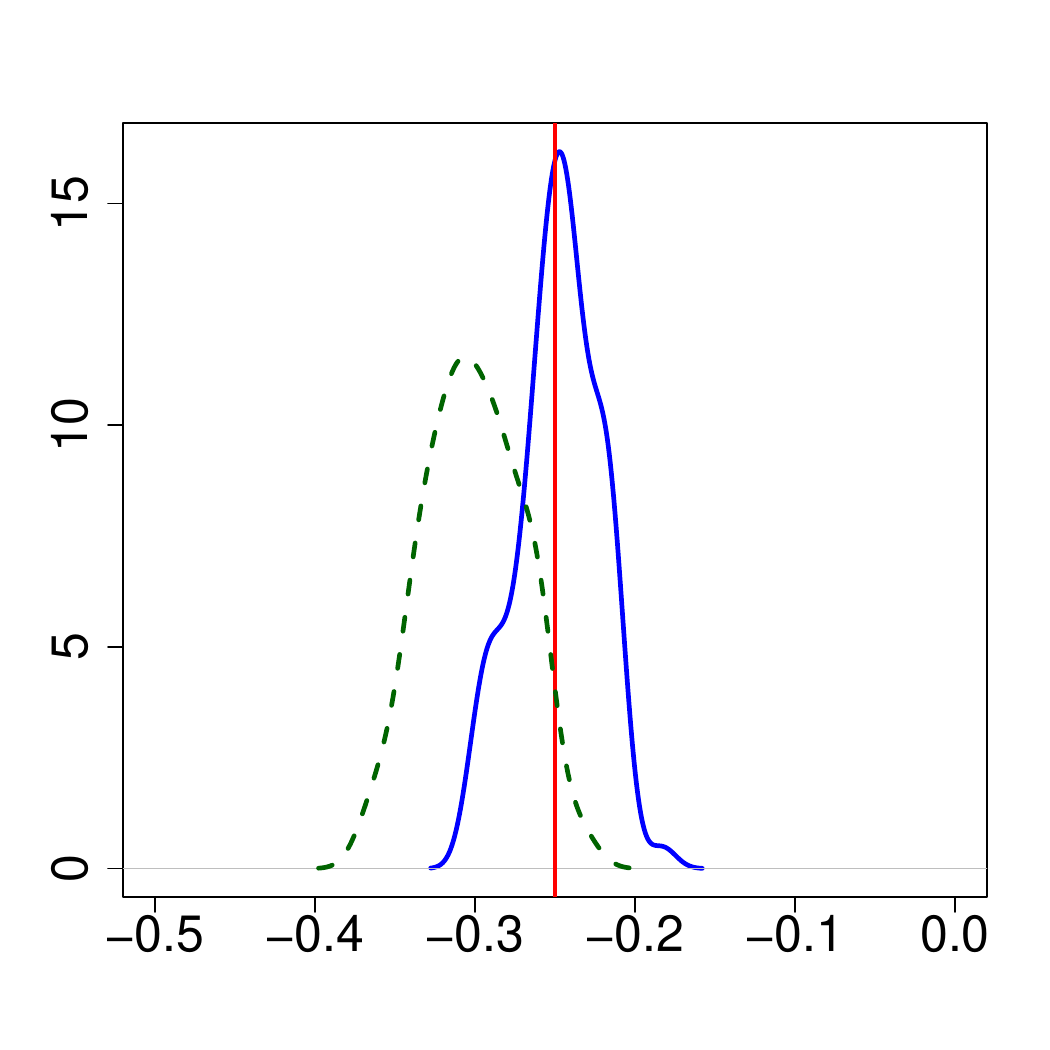}} &
        \adjustbox{valign=m,vspace=1pt}{\includegraphics[width=0.2\linewidth]{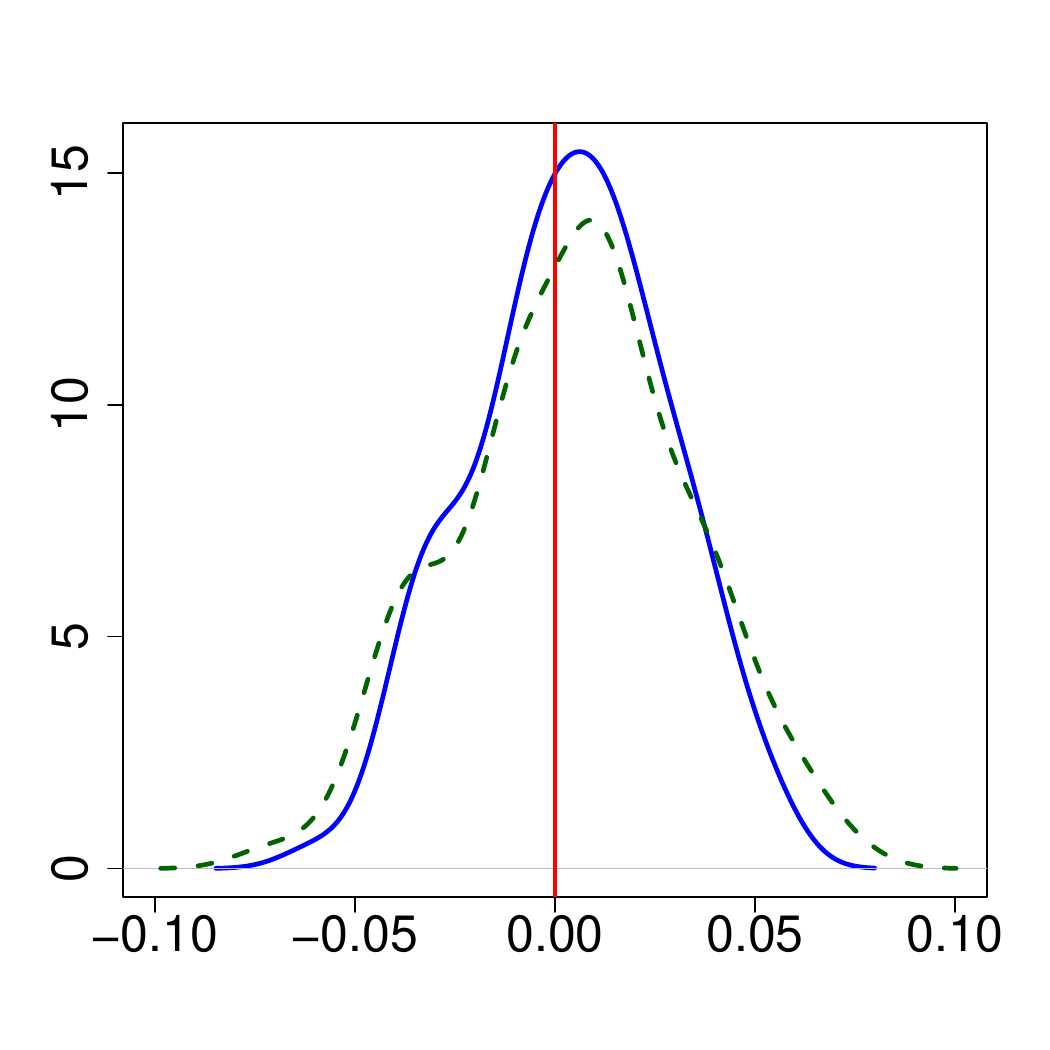}}\\[\bsl]
  \end{tabular}
  \caption{Kernel density estimates of the estimated parameters using
    composite likelihood (blue) and INLA (dark green) for recruits
    (upper row) and deaths (lower row) on the window $\widetilde W_2$. In case of recruits, the solid blue curve is for composite likelihood using high resolution covariates and the dashed blue curve is for composite likelihood using coarsened covariates on the 50$\times$100 grid used for the INLA.}\label{fig:INLA}
\end{figure}

Figure~\ref{fig:INLA} shows kernel density estimates of composite likelihood and INLA posterior mean estimates in case of the large window $\widetilde W_2$. For some parameters, the composite likelihood and INLA estimates agree very well being close to unbiased and with similar estimation variances. However, in case of recruits, the INLA posterior means are strongly biased for $\beta_{0\b}$ and $\gamma_{2\b}$ and some bias is also visible for the INLA estimates of the death parameters $\beta_{0\dd}$ and $\gamma_{1\dd}$. We conjectured that this could be due to discretization error when using the $50 \times 100$ grid for INLA. For the composite likelihood estimation of the recruits parameters we also tried out the coarsened covariates on the $50\times 100$ grid and the resulting estimates were biased in the same direction as the INLA estimates. For $\gamma_{2\b}$ in particular, the composite likelihood estimates with the $50\times 100$ grid agree very well with the INLA posterior means. For deaths, some bias can be expected for INLA, cf.\ the final remark in Section~\ref{sec:inla}.

\subsection{Supplementary figures for the  BCI data section}\label{sec:bciapplication}

Table~\ref{tab:census} summarizes the numbers of recruits and deaths in
each census. The population of {\em Capparis} trees seems to be
declining with a decreasing trend regarding number of recruits and increasing
trend regarding number of deaths.
\begin{table}
\centering
\caption{For each census conducted in the Barro Colorado Island plot:
  number of \emph{Capparis frondosa} trees, their mean diameter at
  breast height, and number of recruits and dead trees relative to the
  previous census.}
\label{tab:census}
\begin{tabular}{ccccc}
$k$  & no.\ trees & mean dbh & no.\ recruits & no.\ deaths  \\\hline
census 0 & 3536 & 21.2   & ---  & --- \\
census 1 & 3823 & 21.9  & 401 & 114 \\ 
census 2 &  3823 & 24.5  & 252 & 252  \\
census 3 &  3822 & 25.10  & 156 & 157  \\
census 4 &  3581 & 26.3  & 82  & 323  \\
census 5 &  3410 & 26.9  & 83  & 254  \\
census 6 &  3107 & 28.1  & 52 &  355 \\
census 7 &   2840 & 28.7  & 59 &  326 \\\hline
\end{tabular}
\end{table}

\subsection{Comparison with existing BCI analyses}\label{sec:existing}

The rich BCI data is exploited in a large number of ecological publications. Many of these concern mortality and recruitment from a wide
range of perspectives and using a variety of tools from the
statistical toolbox. Similar to our approach for mortality, several
papers consider binary observations of death/survival of trees within one or several time intervals
between BCI
censuses. These papers also consider a variety of spatial covariates including
covariates modeling positive or negative influence of conspecific and
heterospecific neighbours in the spirit of our stochastic covariates. \cite{gilbert2001effects,comita2009local,johnson2017abiotic,zhu2018density,zuleta2022individual}
are examples of papers that use logistic regression or similar generalized linear
models (GLMs).

The
first of these papers does not take into account spatial dependence
which may be justified by the small 1 ha study region (subset of BCI)
considered in that paper. The second to fourth papers acknowledge the need to
account for random spatial dependence. The second paper uses a block
bootstrap to estimate parameter standard errors that take into
account spatial dependence. The third and fourth
paper instead model random spatial variation in a generalized linear mixed model (GLMM) framework by
introducing independent random effects associated to  cells of a grid
partitioning the BCI study region. The fifth paper only includes species
specific random effects. The generalized linear mixed models with
spatial random effects can be
viewed as simple examples of our INLA model for mortality which has
spatio-temporally correlated random effects rather than independent
spatial random effects.

Issues with GLMMs and
bootstrap are computational expense and the need to choose grid
resolution for spatial random effects or bootstrap blocks. As an
alternative to random effects modeling, \cite{hubbell2001local} use an
autologistic model to model probabilities of tree death over one long
time-interval and implement approximate maximum likelihood estimation using Markov chain Monte
Carlo (MCMC). However, MCMC is computationally costly and entails issues with ascertaining convergence of MCMC samples.

Instead of logistic regression,
\cite{camac2018partitioning,chen2019effects} model death probabilities
in terms of an underlying continuous time hazard function (additive or
Weibull) and implement Bayesian inference using MCMC. The second of
these papers introduce spatial random effects as in the GLMMs
mentioned in the previous paragraph. Using underlying continuous time
hazard functions is advantageous if time-intervals of different
lengths are considered (for example, for BCI the first between census time interval is shorter than
the remaining 6). In principle, we could also parametrize our death
probabilities using an underlying continuous time hazard function but at
the expense of losing the computational advantages of logistic
regression.

Concerning recruitment, many BCI ecological papers define recruits in
the same manner as ours, as trees that emerge between two censuses.
\cite{wiegand2009recruitment,getzin2014stochastically}  use spatial point process summary
statistics to investigate associations between recruits and adults and
test independence by randomization of the recruits. Regression modeling is not used in these papers. In contrast
\cite{ruger2009response} use a negative binomial regression to model
effects of light availability on grid cell
counts of recruits with inference implemented using MCMC. Rather than
relying on grid cell counts with possible sensitivity to choice of
grid cell size we use the precise locations of recruits as in
\cite{wiegand2009recruitment,getzin2014stochastically}. However, we
study associations with adult trees within a regression modeling framework while avoiding the complexities of MCMC.

\label{lastpage}

\end{document}